\documentclass[a4paper,11pt]{article}
\usepackage{heppub}
\pdfoutput=1
\newcommand{\rescalethreeplots}{0.32\textwidth}

\newcommand{\abs}[1]{\lvert#1\rvert}
\newcommand{\ord}[1]{\mathcal{O}(#1)}

\newcommand{\df}{\mathrm{d}}
\newcommand{\img}{\mathrm{i}}

\newcommand{\GeV}{\,\mathrm{GeV}}
\newcommand{\TeV}{\,\mathrm{TeV}}

\newcommand{\nn}{\nonumber}

\newcommand{\bt}{{\vec b}_T}
\newcommand{\qt}{{\vec q}_T}

\newcommand{\tB}{\tilde{B}}
\newcommand{\tS}{\tilde{S}}

\newcommand{\tf}{\tilde{f}}
\newcommand{\talpha}{\tilde{\alpha}}

\newcommand{\as}{\alpha_s}
\newcommand{\Ecm}{E_\mathrm{cm}}
\newcommand{\GammaC}{\Gamma_\mathrm{cusp}}
\newcommand{\lqcd}{\Lambda_\mathrm{QCD}}

\newcommand{\zero}{{(0)}}

\newcommand{\nonp}{\mathrm{np}}

\newcommand{\scetlib}{{\textsc{SCETlib}}}

\newcommand{\minuit}{{\textsc{Minuit}}}
\newcommand{\dyturbo}{{\textsc{DYturbo}}}

\renewcommand{\arraystretch}{1.2}
\title{\boldmath Theory Uncertainties in the Extraction of $\alpha_s$ from Drell-Yan at Small Transverse Momentum}

\author[a]{Thomas Cridge,}
\emailAdd{thomas.cridge@uantwerpen.be}

\author[b]{Giulia Marinelli,}
\emailAdd{giulia.marinelli@desy.de}

\author[b]{and Frank J.~Tackmann}
\emailAdd{frank.tackmann@desy.de}

\affiliation[a]{Elementary Particle Physics, University of Antwerp,\\Groenenborgerlaan 171, 2020 Antwerp, Belgium}
\affiliation[b]{Deutsches Elektronen-Synchrotron DESY, Notkestr. 85, 22607 Hamburg, Germany}

\abstract{%
We perform a detailed pseudodata study to estimate the expected theory uncertainty
in the extraction of the strong coupling constant, $\as(m_Z)$, from a fit to the measured
Drell-Yan transverse momentum ($q_T$) spectrum at small $q_T \ll m_Z$. We consider two
approaches to estimate the dominant perturbative uncertainties. We first discuss that
the traditional approach based on varying unphysical scales is insufficient here
because it cannot correctly account for bin-by-bin theory correlations
in the $q_T$ spectrum, which are critically important in this case.
We then use this case as a nontrivial application of a new approach based on
theory nuisance parameters (TNPs), which encodes
the correct theory correlations by construction. Moreover, the TNPs can be profiled
in the fit thereby allowing the data to constrain the theory uncertainties
in a consistent manner.
We furthermore discuss the interplay with nonperturbative effects in the peak
region $q_T \lesssim 10\GeV$, from where most of the $\as$ sensitivity originates.
The associated nonperturbative uncertainties on $\as$ when fitting only the $q_T$ spectrum are
large. They can in principle be reduced by including additional constraints on
the nonperturbative Collins-Soper kernel from lattice QCD calculations.
We find that these improvements in the treatment of perturbative and nonperturbative
uncertainties and their correlations will enable a competitive $\as$ extraction from
Drell-Yan data at small $q_T$.
We also discuss the implications of our findings, calling into question a recent
$\as$ extraction from the $Z$ $q_T$ spectrum by the ATLAS experiment.
}

\date{June 16, 2025}

\preprint{\vbox{%
\hbox{DESY-25-049}}
}

\arxivnumber{2506.13874}

\begin{document}

\maketitle

\section{Introduction}
\label{sec:intro}

The strong coupling constant $\alpha_s$, which parameterizes the strength of the strong
interactions, is the least well known among the fundamental forces of nature.
Its precise determination has thus been of great interest with many complementary
determinations from different sources being pursued over the years~\cite{dEnterria:2022hzv}.
The current world average of $\alpha_s(m_Z) = 0.1180(9)$~\cite{ParticleDataGroup:2024cfk} has
a relative uncertainty of $0.8$\%, with the most
precise result stemming from the average of lattice QCD determinations
$\alpha_s(m_Z) = 0.1183(7)$ ($0.6\%$)~\cite{FlavourLatticeAveragingGroupFLAG:2024oxs}.

One of the currently most precise determinations from experimental measurements at colliders comes from
fits to hadronic event shapes in $e^+e^-$ collisions~\cite{Kluth:2006bw}, most notably
a determination from the thrust distribution~\cite{Abbate:2010xh, Abbate:2012jh, Benitez:2024nav}
yielding $\alpha_s(m_Z) = 0.1136(12)$ with a precision of about 1\%,
which however has shown a persistent tension with the world average, despite much
theoretical scrutiny~\cite{Becher:2008cf, Hoang:2015hka, Caola:2022vea, Nason:2023asn, Nason:2025qbx, Bell:2023dqs, Aglietti:2025jdj, Benitez:2024nav, Benitez:2025vsp}. An independent determination of similarly high precision at or below the one-percent level from collider measurements is thus of great importance.
A promising candidate for such a high-precision determination is the transverse-momentum ($q_T$)
spectrum in Drell-Yan production, which has been measured to subpercent precision at the
LHC~\cite{Aad:2014xaa, Aad:2015auj, Aad:2019wmn, ATLAS:2023lsr, Chatrchyan:2011wt, Khachatryan:2016nbe, Sirunyan:2019bzr, LHCb:2015mad, LHCb:2016fbk}
and for which theoretical predictions have been reaching the same N$^3$LL$'$ and (approximate) N$^4$LL
resummation as for $e^+e^-$ event shapes resulting in a perturbative
precision at the few-percent level~\cite{Billis:2021ecs, Ju:2021lah, Re:2021con, Chen:2022cgv, Neumann:2022lft, Camarda:2023dqn, Moos:2023yfa, Billis:2024dqq}.

The sensitivity to $\as$ in the $q_T$ spectrum arises solely from higher-order
QCD corrections. A robust theoretical description of the $q_T$
spectrum is thus mandatory to obtain a reliable extraction of $\as(m_Z)$.
It requires in particular a reliable treatment of theory uncertainties and
their correlations.
The dominant $\as$ sensitivity is a subtle shape effect in the peak of the spectrum --
a change in $\as$ effectively shifts the peak of the spectrum.
Since the $\as$ dependence is intimately tied to the perturbative expansion itself,
missing higher-order terms can closely mimic the effect of changing the value of $\as$.
Furthermore, nonperturbative effects induce a similar shape effect in the peak
of the spectrum. Hence, for a reliable $\as$ determination it is vital to reliably
distinguish the shape effects induced by changing $\as$
from those induced by missing higher-order corrections or by nonperturbative effects,
so we must be able to correctly account for the latter two. This means
it is critically important to account for the correct point-by-point (or bin-by-bin)
theory correlations in the $q_T$ spectrum, which are determined by the precise shape
of the individual components of the theory uncertainty.

The traditional approach to estimate perturbative theory uncertainties based on
scale variations suffers from well-known limitations: The estimated uncertainties
are not very meaningful, and in particular they fundamentally lack a proper account of
theory  correlations. As discussed in \sec{scale_variations}, the underlying reason
is that the scales that are being varied are unphysical and not actual parameters
of the calculation. Their variation cannot be interpreted like that of an ordinary
parameter whose uncertainty is being propagated. One consequence is that scale
variations do not provide the correct shape of the missing higher-order terms.
Given the critical importance of accounting for the correct point-by-point theory correlations,
the scale-variation approach is basically ill-suited for this application.
We find that the perturbative uncertainty on $\as(m_Z)$ one might derive
based on scale variations is indeed not reliable and quite arbitrary due to it
being subject to uncontrolled correlation assumptions.
For example, a recent extraction from Drell-Yan measurements at small $q_T$
claimed a record precision of $\as(m_Z) = 0.1183(9)$~\cite{ATLAS:2023lhg},
as precise as the current world average. However, as discussed in detail in
\app{atlas}, it relies on a particular scale-variation recipe as well as other
questionable assumptions, so the quoted uncertainty
on $\as(m_Z)$ cannot be taken at face value.

The new approach of theory nuisance parameters (TNPs) introduced in \Refcite{Tackmann:2024kci}
overcomes the limitations of scale variations and provides meaningful theory
uncertainties including in particular correct theory correlations.
The purpose of our paper is thus twofold:
Our first goal is to demonstrate the performance and robustness of TNPs in
a real-world application that demands meaningful theory uncertainties and
correlations. We find that the TNPs perform as advertised.
Our second goal is to estimate
the theory uncertainties in $\as(m_Z)$ we can expect to achieve. To do so we
perform a detailed study using so-called Asimov fits to (unfluctuated) pseudodata.
This is a standard procedure to estimate expected uncertainties
in a clean and controlled environment that is unobscured by statistical fluctuations in the data
and unbiased by subleading effects that are present in the real data and not (yet) accounted
for in the theoretical description. In our case, it allows us to study in a theoretically
fully controlled environment the
dominant perturbative and nonperturbative uncertainties in the $q_T$ spectrum
and their propagation to and impact on $\as(m_Z)$.
In particular, by profiling the TNPs in the fit and by incorporating
lattice QCD constraints on nonperturbative parameters, we expect that it will be possible in the
future to achieve sufficiently small theory uncertainties to allow
for an extraction of $\as(m_Z)$ from the Drell-Yan $q_T$ spectrum that is
competitive with other measurements.
It remains to be seen whether or not it will be possible to reach
or beat the precision of lattice QCD determinations or the world average.

We use the binning and experimental
uncertainties of the recent ATLAS 8 TeV inclusive measurement~\cite{ATLAS:2023lsr}
as a representative example for our Asimov study. This choice is partially motivated
by the fact that the same dataset was used in \Refcite{ATLAS:2023lhg}.
Our exact numerical results of course depend on this choice
and will change by including other datasets. However, our overall qualitative
findings on the robustness (or lack thereof) and performance of the different
theory uncertainty approaches hold in general.

The dominant perturbative and nonperturbative uncertainties we wish to address
enter via the dominant leading-power resummed
contribution to the small-$q_T$ spectrum, which is therefore the focus of our
investigations. In the appendix, we also briefly consider subdominant effects from
power corrections suppressed by $\ord{q_T^2/m_Z^2}$ and finite quark masses
relevant for $q_T^2 \sim m_c^2, m_b^2$. While it is important to
include these effects to obtain a percent-level or better description of the
real data, we can drop them in our main study since we consistently drop them in both
our pseudodata and our fitted theory model.

Another important source of uncertainty is due to the limited knowledge of parton distribution functions (PDFs), which are required inputs for such hadron collider determinations of $\as(m_Z)$. The PDFs themselves have a dependence on $\as$~\cite{Hou_2021,Forte:2020pyp,Ball_2022,Cridge:2021qfd,Cridge:2024exf}, and this non-trivial PDF-$\as$ correlation must be propagated faithfully to the extracted output $\as(m_Z)$. Whilst the propagation of these uncertainties is in principle
a more straightforward exercise, in practice it faces some nontrivial caveats. We therefore
do not consider them here but will address them in a dedicated paper~\cite{pdf_paper}.

The remainder of this paper is organized as follows.
In \sec{theory_unc}, we discuss in more detail the general issue of theory uncertainties
and their correlations for differential spectra, the limitations of scale variations,
and the general ideas behind the TNP approach.
In \sec{theory}, we summarize our theory and pseudodata setup.
We then discuss in \sec{perturbative} the treatment of perturbative uncertainties in the extraction of
$\as$ and including nonperturbative uncertainties in \sec{nonperturbative}.
We conclude in \sec{conclusions}.
\App{atlas} provides a detailed discussion of the estimation of theory uncertainties
in \Refcite{ATLAS:2023lhg} in comparison to our own analysis.
In \app{addresult} we collect some additional numerical results.

\section{Theory Uncertainties and Correlations for Differential Spectra}
\label{sec:theory_unc}

In this section, we provide a brief general discussion of perturbative theory
uncertainties for the case of a differential spectrum. The discussion here is
largely adapted from sections 2 and 3 of \Refcite{Tackmann:2024kci}, to which we
refer for a more in-depth discussion. In \sec{correlations}, we highlight the
importance of theory correlations. In \sec{scale_variations} we discuss the
limitations of scale variations, and in \sec{tnps} we discuss the approach of
theory nuisance parameters.

We consider a differential quantity $f(x)$ that depends on some variable $x$,
and its perturbative expansion in a small parameter $\alpha$,
\begin{equation} \label{eq:f_series}
f(x, \alpha) = f_0(x) + f_1(x)\,\alpha + f_2(x)\,\alpha^2 + f_3(x)\,\alpha^3 + \ord{\alpha^4}
\,.\end{equation}
We will generically use $f_n(x)$ to denote the expansion coefficients and specifically
$\hat f_n(x)$ to denote their true values. Calculating the first few true coefficients,
we obtain a theory prediction for $f(x)$ at leading order (LO), next-to-leading order
(NLO), next-to-next-to-leading order (NNLO), and so on,
\begin{alignat}{9} \label{eq:f_pred}
\text{LO:}\qquad && f(x, \alpha) &= \hat f_0(x)
\,,\nn\\
\text{NLO:}\qquad && f(x, \alpha) &= \hat f_0(x) + \hat f_1(x)\,\alpha
\,,\nn\\
\text{NNLO:}\qquad && f(x, \alpha) &= \hat f_0(x) + \hat f_1(x)\,\alpha + \hat f_2(x)\,\alpha^2
\,.\end{alignat}
The perturbative theory uncertainty we wish to discuss in the following is due
to the fact that the predictions are only approximations to the exact result
because of truncating the series at a certain order.

\subsection{Theory correlations}
\label{sec:correlations}

Theory correlations, namely the correlations in the theory uncertainties of
different predictions, are in principle required whenever several
predictions are used at the same time, for example when
one performs a simultaneous interpretation of several measurements. A standard
example is the interpretation of a differential spectrum, which requires
point-by-point (or bin-by-bin) theory correlations across the spectrum.

\begin{table}
\centering
\renewcommand{\arraystretch}{1.3}
\begin{tabular}{c|cccc}
\hline\hline
$\rho$ & 99.5\% & 98\% & 95.5\% & 87.5\%
\\
$\delta_{f/g}/\delta$ & $0.1$ & $0.2$ & $0.3$ & $0.5$
\\\hline\hline
\end{tabular}
\caption{Dependence of the relative uncertainty of the ratio $f/g$ on
the correlation $\rho$, see text for details.}
\label{tab:correlations}
\end{table}

To appreciate the potential importance of correlations, first consider the case of
two observables $f$ and $g$ that have both a relative
uncertainty $\delta_f = \delta_g = \delta$ with correlation $\rho$. The relative uncertainty of their ratio,
$\delta_{f/g}$, as a function of $\rho$ is given by
\begin{equation}
\delta_{f/g} = \delta \sqrt{2(1-\rho)}
\,.\end{equation}
When $\rho$ is close to 1, i.e., in the limit of strong correlation, $\delta_{f/g}$
is very sensitive to the precise
value of $\rho$, as illustrated in \tab{correlations}, because the square root
becomes infinitely steep for $\rho \to 1$.

The different points (or bins) of a differential spectrum $f(x)$ are a priori separate
observables. When obtaining a theory prediction for $f(x)$ we
are actually obtaining predictions simultaneously for many separate (though closely related)
observables $f(x_i)$. The question of how well the \emph{shape} of $f(x)$
is known then corresponds to the question of how the theory uncertainties
at different points in the spectrum are correlated. That is, $f \equiv f(x_i)$ and $g \equiv f(x_j)$
now correspond to the prediction of the spectrum at any two points $x_i$ and $x_j$,
and the shape of the spectrum is equivalent to the ratio $f(x_i)/f(x_j)$.

As an illustration, consider the case where we have the same relative uncertainty $\delta$
at all points. Let $\rho_{ij}$ denote the correlation between the uncertainties
of $f(x_i)$ and $f(x_j)$. In the extreme case
where the uncertainties at all points are 100\% correlated, $\rho_{ij} = 1$,
the shape of the spectrum would be known exactly with vanishing uncertainty
and $\delta$ would correspond to a pure overall normalization uncertainty.
In the other extreme where the uncertainties
are completely uncorrelated among all points, $\rho_{ij} = 0$, the shape would be completely unknown
within an overall band of relative size $\delta$. In practice, the theory predictions for
neighboring points (or bins) are closely related, so we naturally expect their
uncertainties to be strongly correlated.
We are thus precisely in the limit of strong correlations, and so the precise
point-by-point (or bin-by-bin) correlations are crucial to correctly account for
the shape uncertainty. Just like for the ratio $f/g$ in \tab{correlations}, tiny
differences in the
point-by-point correlations can lead to dramatically different shape uncertainties.

It is important to realize that different quantities, including neighboring points or bins
in a spectrum, do not by themselves have a notion of being correlated with each other.
The only thing that can be correlated are their uncertainties. More precisely,
the impact of a common source of uncertainty is fully correlated between different
quantities that depend on that same source. This is fundamentally the only way a
correlation can arise. For example,
neighboring bins measured by strictly independent experiments will have
uncorrelated experimental uncertainties, because they do not share a common source
of uncertainty.
Since two neighboring bins tend to be very similar observables, their theory predictions
tend to involve common ingredients with a similar impact on both bins.
This is the (only) reason why we expect their theory uncertainties
to be strongly correlated.

More generally, when different quantities depend on several independent
sources of uncertainty, the correlation of their total uncertainty depends on the
relative size of the various fully correlated impacts from each source. In terms
of covariance matrices, the total covariance matrix is the sum of several
100\% correlated ones, which is in general not 100\% correlated anymore.
It then follows that to obtain the correct theory correlations for a differential spectrum
we have to break down the total theory uncertainty
into (mutually independent) components, where each component
corresponds to a well-defined common source of uncertainty so it is fully correlated
across the spectrum. The shapes of the components, reflecting the impact of their respective
sources across the spectrum, can be different and their combination then
determines the net point-by-point correlations of the total theory uncertainty.
As we will discuss in \sec{tnps}, theory nuisance parameters are constructed exactly
to provide this breakdown. In contrast, as discussed next, scale variations
do not provide the means for such a breakdown.

\subsection{Limitations of scale variations}
\label{sec:scale_variations}

The series coefficients in \Eq{f_series} depend on the precise way of performing
the expansion, i.e., the perturbative scheme we use, which corresponds to the precise
choice or definition of the expansion parameter $\alpha$.
We can define a new scheme by defining a different expansion parameter $\talpha$,
\begin{equation} \label{eq:talpha}
\talpha(\alpha)
= \alpha \bigl[1 + b_0\,\alpha + b_1\,\alpha^2 + b_2\,\alpha^3 + \ord{\alpha^4} \bigr]
\,,\end{equation}
which is fully specified by the coefficients $b_k$.
For scale variations in QCD, $\alpha \equiv \alpha_s(\mu_0)$ corresponds to the strong coupling at
the chosen central scale $\mu_0$, $\talpha \equiv \alpha_s(\mu)$ is the
coupling at some other scale $\mu$, and the $b_k$ are given in terms of the QCD
beta function coefficients $\beta_k$ as
\begin{alignat}{9} \label{eq:bk_L}
b_0 &= \frac{\beta_0}{2\pi} \ln\frac{\mu_0}{\mu}
&&= 0.85\, L
\,,\nn\\
b_1
&= \frac{\beta_0^2}{4\pi^2} \ln^2\frac{\mu_0}{\mu}
   + \frac{\beta_1}{8\pi^2} \ln\frac{\mu_0}{\mu}
&&= 0.72\,L^2 + 0.34\,L
\,,\nn\\
b_2
&= \frac{\beta_0^3}{8\pi^3} \ln^3\frac{\mu_0}{\mu}
   + \frac{5\beta_0\beta_1}{32\pi^2} \ln^2\frac{\mu_0}{\mu}
   + \frac{\beta_2}{32\pi^3} \ln\frac{\mu_0}{\mu}
&&= 0.61\,L^3 + 0.72\,L^2 + 0.13\,L
\,.\end{alignat}
For illustration, in the second equalities on the right-hand side we plugged in
$n_f = 5$ and used $L \equiv \ln(\mu_0/\mu)/\ln 2$, so the usual convention
corresponds to varying $L = \pm 1$.

As long as the $b_k$ in \Eq{talpha} are $\ord{1}$, $\talpha$ only differs from $\alpha$ by
higher-order terms and should thus provide a similarly good expansion parameter.
Using $\talpha$ to perform the perturbative expansion, we obtain different
truncated predictions in terms of different expansion coefficients $\tilde f(x)$.
For example, at NLO we have
\begin{alignat}{9} \label{eq:tf_pred}
\text{NLO:}\quad && \tf(x, \talpha)
&= \hat \tf_0(x) + \hat \tf_1(x)\,\talpha
\nn\\ &&
&= \hat f_0(x) + \hat f_1(x)\, \alpha + b_0 \hat f_1(x)\,\alpha^2 + b_1 \hat f_1(x)\,\alpha^3 + \ord{\alpha^4}
.\end{alignat}
In the second line we rewrote the predictions in terms of the original $\hat f_n(x)$
and $\alpha$. As expected, to the order one is working, the predictions in different
schemes agree but they differ by higher-order terms beyond the nominal working order.

To all orders, the different expansions must give exactly the same result,
$f(x,\alpha) = \tilde f(x, \talpha)= f(x)$. The dependence on the scheme is thus
an artifact of truncating the series at a finite order.
In the scale-variation approach, the residual scheme differences
are then exploited to provide an estimate of the theory uncertainty by
taking $\Delta f(x,\alpha) = \tf(x,\talpha) - f(x, \alpha)$. For example, at NLO and NNLO
this yields
\begin{alignat}{9} \label{eq:Deltaf}
\text{NLO:}\qquad &&
\Delta f(x, \alpha) &= b_0 \hat f_1(x)\,\alpha^2 + b_1 \hat f_1(x)\,\alpha^3 + \ord{\alpha^4}
\,,\nn\\
\text{NNLO:}\qquad &&
\Delta f(x, \alpha) &= \bigl\{2b_0 [\hat f_2(x) - b_0 \hat f_1(x)] + b_1 \hat f_1(x)\bigr\}\alpha^3 + \ord{\alpha^4}
\,.\end{alignat}
The resulting uncertainty estimate is automatically of the correct one higher
order, $\Delta f(x, \alpha) \sim \ord{\alpha^{n+1}}$ at N$^n$LO,
which makes this approach very convenient to use in practice.

However, the scale-variation approach suffers from well-known limitations.
The parameters $b_k$, or in case of
QCD the single parameter $L$ in \Eq{bk_L}, are a priori arbitrary and contain
no information on the actual missing higher-order terms. At NLO for example
there is nothing that guarantees that $b_0 \hat f_1(x)$ is a good estimate
of $\hat f_2(x)$ in any way. In fact, typically $f_2(x)$ will have a (much) more
complicated $x$ dependence than $f_1(x)$ and will not just be proportional to it.
As a result, \Eq{Deltaf} does not provide a very reliable uncertainty estimate.
Typical pitfalls leading to an underestimation are when a lower-order $\hat f_{\leq n}(x)$
happens to be accidentally small (e.g. in the vicinity of zero-crossings) or when
$f_{n+1}(x)$ contains new structures not present in $f_{\leq n}(x)$.

While one might be able to avoid these pitfalls of underestimation, there is
a more severe  limitation of scale variations, which is fundamental and unavoidable:
The $b_k$ (or $L$) are unphysical parameters. That is, they are artificially
introduced via the scheme change, but they are not actual parameters of $f(x)$
with an existing but unknown true value.
While there might be a certain value of $L$ for which
$\Delta f(x_i, \alpha)$ coincides with the true value of the missing
higher-order term(s) for a specific $x_i$, this value of $L$ will be different
for different $x_i$. There may also be no (reasonable) value of $L$ for
which this is the case.
Therefore, the resulting variation $\Delta f(x, \alpha)$
in \Eq{Deltaf} \emph{cannot} be interpreted
as a fully correlated uncertainty like the variation due to an ordinary
parameter whose uncertainty is being propagated.
In other words, \Eq{Deltaf} does not provide a correct parameterization of
the $x$ dependence of the missing higher-order terms, which however
would be necessary to obtain the correct theory correlation structure in
$x$~\cite{Tackmann:2024kci}.

Oftentimes, including for our case of interest, predictions involve multiple
scale parameters (e.g.\ due to making separate
choices of the expansion parameter in different parts of the calculation).
The standard approach to mitigate the danger of misestimation is by considering
a variety of different variations of all relevant scales in some fashion to obtain a
variety of different results $\tilde f_k(x)$. The uncertainty estimate is
then obtained by taking the envelope over all variations%
\footnote{Here we take the default result as the central value and the maximum
absolute difference as its symmetric uncertainty. Other ways to construct the
envelope are also used, further demonstrating the arbitrary nature of these
prescriptions. However, the precise way of doing so is irrelevant for our discussion.}
\begin{equation}
\Delta f_k(x) = \tf_k(x) - f_{\rm default}(x)
\,,\qquad
\Delta f(x) = \max_{k} \bigl\lvert \Delta f_k(x) \bigr\rvert
\,.\end{equation}
The hope is that at any given $x_i$ there will be at least
one variation $\Delta f_k(x)$, not necessarily the same for all $x_i$, that gives a
reasonable upper limit for the possible size of the missing higher-order terms.
Taking the envelope then ensures that the total uncertainty does not scale
with the number of considered variations, which is in principle arbitrary.
Performing as many as possible different types of variations
(as opposed to merely increasing the size of variation) will then generically
improve the reliability of the estimate.
This is particularly true in situations (including our case of interest)
where individual $\Delta f_k(x)$ tend to cross through zero at some $x_i$ and thus
individually grossly underestimate the uncertainty around that point.

However, even assuming that the final envelope $\Delta f(x)$ provides a reasonable
theory uncertainty for the spectrum at any given $x_i$, the key question is how
to propagate this uncertainty from the spectrum to the parameter of interest
when performing an interpretation, which as discussed in \sec{correlations}
requires the correct point-by-point correlations across the spectrum, i.e.,
the correlation between $\Delta f(x_i)$ and $\Delta f(x_j)$ for any two $x_i$
and $x_j$. This is where we run afoul of the fundamental lack of correlations
in the scale-variation approach.

\begin{figure}
\renewcommand{\arraystretch}{1.1}
\parbox[c]{0.37\textwidth}{%
\includegraphics[width=0.38\textwidth]{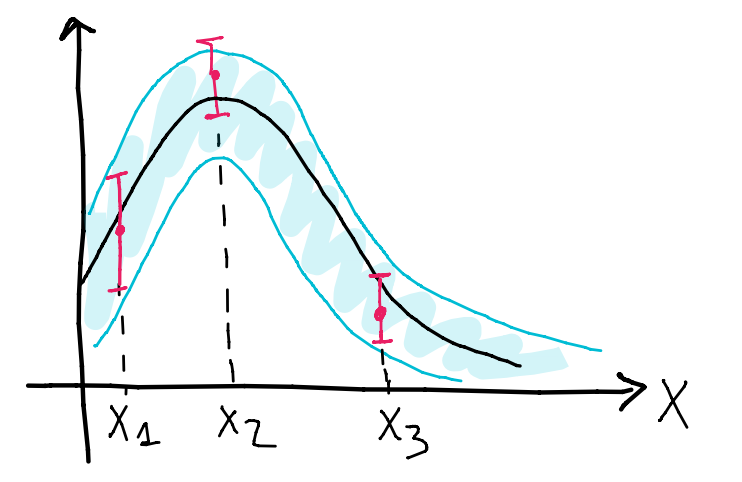}%
}%
\parbox[c]{0.37\textwidth}{%
\includegraphics[width=0.38\textwidth]{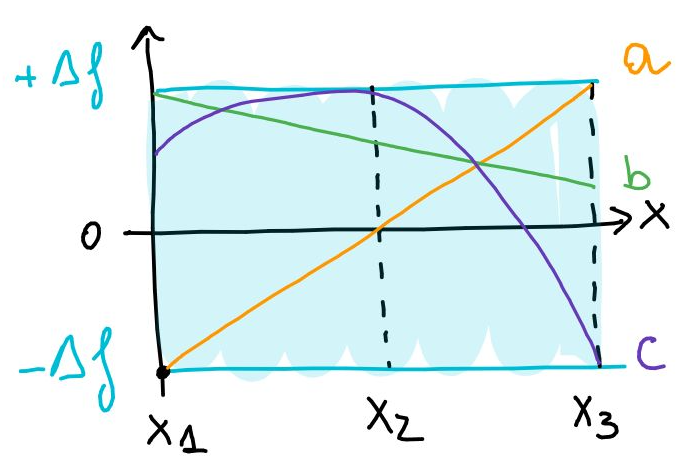}%
}%
\hfill%
\begin{tabular}{c|ccc}
\hline\hline
& $\rho_{12}$ & $\rho_{13}$ & $\rho_{23}$
\\\hline
$a$ & $ 0$ & $-1$ & $ 0$
\\
$b$ & $+1$ & $+1$ & $+1$
\\
$c$ & $+1$ & $-1$ & $-1$
\\\hline\hline
\end{tabular}
\caption{Interpreting a differential spectrum by scanning scale variations.
Left: The theory prediction (black line) with a theory uncertainty band (blue) from
the envelope of scale variations compared to measurements (red) at
three points $x_{1,2,3}$.
Middle: The normalized uncertainty band (blue) with three possible scale variations
(orange, green, violet) filling out the band.
Right: The assumed correlations $\rho_{ij}$ between the uncertainties at the
points $x_i$ and $x_j$ resulting from a given scale variation.
Note that for $\rho_{ij} = 0$ the uncertainty at either $x_i$ or $x_j$ must vanish.
}
\label{fig:cartoons}
\end{figure}

A common method to ``propagate'' the scale-variation envelope from the spectrum
to the parameters of interest is by scanning over scale variations. That is,
the interpretation is repeated for all predictions $\tf_k(x)$
that fill out the envelope. The various different results obtained for the parameters
of interest are then enveloped to give an estimate of the theory uncertainty on
the parameters of interest. This method is however insufficient due to the
fundamental limitations discussed above:
Since the underlying scales are unphysical parameters, the resulting
variations $\Delta f_k(x)$ do not correspond to well-defined fully correlated uncertainty
components or sources. Rather they merely provide different ways to probe the possible size of the
total theory uncertainty. This is exactly why we take their envelope
instead of adding them in quadrature.
In other words, by taking an envelope we explicitly acknowledge the fact that the
individual $\Delta f_k(x)$, and the point-by-point correlation they would
imply, are not particularly meaningful
and that they do not correspond to well-defined uncertainty components.
Otherwise we would combine them in quadrature.
Therefore, scanning over scale variations does not provide the necessary breakdown of
the total uncertainty into uncertainty components.

Instead, this scanning over enveloped variations amounts to
trying out various correlation models for the same total uncertainty band,
as illustrated in \fig{cartoons}.
We might then wonder what we can learn from this.
First, it is important to realize that none of the trial variations actually
provides a realistic correlation model, because as seen in \fig{cartoons},
they are by construction always fully correlated, only allowing correlations
$\rho_{ij} = \pm 1$ (or $\rho_{ij} = 0$ in case $\Delta f(x_{i,j}) = 0$).
This is clearly problematic when one is sensitive to shape effects, where as
discussed in \sec{correlations}, the precise correlation near $\rho = 1$ completely
determines the resulting shape uncertainty and thus the impact on the parameter
of interest. Therefore any given variation with
its implied correlation model can strongly underestimate (or overestimate) the
resulting theory uncertainty on the parameter of interest.
One might be ``lucky'' such that one trial variation sufficiently closely mimics
the shape effect of the parameter of interest so as to yield a ``conservative''
uncertainty and avoid underestimation. However, this might just as well be
very ``unlucky'' and yield a substantial overestimation.

In summary, scale variations provide no means to correctly address the problem
of theory correlations. Therefore, when theory correlations are important, or
even critical as in cases like ours when looking for shape effects,
they cannot be used to reliably propagate the theory uncertainties
from the original spectrum to the parameters of interest.
One might wonder why this fundamental lack of theory correlations in scale variations
is then not discussed more often. The main reasons might be that theory correlations are often either
not as important or not recognized to be important, or simply because there has
not been an alternative so far. There are however other
cases, some of which quite similar to ours, where the lack of theory correlations
in scale variations has been well recognized to be problematic or limiting.
To mitigate these problems, the best that can be done with scale variations is to
impose an explicit and case-specific correlation model on the theory
uncertainty, see e.g.~\Refscite{Berger:2010xi, Stewart:2011cf, Banfi:2012yh,
Gangal:2013nxa, Stewart:2013faa, LHCHiggsCrossSectionWorkingGroup:2016ypw, Proceedings:2018jsb, Lindert:2017olm, Harland-Lang:2018bxd, NNPDF:2019ubu, Bizon:2019zgf, Ebert:2020dfc, Benitez:2024nav}.
In comparison, by scanning over scale variations one basically relies on
some implicit ad hoc theory correlation model.

\subsection{Theory nuisance parameters}
\label{sec:tnps}

The limitations of scale variations have led to a variety of alternative
methods being explored over the years~\cite{Cacciari:2011ze, Bagnaschi:2014wea,
Bonvini:2020xeo, Duhr:2021mfd, David:2013gaa, Ghosh:2022lrf} including the
TNP approach~\cite{McGowan:2022nag, Dehnadi:2022prz, Cal:2024yjz, Tackmann:2024kci, CMS:2024lrd,
Lim:2024nsk, Clark:2025riz}. Among the motivations for the latter is to allow one to properly
account for theory correlations.

The fundamental source of the perturbative theory uncertainty is not the choice of scale or
perturbative scheme. Rather, the actual
sources of uncertainty in any particular scheme are its missing unknown $\hat f_n(x)$.
To be able to correctly account for the theory uncertainty, we have to actually
include the (leading) source(s) of uncertainty in our predictions. To do so,
compared to \Eq{f_pred} we have to include (at least) the next term in the series,
which contains the dependence on the unknown series coefficient $f_n(x)$. This is the
starting point of the TNP approach~\cite{Tackmann:2024kci}.
To obtain a perturbative prediction for $f(x)$ at order
N$^{m+k}$LO, we include the true values of the first $m$ series
coefficients and in addition the next $k \geq 1$ terms
whose coefficients are unknown. We are typically in a regime where the next order
is the dominant source of uncertainty, in which case $k=1$, so the predictions in
\Eq{f_pred} turn into
\begin{alignat}{9} \label{eq:f_pred_tnps}
\text{N$^{0+1}$LO:}\qquad && f(x, \alpha, \theta_1) &= \hat f_0(x) + f_1(x, \theta_1)\,\alpha
\,,\nn\\
\text{N$^{1+1}$LO:}\qquad && f(x, \alpha, \theta_2) &= \hat f_0(x) + \hat f_1(x)\,\alpha + f_2(x, \theta_2)\,\alpha^2
\,,\nn\\
\text{N$^{2+1}$LO:}\qquad && f(x, \alpha,\theta_3) &= \hat f_0(x) + \hat f_1(x)\,\alpha + \hat f_2(x)\,\alpha^2
+ f_3(x, \theta_3)\,\alpha^3
\,.\end{alignat}
Here, the (considered to be) unknown series coefficients $f_n(x, \theta_n)$
are parameterized in terms of \emph{theory nuisance parameters} $\theta_n$,
where for simplicity we let $\theta_n \equiv \{\theta_{n,i}\}$ denote an appropriate
set of parameters $\theta_{n,i}$ that are by construction scalars~\cite{Tackmann:2024kci}.

A key condition that the TNP parameterization $f_n(x, \theta_n)$ has to satisfy
is that there must exist true values $\hat\theta_n$ for which the true value
$\hat f_n(x)$ is reproduced,
\begin{equation} \label{eq:tnp_condition}
\hat f_n(x) = f_n(x, \hat\theta_n)
\,.\end{equation}
This condition implies that the TNPs are true parameters of the perturbative series.
As for any other ordinary parameter, we can assign to them a best estimate with
some uncertainty,
\begin{equation}
\theta_n = u_n \pm \Delta u_n
\,,\end{equation}
which can come from theory considerations or experimental constraints or both.
Without loss of generality we will assume that, prior to applying any experimental
constraints, the $\theta_n$ are chosen
such that based on theory information only we have $u_n = 0$ and
$\Delta u_n \equiv \Delta \theta_n = 1$.%
\footnote{Starting from a theory constraint $u_n \pm \Delta u_n$ for $\theta_n$,
we can always reparameterize it as $\theta_n' = (\theta_n - u_n)/\Delta u_n$, such
that the corresponding constraint on $\theta_n'$ is $0\pm 1$.}

The central prediction then corresponds to setting the $\theta_n$ to their
central values. Note that in general we can have $f_n(x, \theta_n = 0) \neq 0$,
so the central prediction at say N$^{2+1}$LO does not necessarily coincide with
the previous N$^2$LO prediction without TNPs. This is not a bug but a feature, because
the TNP parameterization allows us to correctly include contributions
that are already known, e.g.\ from lower-order terms. On the other hand,
\Eq{tnp_condition} implies that when evaluated at the true values $\hat\theta_n$,
the N$^{2+1}$LO result agrees with the N$^3$LO result.

The condition in \Eq{tnp_condition} ensures that the $\theta_n$ are well-defined common
parameters for the predictions at different $x$. Their unknown value therefore
represents a common source of uncertainty whose impact is 100\% correlated
across all $x$. At the same time, different individual $\theta_{n,i}$ are
a priori mutually independent, so they can be varied separately and the impacts
of different $\theta_{n,i}$ are a priori fully uncorrelated.
The theory uncertainty can thus be evaluated by propagating the
uncertainties $\Delta u_n$ through the interpretation to the parameters of
interest using any desired standard method for error propagation and combination.
In particular, they can be consistently treated and profiled in fits to data
like any other nuisance parameters. In this case, their post-fit uncertainties
can of course become correlated, which one has to
properly account for when combining uncertainties.

Theory-based constraints inevitably involve some human prejudice (unless they come
from calculations that are either exact or numerically approximate with a well-defined
numerical uncertainty).
As shown in \Refcite{Tackmann:2024kci}, it is nevertheless possible to obtain reliable and
statistically meaningful theory constraints.
When the TNPs are constrained by measurements, we can furthermore decide how much
to rely on any given theory constraint by choosing how much to impose it as an additional
constraint on the fit.

In summary, the TNP approach provides truly parametric theory uncertainties.
As a result, it correctly captures the point-by-point correlations
in $x$ by providing the necessary breakdown of the total theory uncertainty
into mutually independent uncertainty components.
In essence, by requiring \Eq{tnp_condition},
the TNP parameterization must encode the correct dependence of the missing
higher order on the observable $x$, which is precisely what is needed
to encode the correct correlation structure.
The general principles and strategies for constructing suitable parameterizations
are discussed in \Refcite{Tackmann:2024kci}.
The relevant TNPs for our case at hand are discussed in \sec{tnps_for_qT}.

\section{Theory Description and Pseudodata Setup}
\label{sec:theory}

In this section, we summarize the theoretical description of the $q_T$
spectrum which we use for our study. After discussing the overall theory
requirements in \sec{theory_inputs}, we discuss the relevant perturbative
contributions in \sec{qT_resummation} and the nonperturbative effects
in \sec{nonp_theory}. In \sec{asimov_setup}, we discuss our Asimov
fit setup.

\subsection{Theory requirements}
\label{sec:theory_inputs}

The transverse momentum ($q_T$) distribution of dilepton pairs from Drell–Yan
production is one of the most precisely measured observables at hadron colliders,
with the latest LHC measurements reaching subpercent precision
over a broad range of $q_T$ values.
This high level of experimental precision puts high demands on the theoretical
description. For a reliable (sub)percent-level extraction of $\as(m_Z)$ from the small-$q_T$
spectrum, not only must the theory predictions be accurate
to the (sub)percent level, but as already discussed the theory uncertainties and
correlations must also be well understood and reliably quantified.

The differential cross section in $q_T$ can be written as
\begin{equation} \label{eq:matching}
\frac{\df \sigma}{\df q_T}
= \frac{\df \sigma^\zero}{\df q_T} + \frac{\df \sigma_{\rm nons}}{\df q_T}
\,.\end{equation}
Here, $\df \sigma^\zero$ denotes the leading-power contribution, which dominates
the spectrum in the small-$q_T$ region, $q_T \ll Q$, where $Q \equiv \sqrt{q^2} \sim m_Z$
is the dilepton invariant mass.
The nonsingular term, $\df \sigma_{\rm nons}$, contains
all remaining contributions, which are power-suppressed by $\ord{q_T^2/Q^2}$
and thus subdominant relative to $\df\sigma^\zero$.

The largest sensitivity to $\as(m_Z)$ comes from the small-$q_T$ region. Since $\df\sigma^\zero$
dominates here, it is also the dominant source of perturbative and nonperturbative uncertainties
and therefore the main focus of our study.
At each order in $\as$, $\df\sigma^\zero$ contains logarithms $\ln^n(q_T/Q)$, which must be resummed
to obtain precise and perturbatively stable predictions. 
In parallel, it is affected by nonperturbative corrections of $\ord{\lqcd^n/q_T^n}$,
which shape the $q_T$ spectrum at very low $q_T \lesssim 10\GeV$.
Their associated uncertainties therefore play an important role for extracting $\as(m_Z)$.
A complete treatment of perturbative uncertainties and a proper parameterization of
nonperturbative effects in the leading-power contribution are thus indispensable for
a reliable $\as$ extraction at low $q_T$.

Another key ingredient is the treatment of parton distribution functions (PDFs).
PDFs influence both the normalization and shape of the spectrum. They are
extracted from global fits that themselves depend on assumptions about $\as$.
This introduces a nontrivial correlation between the value of $\as(m_Z)$ one
extracts and the underlying PDFs used in the prediction \cite{Hou_2021,Forte:2020pyp,Ball_2022,Cridge:2021qfd,Cridge:2024exf}.
Thus, a consistent extraction must account for PDF uncertainties and, ideally,
include the interplay between PDFs and $\as$ in a consistent fashion.
A more detailed study of PDF uncertainties in the extraction
of $\as(m_Z)$ from the $Z$ $q_T$ spectrum will be presented in a forthcoming
paper~\cite{pdf_paper}.

Beyond these leading contributions, achieving a (sub)percent-level theoretical description
requires incorporating various subleading effects.
One such effect is the nonsingular contribution in \Eq{matching}.
Similarly, quark mass corrections scaling as $m_q^2/q_T^2$, and QED and electroweak
effects become relevant at this level of precision (see e.g. \Refscite{Pietrulewicz:2017gxc, flavorthr_cs, Cieri:2018sfk, Billis:2019evv, Autieri:2023xme, Buonocore:2024xmy}). Neglecting these subleading
effects can cause a substantial bias in the extracted value of $\as(m_Z)$, so it is important
to include them in a fit to real data. However, once they are included, we do not expect them to
have a major influence on the uncertainty in $\as(m_Z)$.
We can therefore
consistently neglect them in our study as further discussed in \sec{asimov_setup}.

Summarizing, for our study we focus on the leading-power contribution,
its resummed perturbative prediction and associated uncertainty as discussed
in \sec{qT_resummation}, and its nonperturbative effects and associated
uncertainty as discussed in \sec{nonp_theory}.
Among the subleading effects, the nonsingular contribution is the largest. We therefore briefly
study its impact in \app{nonsingular}, showing that it can indeed be neglected
for the purposes of our main analysis.
In \app{quark_masses} we also briefly comment on the potential bias from neglecting
quark masses.

\subsection{Resummation of perturbative contributions at small \texorpdfstring{$q_T$}{qT}}
\label{sec:qT_resummation}

We use the SCET resummation framework including TNPs of
\Refscite{Ebert:2020dfc, Billis:2024dqq, Tackmann:2024kci}.
In the following, we focus on the aspects relevant to our study and refer to
these references for further details.
The numerical results for our study are obtained with the help of
\scetlib~\cite{scetlib} and its implementation of $q_T$ resummation with
TNPs~\cite{Billis:2019vxg, Ebert:2020dfc, Billis:2021ecs, Billis:2024dqq, Tackmann:2024kci}.

We denote the
four-momentum of the vector boson by $q^{\mu}$, its invariant mass by $Q \equiv \sqrt{q^2}$, its total rapidity by $Y$, and its transverse momentum by $q_T = \abs{{\vec q}_T}$.
We are interested in the leading-power cross section
fully differential in $Q$, $Y$ and $\vec q_T$, collectively denoted as $\df\sigma^\zero/\df^4 q$.
In our framework, its resummation is based on
its factorization theorem~\cite{Collins:1981uk, Collins:1981va, Collins:1984kg, Collins:1350496, Bauer:2000ew, Bauer:2000yr, Bauer:2001yt, Bauer:2002nz, Becher:2010tm, GarciaEchevarria:2011rb, Chiu:2012ir, Li:2016axz},
which is written as
\begin{align} \label{eq:factorization}
\frac{\df \sigma^\zero}{\df^4 q}
&= \frac{1}{2\Ecm^2}\, L_{VV'}(q^2) \,
\sum_{a,b} H_{VV'\,ab}(q^2, \mu)
\\\nn & \quad\times
\int\!\!\frac{\df^2\bt}{(2\pi)^2} \, e^{\img \bt \cdot \qt} \,
   \tB_a(x_a, b_T, \mu, \nu/Q) \, \tB_b(x_b, b_T, \mu, \nu/Q)
   \tS(b_T, \mu, \nu)
\,.\end{align}
Here, $E_{\rm cm}$ is the hadronic center-of-mass energy and $L_{VV'}(q^2)$ denotes the leptonic
tensor associated with the final-state vector boson
$VV' = \{ \gamma \gamma, \gamma Z, Z \gamma, ZZ\}$,
which is not affected by QCD corrections.
The sum runs over parton flavors $a,b$, and the remaining ingredients are the hard function
$H_{VV'\,ab}(q^2, \mu)$ , the beam functions $\tB_{a,b}(x_{a,b}, b_T, \mu, \nu/Q)$, and
the soft function $\tS(b_T, \mu, \nu)$.
The variable $b_T = \abs{\bt}$ is the Fourier-conjugate of $q_T$ and
$x_{a,b} = (Q/\Ecm) \exp^{\pm Y}$ encode the dependence on the rapidity and collider energy.

\subsubsection{Resummation}
\label{sec:profile_scales}

Each of the perturbative ingredients in \Eq{factorization} obeys a
renormalization group evolution (RGE), which determines its scale and kinematic dependence to
all orders in $\as$.
As a result, the factorization theorem predicts the full $q_T$ dependence and,
for fixed $x_{a,b}$, also the $Q$ dependence.
Although it does not predict the full functional form in $x_{a,b}$, it simplifies this dependence
to the product of two one-dimensional beam functions.
We now describe each ingredient in turn.

The \textbf{hard function} $H_{VV'\,ab}$ encodes virtual corrections to the underlying
hard partonic interaction process $ab \to V$. It
can be extracted as the infrared-finite part of the
corresponding form factors. Explicit expressions for $H_{VV'\,ab}$ and $L_{VV'}$ in terms of the
underlying SCET Wilson coefficients $C_q(q^2, \mu)$ are given in~\Refcite{Ebert:2020dfc}.
The $C_q$'s RGE equation is given by
\begin{align}\label{eq:Cq_RGE}
\mu\frac{\df}{\df\mu} \ln C_q(q^2,\mu)
&= \GammaC^q[\as(\mu)] \ln\frac{-q^2 - \img 0}{\mu^2} + 2\gamma_C^q[\as(\mu)]
\,,\end{align}
where $\GammaC^q (\as)$ and $\gamma_C(\as)$ are the cusp and noncusp anomalous
dimensions, respectively.

The \textbf{soft function} captures wide-angle soft radiation. It additionally
depends on the rapidity renormalization scale $\nu$.
Its scale dependence is governed by the following RGE system
\begin{align}\label{eq:S_RGEs}
\mu\frac{\df}{\df\mu} \ln\tS(b_T,\mu,\nu)
&= 4 \GammaC^q [\as(\mu)] \ln\frac{\mu}{\nu} + \tilde\gamma_S[\as(\mu)]
\,,\nn\\
\nu\frac{\df}{\df\nu} \ln\tS(b_T,\mu,\nu)
&= \tilde\gamma_\nu(b_T,\mu)
\,,\nn\\
\mu \frac{\df}{\df \mu} \tilde\gamma_\nu(b_T, \mu)
&= - 4\GammaC^q[\as(\mu)]
\,.\end{align}
The rapidity anomalous dimensions $\tilde\gamma_\nu(b_T, \mu)$ encodes the nontrivial
$b_T$ dependence, which in turn is governed by its own $\mu$ RGE.

The \textbf{beam functions} $\tB_i(x, b_T, \mu, \nu/Q)$ describe the extraction of a parton $i$ carrying
momentum fraction $x$ at some transverse-momentum conjugate $b_T$ from an
unpolarized hadron. They evolve according to
\begin{align}\label{eq:B_RGEs}
\mu\frac{\df}{\df\mu} \ln\tB_q(x,b_T,\mu,\nu/Q)
&= 2 \GammaC^q[\as(\mu)] \ln\frac{\nu}{Q} + \tilde\gamma_B[\as(\mu)]
\,,\nn\\
\nu\frac{\df}{\df\nu} \ln\tB_q(x,b_T,\mu,\nu/\omega)
&= -\frac{1}{2}\tilde\gamma_\nu(b_T,\mu)
\,,\nn\\
\mu \frac{\df}{\df \mu} \tilde\gamma_\nu(b_T, \mu)
&= - 4\GammaC^q[\as(\mu)]
\,.\end{align}
While their RGE fully determines their $b_T$ and $Q$ dependence, their dependence
on $x$ is not predicted and only arises via the boundary condition at
canonical scales $\mu = b_0/b_T$ and $\nu = Q$, defined as
\begin{equation} \label{eq:B_boundary_condition}
\tilde b_i(x, \alpha_s) \equiv \tB_i(x, b_T, \mu = b_0/b_T, \nu/Q = 1)
\,.\end{equation}
In the perturbative region $q_T \sim 1 / b_T \gg \lqcd$,
this boundary condition is computed in terms of standard PDFs at
the canonical scale $\mu_f^{\rm can} = b_0/b_T$,
\begin{align} \label{eq:ope_beam_soft_tmd}
\tilde b_{i,n}(x)
&= \sum_j \int\! \frac{\df z}{z} \, \tilde I_{ij,n}(z) \, f_{j}\Bigl( \frac{x}{z}, \mu_f^{\rm can} \Bigr)
\,,\end{align}
where $\tilde I_{ij,n}(z)$ are perturbative matching kernels. 
The $x$ dependence of the beam function is thus given by
the Mellin convolution of these kernels with the PDFs.

The resummation proceeds by evaluating each of the perturbative ingredients
in \Eq{factorization} in fixed order at their natural (canonical)
$\mu$ and $\nu$ scales, where they are free of large logarithmic corrections.
These are called the boundary conditions. Starting from their boundary
conditions, the functions are then evolved to common overall $\mu$ and $\nu$ scales
by solving their respective RGEs. In the combination of all RGEs and upon Fourier-transforming
back to momentum space, the logarithms of $q_T/Q$ are resummed
to all orders in perturbation theory.
The canonical scales in $b_T$ space are given by
\begin{align} \label{eq:canonicalscales}
\mu_B^{\rm can} = \mu_S^{\rm can} = \nu_S^{\rm can} = \mu_0^{\rm can} = \mu_f^{\rm can} = \frac{b_0}{b_T} \, , \qquad
\nu_B^{\rm can} = \mu_H^{\rm can} = Q
\,,\end{align}
where $\mu_{S,B}$ and $\nu_{S,B}$ are the scales of the soft and beam functions,
$\mu_0$ is the scale at which the boundary condition of the rapidity anomalous
dimension is evaluated, $\mu_f$ is the scale at which the matching to PDFs
in \Eq{ope_beam_soft_tmd} is performed,
and $b_0 = 2 e^{-\gamma_E} \approx 1.12292$ is a conventional factor.

Since we do not include nonsingular power corrections in our main study,
we do not discuss here the procedure for switching off the resummation at large
$q_T$ to recover the fixed-order result.
For that purpose, the canonical scales in \Eq{canonicalscales} are modified
using hybrid profile scales~\cite{Lustermans:2019plv} (see \Refcite{Billis:2024dqq} for details), defined as
\begin{alignat}{9} \label{eq:profiled_scales}
\mu_H &= Q
\,, \qquad \nn \\
\mu_B &= Q\, f_{\rm run} \biggl( \frac{q_T}{Q}, \frac{1}{Q} \mu_{*}
\Bigl( \frac{b_0}{b_T}, \mu_B^{\rm min} \Bigr) \biggr)
\,, \qquad &
\nu_B &= Q
\,, \nn \\
\mu_S &= Q\, f_{\rm run} \biggl( \frac{q_T}{Q}, \frac{1}{Q} \mu_{*}
\Bigl( \frac{b_0}{b_T}, \mu_S^{\rm min} \Bigr) \biggr)
\,, \qquad&
\nu_S &= Q\, f_{\rm run} \biggl( \frac{q_T}{Q}, \frac{1}{Q} \mu_{*}
\Bigl( \frac{b_0}{b_T}, \nu_S^{\rm min} \Bigr) \biggr)
\,, \nn \\
\mu_0 &= \mu_{*} \Bigl( \frac{b_0}{b_T}, \mu_0^{\rm min} \Bigr)
\,, \qquad &
\mu_f &= Q\, f_{\rm run} \biggl( \frac{q_T}{Q}, \frac{1}{Q} \mu_{*}
\Bigl( \frac{b_0}{b_T}, \mu_f^{\rm min} \Bigr)  \biggr)
\,,\end{alignat}
where the profile function $f_{\rm run}(x, y)$ ensures a smooth transition between
the canonical scales in the resummation region, $x = q_T/Q\ll 1$, and the
fixed-order region, $x = q_T/Q\sim 1$.
The most relevant point for our purposes is that the transition only starts
at a transition point $x_1$ (by default $x_1 = 0.3$) and below it
$f_{\rm run}(x \leq x_1, y) = y$, so the default scales in \Eq{profiled_scales}
reduce to
\begin{equation} \label{eq:profiled_scales_smallqT}
\mu_{B,S,0,f} = \mu_*\Bigl( \frac{b_0}{b_T}, \mu_{B,S,0,f}^{\rm min} \Bigr)
\,, \quad
\nu_S = \mu_*\Bigl( \frac{b_0}{b_T}, \nu_S^{\rm min} \Bigr)
\,, \quad
\nu_B = \mu_H = Q
\qquad \text{(for $q_T \leq x_1 Q$)}
\,.\end{equation}
Hence, in this ``canonical'' $q_T$ region, the default scales we use only differ
from their canonical values in \Eq{canonicalscales} by the $\mu_{*}$ function,
which simply enforces a smooth lower bound on the scales to avoid the nonperturbative regime at
large $b_T$, where the QCD Landau pole would be encountered, with
\begin{align} 
\mu_B^{\min} =  \mu_S^{\min} = \mu_0^{\min} = 1 \GeV
\,,\qquad \mu_f^{\min} = 1.4 \GeV
\,,\qquad \nu_S^{\rm min} = 0
\,.\end{align}

\subsubsection{Scale variations}
\label{sec:profile_scale_variations}

We now briefly summarize the scale-variation setup in
\scetlib, referring to \Refcite{Billis:2024dqq} for more details. The central
scale choices are equivalent to the default scales in \Eqs{profiled_scales}{profiled_scales_smallqT}.
The scale variations are performed by varying the scales around their central
values with multiplicative factors. In the canonical region $q_T\leq x_1 Q$,
these are
\begin{alignat}{9}
\mu_H &= 2^{w_{\rm FO}}\, Q
\,, \nn \\
\mu_B &= 2^{w_{\rm FO}} \, f_{\rm vary}^{v_{\mu_B}} \,
\mu_{*}\Bigl( \frac{b_0}{b_T}, \frac{\mu_B^{\rm min}}{2^{w_{\rm FO}} f_{\rm vary}^{v_{\mu_B}}} \Bigr)
\,, \qquad &
\nu_B &= 2^{w_{\rm FO}} \, f_{\rm vary}^{v_{\nu_B}} \, Q
\,, \nn \\
\mu_S &= 2^{w_{\rm FO}} \, f_{\rm vary}^{v_{\mu_S}} \, \mu_{*}
\Bigl( \frac{b_0}{b_T}, \frac{\mu_S^{\rm min}}{2^{w_{\rm FO}} f_{\rm vary}^{v_{\mu_S}}} \Bigr)
\,, \qquad &
\nu_S &= 2^{w_{\rm FO}} \, f_{\rm vary}^{v_{\nu_S}} \, \mu_{*}
\Bigl( \frac{b_0}{b_T}, \frac{\nu_S^{\rm min}}{2^{w_{\rm FO}} f_{\rm vary}^{v_{\nu_S}}} \Bigr)
\,, \nn \\
\mu_f &= 2^{v_{\mu_f}} \, \mu_{*}
\Bigl( \frac{b_0}{b_T}, \frac{\mu_f^{\rm min}}{2^{v_{\mu_f}}} \Bigr)
\,.\end{alignat}
The same overall factors for each scale also appear inside its $\mu_{*}$ function,
which ensures that the effective minimum scales at large $b_T$ are unaffected by
the variations.
The boundary scale $\mu_0$ is held fixed, as its variation is effectively captured
already by those of $\nu_{B,S}$.
The different variation factors are used to perform different classes of variations,
which we describe in turn.

The \textbf{resummation uncertainty $\Delta_{\rm resum}$} is obtained by varying the
beam and soft scales via $v_{\mu_B}, v_{\nu_B}, v_{\mu_S}, v_{\nu_S}$ by taking
$v_i = \{ -1, 0, +1 \}$ with $v_i = 0$ corresponding to the central scale.
Here, $f_{\rm vary} \equiv f_{\rm vary}(q_T/Q)$ is a function of $q_T/Q$ that approaches
$2$ for $q_T \to 0$ and $1$ for $q_T/Q\to 1$, i.e., it smoothly turns off the variation
outside the resummation region.
We perform 36 variations of suitable combinations of the $v_i$ and then take their maximum envelope
as the resulting final uncertainty $\Delta_{\rm resum}$.

For the \textbf{fixed-order uncertainty $\Delta_{\rm FO}$}, we simultaneously vary
the scales of the hard, beam, and soft functions by an overall factor of 2 by
taking $w_{\rm FO} = \{ -1, 0, +1 \}$.
The final uncertainty is again taken as the maximum envelope of these two variations.
This variation only probes the fixed-order boundary conditions without changing
the resummed logarithms, hence its name.

The \textbf{uncertainty related to the DGLAP running of the PDFs $\Delta_f$} is obtained
by varying the PDF scale $\mu_f$ by taking $v_{\mu_f} = \{ -1, 0, +1 \}$ and taking the
maximum envelope of the two variations.

Finally, the \textbf{matching uncertainty $\Delta_{\rm match}$} is given by the maximum
envelope of the variations of the transition points $x_i$ inside the $f_{\rm run}$ function
determining the transition from the canonical to the fixed-order region.
There are three points $(x_1, x_2, x_3)$ determining the start, midpoint, and endpoint of the
transition. Their central values are $(0.3, 0.6, 0.9)$ and their variations are
\begin{equation}
(x_1, x_2, x_3 ) \in \{ (0.4, 0.75, 1.1), (0.2, 0.45, 0.7), (0.4, 0.55, 0.7), (0.2, 0.65, 1.1) \}
\,.\end{equation} 

These different classes of variations are specifically chosen to probe separate
aspects of the perturbative series. They are then considered as independent
uncertainties and added in quadrature to obtain the total perturbative
uncertainty,
\begin{align}
\Delta_{\rm pert} = \sqrt{\Delta_{\rm FO}^2 + \Delta_f^2 + \Delta_{\rm match}^2 + \Delta_{\rm resum}^2}
\,.\end{align}
In particular, the $\mu_f$ variation is not enveloped with the
other scale variations, but is instead treated as a separate component. The reason
is that it probes the uncertainty due to DGLAP running, which
can be considered as a separate source of perturbative uncertainty.

\subsubsection{TNPs for \texorpdfstring{$q_T$}{qT} resummation}
\label{sec:tnps_for_qT}

We now briefly summarize the theory nuisance parameters for $q_T$ resummation,
referring to section 6 of \Refcite{Tackmann:2024kci} for more details.
The leading-power spectrum in \Eq{factorization} is fully determined by the hard,
beam and soft functions, $F = \{ H,B,S \}$.
Each of these function satisfies a coupled system of RGEs, given in~\Eq{Cq_RGE},~\Eq{S_RGEs}
and~\Eq{B_RGEs}. Their solution can be written schematically as
\begin{align}
\label{eq:func_depend}
F(\as^{\rm can}, L) = F(\as^{\rm can}) \exp \biggl\{ \int_{0}^{L} \df L'
\Bigr[ \GammaC\bigl(\as(L')\bigr) L' + \gamma_F\bigl( \as (L') \bigr) \Bigl] \biggr\}
\,,\end{align}
where $L \equiv \ln(\mu/\mu_F^{\rm can})$ and $\as^{\rm can} \equiv \as(\mu^{\rm can}_F)$.
It depends on the boundary condition $F(\as)$ (defined at strictly canonical scale for $L = 0$)
as well as the cusp and noncusp anomalous dimensions, $\GammaC(\as)$ and $\gamma_F(\as)$.
These perturbative series contain the complete perturbative information
necessary for the leading-power spectrum.
We can therefore treat each of them following the TNP approach of \Refcite{Tackmann:2024kci}
and as outlined in \sec{tnps}, parameterizing their individual missing higher-order
coefficients with TNPs. Since the $q_T$ dependence is completely predicted
by the factorization theorem in terms of these, this treatment correctly encodes the
point-by-point correlations in $q_T$.

Denoting generically all boundary conditions as $F(\as)$ and all anomalous dimensions as $\gamma (\as)$,
we write
\begin{align} \label{eq:scalar_series}
F(\as) &= 1 + \sum_{n=1} F_n \Bigl( \frac{\as}{4 \pi} \Bigr)^n
\, , \nn \\
\gamma(\as) &= \sum_{n=0} \gamma_n \Bigl( \frac{\as}{4 \pi} \Bigr)^{n+1}
\,,\end{align}
and parametrize each coefficient (for fixed $n_f = 5$) as
\begin{align}
F_n (\theta_n^F) &= 4^n C_n  (n-1)! \, \theta^F_n
\,,\\
\gamma_n (\theta_n^\gamma) &= C_{n+1} 4^{n+1} \, \theta^{\gamma}_n
\,.\end{align}
Here, $C_n$ is the relevant leading $n$-loop coefficient, which in our case is
given by $C_n = C_F C_A^{n-1}$, and $\theta_n^F$, $\theta_n^{\gamma}$ are the TNPs.

The generalized order counting for resummation including TNPs is denoted as
N$^{m+k}$LL and defined in analogy to N$^{m+k}$LO in \sec{tnps} as follows:
A given perturbative resummation order, N$^n$LL, is uniquely and well defined
by including all underlying perturbative series in the RGE to a certain order in
$\alpha_s$, namely the boundary conditions to $(n-1)$ loops, the noncusp anomalous
dimensions to $n$ loops, and the cusp anomalous dimensions and QCD beta functions
to $n+1$ loops. At N$^{m+k}$LL we use the complete resummation
structure at N$^n$LL with $n = m + k$ with the true values for all
coefficients relevant up to N$^m$LL and the TNP parameterization for the last
$k$ coefficients of each series. For example, \Eq{func_depend} at N$^{3+1}$LL
takes the form
\begin{alignat}{9} \label{eq:F_n3p1ll}
\text{N$^{3+1}$LL:}\qquad &&
F(\as^{\rm can}, L)
&= \biggl[1 + \sum_{k=1}^2 \hat F_k \Bigl(\frac{\as^{\rm can}}{4\pi}\Bigr)^k
   + F_3(\theta_3^F) \Bigl(\frac{\as^{\rm can}}{4\pi}\Bigr)^3 \biggr]
\exp \biggl\{ \int_{0}^{L} \df L' \biggr[
\nn\\ &&& \qquad
\sum_{k = 0}^3 \hat\Gamma_k \Bigl[\frac{\as(L')}{4\pi}\Bigr]^{k+1} L'
+ \Gamma_4(\theta_4^\Gamma) \Bigl[\frac{\as(L')}{4\pi}\Bigr]^5 L'
\nn\\ &&& \quad
+ \sum_{k=0}^2 \hat\gamma_{F,k} \Bigl[\frac{\as(L')}{4\pi}\Bigr]^{k+1}
+ \gamma_{F,3}(\theta_3^{\gamma_F}) \Bigl[\frac{\as(L')}{4\pi}\Bigr]^4
\biggl]\biggr\}
\,.\end{alignat}
Furthermore, in the full resummed cross section, the product of all boundary
conditions is expanded in $\as$ including cross terms up to $\ord{\as^3}$ as
relevant for N$^4$LL.
Note that despite the fact that the highest $k$ coefficients at N$^{m+k}$LL are treated
as unknown, due to having the complete $(m+k)$th-order structure, this order still
contains more perturbative information than N$^m$LL, so the perturbative accuracy
at N$^{m+k}$LL is generically higher than at N$^m$LL.

For our analysis of the Drell-Yan $q_T$ spectrum, we consider a minimal
set of seven independent TNPs,
\begin{equation}
\theta_n^\gamma: \gamma \in \{\Gamma, \gamma_\mu, \gamma_\nu\}
\,,\qquad
\theta_n^F: F \in \{H, S, B_{qq}, B_{qg} \}
\,,\end{equation}
associated with the following independent perturbative series and thus sources of uncertainty:
The cusp anomalous dimension ($\GammaC$), the noncusp anomalous dimensions
($\gamma_{\mu}$, $\gamma_{\nu}$), and the boundary conditions of the hard ($H$), soft ($S$) and beam
($B_{qq}, B_{qg}$) functions. Their detailed normalization conventions are discussed
in \Refcite{Tackmann:2024kci}. For the beam functions in particular,
based on \Eq{ope_beam_soft_tmd}, we adopt the parameterization
\begin{align}
\tilde I_{ij,n}(z,\theta_n^{B_{ij}})
= \frac{3}{2}\, \theta_n^{B_{ij}}\, \hat{\tilde{I}}_{ij,n}(z)
\,,\end{align}
which uses their known $z$ dependence multiplied by an overall prefactor of
$3/2$ to be conservative. With this parameterization we effectively treat their shape
as known and their overall normalization as unknown. This choice is a compromise. It
means while we do not yet account for the uncertainty associated with their shape,
we have the correct correlations in $z$ and thereby in $Y$ for the dominant
overall normalization uncertainty we do consider.
Since the dominant partonic channels for $Z$ boson production are
$ij = \{ qq, qg\}$, we introduce only two TNPs for the beam boundary conditions.
For simplicity, singlet contributions that appear at higher orders and only have
a very minor impact are not considered separately. Specifically, we use a single
effective TNP, $\theta_n^{B_{qq}}$, which varies all $qq$ channels together,
\begin{align}
\theta_n^{B_{qq}} \equiv \theta_n^{B_{qqV}} \equiv \theta_n^{B_{q\bar{q}V}} \equiv \theta_n^{B_{qqS}} \equiv \theta_n^{B_{qq\Delta S}}
\,,\end{align}
and a separate TNP, $\theta_n^{B_{qg}}$, for the $qg$ channel.

As explained in \sec{tnps} and following \Refcite{Tackmann:2024kci}, each nuisance parameter
$\theta_n$ is modelled as a Gaussian-distributed random variable which by default
has zero mean,
\begin{equation}
\theta_n = 0 \pm \Delta \theta_n
\,,\end{equation}
with $\Delta \theta_n$ setting the width of the distribution and hence the size
of the variation. Unless otherwise stated, we take $\Delta \theta_n = 1$ as our
default theory constraint, corresponding to a $68\%$ theory confidence level.
We also explore the impact of relaxing the theory constraint in \sec{relaxing_constraints}.
The default (pre-fit) result at N$^{m+k}$LL is then given by the default resummed
result as discussed in \sec{profile_scales} with all TNPs that enter it
set to zero. Its uncertainty is
obtained by varying each TNP by $\Delta \theta_n$ and adding all
variations in quadrature (since by default all TNPs are considered independent).
After fitting the TNPs to data, the post-fit result for the resummed spectrum is
then given by using the post-fit central values and uncertainties of all TNPs.

Finally, it is important to note that the above minimal set of TNPs does not yet capture
all potential sources of uncertainties. As discussed in \Refcite{Tackmann:2024kci},
additional sources are the full
set of partonic channels in the beam functions as well as their functional
dependence, the DGLAP splitting functions, singlet contributions to the hard function,
and the QCD $\beta$ function. These are currently not accounted for and are left for
future improvements. Among these, the latter two are likely to be irrelevant
while the uncertainties associated with the DGLAP splitting functions may prove
to be the most relevant.

\subsection{Nonperturbative effects}
\label{sec:nonp_theory}

The $q_T$ spectrum is affected by nonperturbative effects that scale as
$\lqcd^2/q_T^2$, and which are expected to become relevant at the few-percent
level for $q_T \lesssim 10$ $\GeV$~\cite{Billis:2024dqq}.
In the literature a variety of phenomenological models have been used to
incorporate nonperturbative effects, typically involving a small number of free
parameters fitted to data.
Since this region of moderately small $q_T$ is also where the dominant sensitivity
to $\as(m_Z)$ comes from, we require a correct parameterization of
nonperturbative effects to ensure a reliable theoretical description in this region
with correct point-by-point correlations of the nonperturbative uncertainties.

In this section, we briefly discuss the nonperturbative approach used in
\scetlib~\cite{Billis:2024dqq}, which is based on performing a systematic
OPE expansion of nonperturbative effects. Next, we discuss the specific
nonperturbative parameterization we employ for our study here.

\subsubsection{General discussion}

The nonperturbative effects enter via the beam and soft functions
appearing in \Eq{factorization}. To discuss their nonperturbative contributions,
it is convenient to combine their product in $b_T$ space into conventional
transverse-momentum dependent PDFs (TMD PDFs), defined as
\begin{equation} \label{eq:tmd_beam_soft}
\tilde{f}_i(x, b_T, \mu, \zeta) = \tB_i( x, b_T, \mu, \nu/\sqrt{\zeta})
\sqrt{\tS (b_T, \mu, \nu)}
\,.\end{equation}
Here, $\zeta$ is the so-called Collin-Soper scale, which now takes over the role of
the rapidity scale $\nu$, whose dependence cancels on the right-hand side.
Similarly, the combined rapidity RGE of the beam and soft functions gets replaced
by the $\zeta$ evolution of the TMD PDF,
\begin{equation} \label{eq:CS_evolution}
\tilde{f}_i(x, b_T, \mu, \zeta)
= \tilde{f}_i(x, b_T, \mu, \zeta_0)
\exp\Bigl[\frac{1}{2} \tilde{\gamma}_{\zeta}(b_T, \mu) \ln\frac{\zeta}{\zeta_0} \Bigr]
\,.\end{equation}
The physical value of the final scale $\zeta$ in the resummed
cross section is $\zeta = Q^2$, as seen in \Eq{factorization}.
The boundary scale $\zeta_0$ is equivalent to $\nu_S^2$ and its canonical
value is $\sqrt{\zeta_0} = b_0/b_T$, which eliminates the potentially
large rapidity logarithms $\ln(\zeta_0 b_T^2/b_0^2)$ in the TMD PDF's boundary
condition.
The associated anomalous dimension is known as the Collins-Soper (CS) kernel,
$\tilde{\gamma}_{\zeta}(b_T, \mu)$,
which is simply related to the rapidity anomalous dimension,
\begin{equation} \label{eq:CS_gammanu}
\tilde{\gamma}_{\zeta}(b_T, \mu) = \frac{1}{2} \tilde{\gamma}_{\nu}(b_T, \mu)
\,.\end{equation}

There are in total two distinct sources of nonperturbative contributions, namely to the
TMD PDF boundary condition and the CS kernel, which can be written as~\cite{Boussarie:2023izj},
\begin{align} \label{eq:pert_nonpert_split}
\tilde{\gamma}_{\zeta}(b_T, \mu)
&= \tilde{\gamma}_{\zeta}^{\rm pert}(b^*(b_T), \mu)
+ \tilde{\gamma}_{\zeta}^\nonp(b_T)
\,,\nn\\
\tilde{f}_i(x, b_T, \mu, \zeta_0)
&= \tilde{f}^{\rm pert}_i(x, b_T, \mu, \zeta_0)
\tilde{f}_i^\nonp (x, b_T, \zeta_0)
\,.\end{align}
Here, the quantities on the left-hand side are defined in full QCD. The
separation into perturbative and nonperturbative parts on the right-hand side is well defined, but
a priori not unique. Namely, the exact definition of the
nonperturbative pieces, $\tilde f^\nonp$ and $\tilde\gamma_\zeta^\nonp$, is
implicitly determined by the exact definition of the perturbative pieces,
$\tilde f^{\rm pert}$ and $\tilde\gamma_\zeta^{\rm pert}$, including all choices
of boundary scales and cutoff prescriptions used to avoid the Landau pole, as
discussed below.

The nonperturbative contribution to the CS kernel,
$\tilde{\gamma}_{\zeta}^\nonp(b_T)$, is universal, like the CS kernel (or
rapidity anomalous dimension) itself, in that it only depends on whether the
considered process is quark or gluon initiated (which for simplicity is
suppressed in our notation). In particular, it is universal across processes
such as Drell-Yan production (including $Z$ and $W$ bosons) and semi-inclusive
deep inelastic scattering. It can also be accessed via lattice QCD
calculations~\cite{Schlemmer:2021aij, LatticePartonLPC:2022eev,
Avkhadiev:2023poz, LatticePartonLPC:2023pdv, Shu:2023cot, Avkhadiev:2024mgd}. In
principle, this allows one to obtain information on
$\tilde{\gamma}_{\zeta}^\nonp(b_T)$ by comparing to the lattice results, for
which a proper treatment of massive quark corrections is however
essential~\cite{flavorthr_cs}.

Unlike the CS kernel, the nonperturbative TMD PDF boundary
condition, $\tilde{f}_i^\nonp (x, b_T)$, is not universal to all quarks. Like the TMD
PDF itself, in addition to $b_T$ it also depends on the flavor $i$ and
Bjorken $x$ of the interacting parton. It can be thought of as describing the
intrinsic transverse momentum of the partons inside the proton.

At large $b_T\sim1/\lqcd$, $\tilde f^\nonp_i(x, b_T)$ and $\tilde\gamma_\zeta^\nonp(b_T)$
are genuine nonperturbative functions of $b_T$. However, the relevant region of
interest for our purposes is the spectrum for $q_T \gtrsim 1\GeV$, which is determined
by much smaller $b_T$, namely where $1/b_T \sim q_T \gg\lqcd$
is still in the perturbative domain. In this region, nonperturbative effects
are not negligible, but we can systematically expand them in an operator
product expansion (OPE) in powers of
$\lqcd b_T \ll 1$~\cite{Collins:1981uw, Collins:1984kg, Collins:1350496,
Collins:2014jpa, Collins:2016hqq, Vladimirov:2020umg, Ebert:2022cku}
(see also \Refcite{FerrarioRavasio:2020guj}).
The OPE of the final TMD PDF entering the factorized cross section takes the
form~\cite{Ebert:2022cku}
\begin{equation} \label{eq:tmdpdf_ope}
\tilde{f}_i(x, b_T, \mu, Q)
= \tilde{f}_i^\zero(x, b_T, \mu, Q)
\biggl\{ 1 + b_T^2\Bigl[\Lambda_{2,i}(x) + \lambda_2^\zeta\,\ln\frac{b_T Q}{b_0} \Bigr]
+ \ord{\lqcd^4 b_T^4} \biggr\}
\,,\end{equation}
where the leading term $\tilde f^\zero_i$ corresponds to the (resummed) purely
perturbative result at canonical boundary scales, and the higher-order terms
arise from expanding the nonperturbative contributions to the CS kernel and
TMD PDF boundary condition (at canonical $\zeta_0 = b_0^2/b_T^2$),
\begin{align} \label{eq:np_ope}
\tilde\gamma^{\nonp}_{\zeta}(b_T)
= \lambda_2^\zeta\, b_T^2 + \ord{\lqcd^4 b_T^4}
\,,\nn \\
\tilde f_i^\nonp(x, b_T) = 1 + \Lambda_{2,i}(x) \, b_T^2 + \ord{\lqcd^4 b_T^4}
\,.\end{align}
Here, $\lambda_2^\zeta$ is a single number related to a gluon vacuum
condensate~\cite{Vladimirov:2020umg}, while $\Lambda_{2,i}(x)$ is in principle
still a function of $x$ and $i$. For the $q_T$ spectrum, $\lambda_2^\zeta$ is the
parametrically leading nonperturbative effect since in \Eq{tmdpdf_ope} it is enhanced
by the large logarithm $\ln(b_TQ)$.

\subsubsection{Nonperturbative parameterization}
\label{sec:nonpert_model}

In our case, the perturbative contributions, $\tilde{f}^{\rm pert}$
and $\tilde\gamma_\zeta^{\rm pert}$ in \Eq{pert_nonpert_split} are given by the default
resummed results discussed in \sec{qT_resummation}. In addition,
the explicit $b_T$ dependence of the perturbative CS kernel is evaluated at $b^*(b_T)$,
as indicated in \Eq{pert_nonpert_split}, to regulate the
nonperturbative large-$b_T$ behaviour. Specifically, we use
\begin{align} \label{eq:bstar_presc}
b^{*}(b_T) = b_T \biggl( 1+ \frac{b_T^6}{b_{\rm max}^6} \biggr)^{-\frac{1}{6}}
\,,\end{align}
which smoothly interpolates between the perturbative small-$b_T$ region
and the nonperturbative large-$b_T$ region, where $b_0/b_{\rm max} = 1 \GeV$
acts as an upper cutoff. That is,
$b^{*}(b_T \ll b_{\rm max}) = b_T$ while $b^{*}(b_T \gg b_{\rm max}) = b_{\rm max}$
with a smooth transition in between.
The advantage of using a higher power than the usual quadratic $b^*$ prescription
is that in this way, the perturbative result differs from the strictly canonical
one only starting at $\ord{b_T^6}$, which means it does not alter the OPE below
$\ord{b_T^6}$~\cite{Ebert:2022cku}.

To include the nonperturbative contributions, we have to choose a nonperturbative
model. We require the model to reproduce the correct OPE in \Eq{np_ope} for $b_T\to 0$.
The OPE does not constrain the behaviour for $b_T\to\infty$. While as mentioned before the
precise functional form at large $b_T$ is of limited relevance, to have
some guidance we require the model to obey the asymptotic limits suggested
in~\Refcite{Collins:2014jpa},
\begin{alignat}{9}
\tilde{\gamma}_{\zeta}^\nonp(b_T \to \infty) &\to -{\rm const}
\,,\nn\\
\ln \tilde{f}^\nonp(x, b_T \to \infty) &\to - {\rm const} \times b_T
\,.\end{alignat}
A convenient choice satisfying the requirements in both limits is as follows,
\begin{align}
2\tilde{\gamma}_{\zeta}^\nonp (b_T) =
\tilde{\gamma}_{\nu}^\nonp (b_T)
&= -\lambda_{\infty} \tanh \biggl( \frac{\lambda_2}{\lambda_{\infty}}\, b_T^2
+ \frac{\lambda_4}{\lambda_{\infty}}\, b_T^4 \biggr)
\\\nn
&=
\begin{cases}
-\lambda_2\, b_T^2 - \lambda_4\, b_T^4 + \ord{b_T^6} \qquad &1/b_T \gg \lqcd
\\
- \lambda_\infty & 1/b_T \ll \lqcd
\,,\end{cases}
\nn\\
\ln \tilde{f}^\nonp (x, b_T) &= - b_T \, \Lambda_{\infty} \tanh \biggl[ \biggl( \frac{\Lambda_2}{\Lambda_{\infty}}
+ \frac{\Lambda_4}{\Lambda_{\infty}} b_T^2 \biggr) b_T
+ \frac{1}{3} \frac{\Lambda_2^3}{\Lambda_{\infty}^3} b_T^3 \biggr]
\\\nn
&= \begin{cases}
- \Lambda_2\,b_T^2 - \Lambda_4\,b_T^4 + \ord{b_T^6} \qquad & 1/b_T \gg \lqcd
\\
-\Lambda_\infty\,b_T & 1/b_T \ll \lqcd
\,,\end{cases}
\end{align}
where we exploited that $\tanh(x\to 0) \to x$ and $\tanh(x\to\infty) \to 1$.
Each model depends on three parameters: $\lambda_{2,4}$ and $\Lambda_{2,4}$
determine the quadratic and quartic terms in the expansion around $b_T = 0$ and
are in one-to-one correspondence with the quadratic and quartic OPE coefficients.
The parameters $\lambda_{\infty}$ and $\Lambda_{\infty}$ set the asymptotic behaviour for
$b_T \to \infty$.

We note that the above model for the TMD PDF does not yet include flavor or $x$ dependence,
which would be required for a more complete treatment.
As we explain in \sec{asimov_setup}, this simplified setup is sufficient for Asimov fits
since we can use the same model in both pseudodata and fitted theory model. However,
it is likely not sufficient for a fit to real data, which in general requires flavor and $x$
dependence~\cite{Bacchetta:2018lna, Moos:2023yfa, Bacchetta:2024qre}. As shown in \Refcite{Billis:2024dqq},
the full flavor and $x$ dependence of the nonperturbative TMD model
can be approximated by an effective flavor-averaged model that only depends
on the rapidity $Y$ and a given resonant boson type and center-of-mass energy.
However, this also implies that once several different rapidity bins are considered,
the effective TMD parameters must at least depend on rapidity. The fits to the $q_T$ spectrum
performed in \Refcite{CMS:2024lrd} found
indeed evidence for the presence of a nontrivial effective rapidity dependence of
$\Lambda_2$.

For the purpose of our Asimov fits, we also need some representative values for the
nonperturbative parameters to be used as true values in our Asimov pseudodata.
As already mentioned above, the nonperturbative CS kernel can be constrained by lattice
QCD data~\cite{LatticePartonLPC:2023pdv,Shu:2023cot,Avkhadiev:2023poz}.
This requires properly accounting for quark flavor thresholds and quark mass effects
as will be discussed in \Refcite{flavorthr_cs}.
Using \Refcite{flavorthr_cs} to fit our nonperturbative parameterization of
the CS kernel to the lattice QCD data, we obtain the following results,
\begin{align} \label{eq:cs_constraint}
\lambda_{\infty} & = 1.6853 \pm 0.5069 \, , \nonumber \\
\lambda_2 & = (0.0870 \pm 0.0332) \GeV^2
 \, , \nonumber \\
\lambda_4 &= (0.0074 \pm 0.0066) \GeV^4
\,,\end{align}
with the resulting correlation matrix
\begin{align} \label{eq:csmatrix}
C =
\begin{pmatrix}
1       & 0.5212  & -0.7249 \\
0.5212  & 1       & -0.9135 \\
-0.7249 & -0.9135 & 1
\end{pmatrix}
\,.\end{align}
We will use these as representative values for the CS kernel parameters of our
model.
For the TMD parameters, due to a lack of robust constraints, we choose a generic size
of $(0.5\GeV)^n$ as representative central values. Specifically, we take
\begin{equation}
\Lambda_2 = 0.25 \GeV^2
\,,\qquad
\Lambda_4 = 0.06 \GeV^4
\,,\qquad
\Lambda_{\infty} = 1 \GeV
\,.\end{equation}
We have checked that the results of our Asimov fits do not depend on the precise
central values chosen.

\subsection{Asimov fit setup}
\label{sec:asimov_setup}

In this section, we outline the methodology and rationale behind the use
of Asimov fits, and provide all relevant details of our setup
used to study the uncertainties in extracting $\as(m_Z)$ from the Drell-Yan
$q_T$ spectrum.

\subsubsection{General discussion}

To estimate the expected uncertainties in the extraction of $\as(m_Z)$ -- or any other
theory parameter of interest -- it is common practice to perform Asimov fits.
These fits serve as controlled tests of the fitting framework and allow a transparent
study of the sensitivity to different theory ingredients.%
\footnote{The rationale behind the Asimov fits in this setup is in many ways
similar to the closure tests performed in other contexts where pseudodata are used
to test the quantification of uncertainties.}

The fits we perform use a standard $\chi^2$ minimization procedure, defined as
\begin{align} \label{eq:chi2}
\chi^2 = \sum_{i,j} (y_i - \lambda_i)^{T}C^{-1}_{ij} (y_j - \lambda_j)
+ \sum_i \frac{(\theta_i - 0)^2}{\Delta\theta_i^2}
\,,\end{align}
where $y_i$ represent the data, $\lambda_i$ the corresponding theory model,
and $C_{ij}$ the covariance matrix
encoding all experimental uncertainties and correlations between the data points.
The $\lambda_i$ implicitly depend upon $\as(m_Z)$ and the parameters of the underlying
theory prediction, i.e.\ the TNPs and nonperturbative parameters discussed in \secs{tnps_for_qT}{nonp_theory}.
Within the fit, $\as(m_Z)$, the TNPs, and nonperturbative parameters are treated as continuous
parameters.
The last term imposes the Gaussian theory constraint on the TNPs, $\theta_i$, about their central values of $0$ with a prior uncertainty of $\Delta\theta_i$, as discussed in \sec{tnps_for_qT}.

What distinguishes an Asimov fit is that the data are replaced by pseudodata -- i.e.,
the (unfluctuated) theory predictions computed at a fixed known value of the parameter
of interest, in our case $\as(m_Z) = 0.118$.
The covariance matrix remains that of the actual experimental measurement, ensuring that the uncertainties
remain realistic. By replacing experimental data with theory-generated pseudodata,
this setup offers multiple advantages:
\begin{itemize}
\item The results are unobscured by statistical fluctuations or potential
outlier behaviour present in the real data.
\item It avoids potential inconsistencies or biases due to subleading effects
present in the real data but unaccounted for in the theory model.
\item The impact of various theory components or uncertainties can be tested
and estimated in isolation. In particular, subleading effects can be consistently
dropped in both pseudodata and theory model.
\item Since the theory model and pseudodata are matched
(except for intentional differences in some tests),
the minimum $\chi^2$ should be close to zero, providing a clean validation test for the fitting procedure.
\end{itemize}

As discussed in \sec{theory_inputs}, we focus on the dominant sources of theory uncertainty,
namely the perturbative resummation and nonperturbative modelling of the dominant leading-power
contribution to the small-$q_T$ spectrum.
Subleading effects such as nonsingular power corrections, finite quark-mass effects,
and QCD+electroweak corrections are known to impact the
shape of the $q_T$ spectrum at the few-percent level but can be
neglected in our Asimov fits as long as we consistently neglect them in both the theory model and pseudodata.
Indeed, being subleading effects, their associated uncertainty will also be subdominant
with respect to the dominant sources of uncertainties.

One might worry that even subleading contributions, by altering the shape of the spectrum,
could subtly affect the estimate of the dominant uncertainties.
Among these, the nonsingular terms are the most relevant, as they contribute up to $\ord{5\%}$ of the spectrum for $q_T \lesssim 30\GeV$~\cite{Ebert:2020dfc}.
For this reason, in \app{nonsingular}, we explicitly verify that including nonsingular terms up to
$\ord{\as^2}$ has a negligible effect on the estimated uncertainty for $\as(m_Z)$.

Nevertheless, we emphasize that in fits to real data, these subleading effects must
of course be included in the fitted theory model, as they are present in the real data.
While they may not substantially affect our estimate of the dominant uncertainties,
they can significantly bias the central value of $\as(m_Z)$ and degrade the fit quality
if not properly accounted for.

\subsubsection{Fit setup}

We perform our study using pseudodata based on the inclusive $Z$-boson $q_T$
spectrum at $\Ecm = 8\, \TeV$ in the dilepton mass window
$80\GeV \leq Q \leq 100\GeV$ as measured by ATLAS in \Refcite{ATLAS:2023lsr}.
Specifically, we include the first nine $q_T$ bins covering the region $q_T \in [0, 29]\GeV$,
for each of eight bins in rapidity $Y \in [0, 3.6]$, yielding 72 data points in total.
We use the full experimental covariance matrix, incorporating all bin-to-bin correlations,
as provided by ATLAS.
On top of this, we include a fully correlated $1.8\%$ luminosity uncertainty.
These are the data also used in \Refcite{ATLAS:2023lhg}.

For our theory predictions, both for pseudodata and for the fitted theory model,
we employ the \scetlib\ framework as discussed in \secs{qT_resummation}{nonp_theory}.
All pseudodata are generated at a fixed value of $\as(m_Z) = 0.118$, and thus all
fits must give a similar value for the fitted $\as(m_Z)$.
When using scale variations, we use the resummation at N$^4$LL and follow the
variation scheme defined in \scetlib~\cite{Billis:2024dqq} as described in
\sec{profile_scale_variations}.

When using the TNP approach, we use the \scetlib\ implementation of
\Refcite{Tackmann:2024kci} as described in \sec{tnps_for_qT}. 
Our nominal theory model is given by N$^{3+1}$LL while for the pseudodata
we use either the default N$^{3+1}$LL result with zero as true values for all TNPs
or alternatively N$^4$LL, which is equivalent to N$^{3+1}$LL with all TNPs set
to their true values. As we will see, the advantage of using N$^{3+1}$LL is that
the perturbative precision is sufficiently high compared to the experimental
precision to obtain nontrivial results. At the same time the true values of all
TNPs are still known, which allows us to perform more nontrivial tests by fitting to N$^4$LL,
simulating a fit to real data.

For the nonperturbative modelling, we always use the same nonperturbative model
for both the theory model and pseudodata, and whenever the nonperturbative parameters
are fixed they are set to their central values given in \sec{nonp_theory}.

Regarding the PDF dependence, we always use the MSHT20aN3LO PDF set~\cite{McGowan:2022nag},
and account for the interplay between $\as(m_Z)$ and the PDFs in the $q_T$ spectrum.
Since $\as$ and the PDFs are strongly correlated, we ensure that the PDF set entering
the prediction is consistent with the value of $\as(m_Z)$ being tested. For this purpose, we
perform a linear interpolation between the available PDF sets for $\as(m_Z) = 0.118$ and $\as(m_Z) = 0.117$
when $\as(m_Z) < 0.118$ and between the PDF sets for $\as(m_Z) = 0.118$ and $\as(m_Z) = 0.119$
when $\as(m_Z) > 0.118$. This interpolation is only used for the PDFs, while
the dependence on $\as(m_Z)$ in the perturbative prediction itself is
always treated exactly.

For minimization, we use the standard \minuit\ fitting tool~\cite{James:1975dr,osti_4252678,James:2004xla}.


\section{Perturbative Uncertainties}
\label{sec:perturbative}

In this section, we present our results obtained using both scale
variation and TNP-based approaches to estimate the expected perturbative uncertainty in the
extraction of $\as(m_Z)$ from the $Z$ $q_T$ spectrum.
We describe and highlight two distinct methods that can be applied with TNPs:
scanning and profiling.
Subsequently, we explore the profiling of TNPs in more detail by performing fits
against higher-order pseudodata and by modifying the prior theory constraint
$\Delta \theta_n$ on the TNPs.

\subsection{Scale variations}
\label{sec:prof_scale_var}

\begin{figure}
\includegraphics[width=0.47\textwidth]{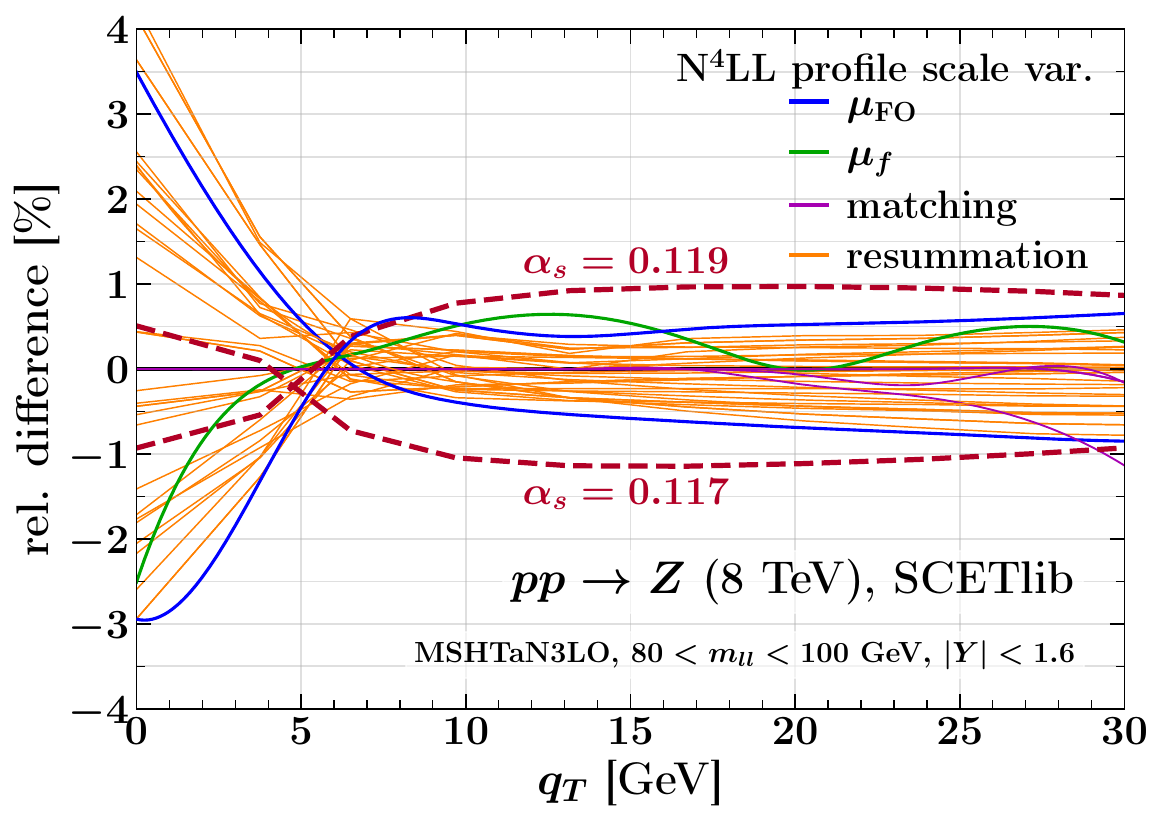}%
\hfill%
\includegraphics[width=0.52\textwidth]{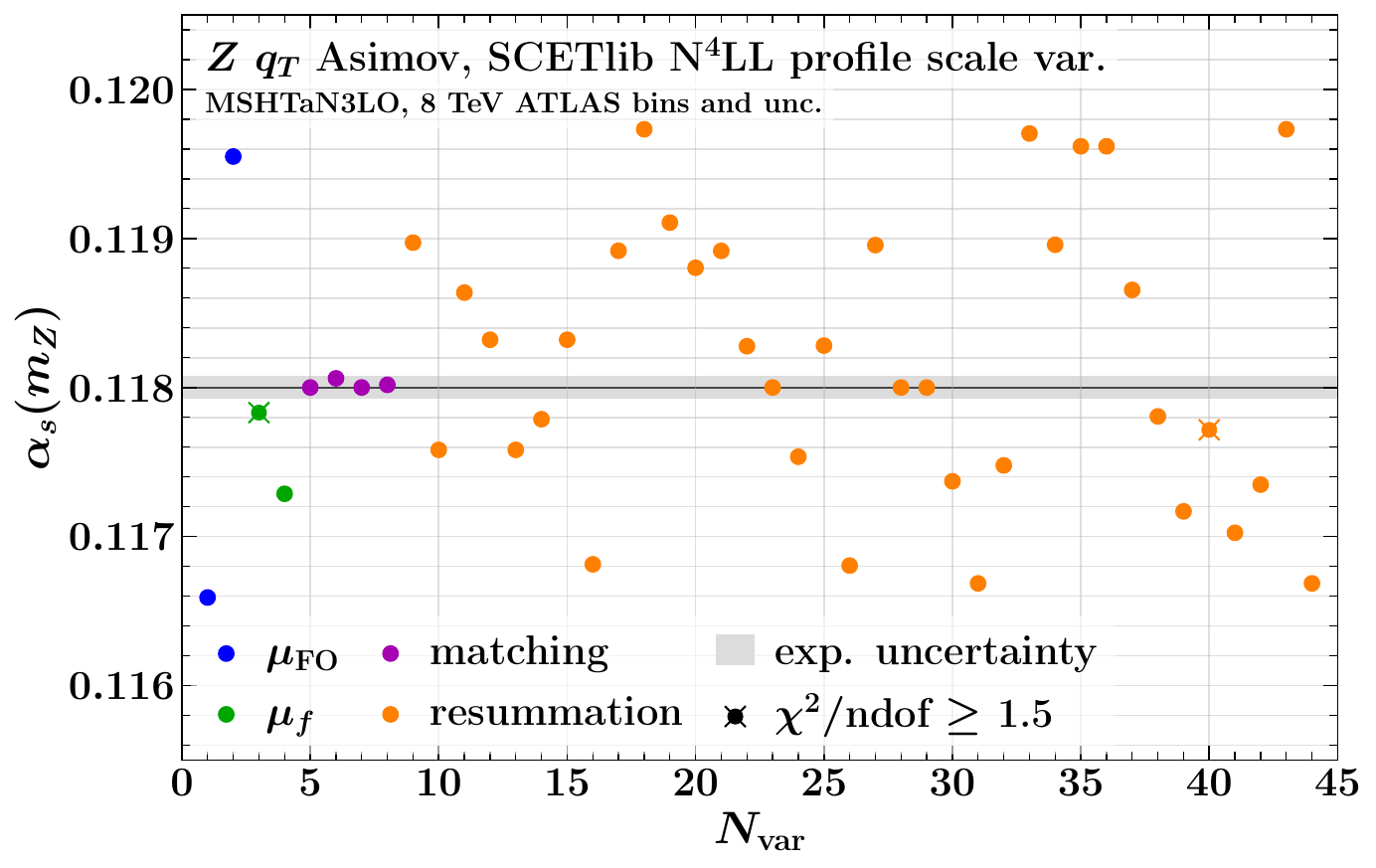}%
\caption{Left panel: Relative differences in the $q_T$ spectrum of individual
scale variations at N$^{4}$LL. Different colored lines show different classes
of variations as defined in \sec{profile_scale_variations}. The red dashed lines show
the relative difference from varying $\as(m_Z)$ for comparison.
Right panel: The resulting $\as(m_Z)$ values when performing the Asimov fit
to the central N$^4$LL prediction for each scale variation. Different
colors correspond to the same classes of variations
as on the left. The grey band shows the fit uncertainty for comparison.
}
\label{fig:N4LL_scanning}
\end{figure}
To estimate the uncertainty due to missing higher-order corrections 
— referred to as the perturbative uncertainty — it is common practice to consider scale variations.
A key argument for this procedure is that the envelope of various scale variations can provide a
measure of the overall perturbative uncertainty in the spectrum.
By taking the envelope, one can mitigate the arbitrariness in choosing the specific number and type of
scale variations to be considered.
As a result, it is a widely-used method for estimating perturbative uncertainties in differential spectra,
which may often provide a reasonable estimate of the approximate size of uncertainties.
The resulting uncertainty band then has limited sensitivity to the individual scale variation shapes.
This is sufficient for the uncertainties at the level of the $q_T$ spectrum where we only require the rough size of the uncertainty band but not the detailed correlations within it.
The inadequacy of scale variations manifests itself when interpreting the spectrum, i.e.,
when the perturbative uncertainty needs to be propagated from the spectrum to the parameter of interest,
in our case $\as(m_Z)$.

One may ask why is it not sufficient to propagate each scale variation to a fit of $\as(m_Z)$ and then take the envelope at the level of the extracted $\as(m_Z)$ value? Any specific scale choice would then correspond to a fully  correlated or anti-correlated assumption, as explained in
\sec{scale_variations}. In taking the envelope of scale variations at the level of the $q_T$ spectrum we explicitly acknowledge that their individual shapes are not physical, and assume complete ignorance regarding the correlations between different bins in the spectrum. However, the propagation
of the uncertainty from the $q_T$ spectrum to $\as(m_Z)$ is crucially sensitive to the
precise point-by-point correlations and the individual theory shapes. Accordingly, any uncertainty resulting from such a scale-variation scanning and enveloping procedure for $\as(m_Z$) should not be
considered as a trustworthy or meaningful uncertainty but merely as an indication of the possible
impact of missing higher-order corrections, i.e., as a means to investigate the
possible theory sensitivity (which could still very well be over- or underestimated).

Despite these known limitations of the scale variation approach, it is still interesting
in this context to use our pseudodata setup to see what happens if scale variations are utilised. For our
purposes they will serve to demonstrate that correlations and the precise shape of the
theory uncertainties are indeed critical for the extraction of $\as(m_Z)$.

In the left panel of \fig{N4LL_scanning}, we show the breakdown of all scale variations
for the $q_T$ spectrum at N$^4$LL from \scetlib\ as discussed in \sec{profile_scales}.
Overlaid on these variations are the central predictions for $\as(m_Z) = 0.117$ and $\as(m_Z) = 0.119$,
illustrating that the sensitivity to $\as(m_Z)$ in the $q_T$ spectrum is primarily a shape effect.
To illustrate the scale-variation scanning in practice, we use the Asimov fit setup described in \sec{asimov_setup}.
Our pseudodata corresponds to the N$^4$LL central scale prediction at $\as(m_Z) = 0.118$.
We then repeat the fit of $\as(m_Z)$ for each scale variation, each representing a different
theory model.
The right panel of \fig{N4LL_scanning} shows the results, where each point
represents the fitted value of $\as(m_Z)$ for one of the scale variations on the left.
Points marked with a cross indicate cases where $\chi^2/{\rm n_{\rm dof}} \geq 1.5$,
signifying a very poor fit to the pseudodata.
The fitted $\as(m_Z)$ values strongly depend on the choice of scale variation.
While some variations have almost no impact on $\as(m_Z)$, yielding values very close
to the nominal value of $0.118$, others yield very different values.
This demonstrates that as anticipated the impact on $\as(m_Z)$ strongly depends on the precise shape
of any given variation.
As a result, the precise choice of scale variations directly dictates the size of the
would-be perturbative uncertainty, making this approach unsuitable for a reliable uncertainty estimate.

Nevertheless, for reference we can obtain a scale-variation based estimate of the
perturbative uncertainty in $\as(m_Z)$
by considering different options to envelope the fitted results.
The envelopes of each different class of scale variations described in \sec{profile_scale_variations}
yield
\begin{alignat}{9} \label{eq:subclass_env}
\Delta_{\rm FO} &= 1.55 \times 10^{-3}
\,, \qquad &&
\Delta_f &= 0.71 \times 10^{-3}
\,, \nn \\
\Delta_{\rm resum} &= 1.73 \times 10^{-3}
\,, \qquad &&
\Delta_{\rm match} &= 0.06 \times 10^{-3}
\,.\end{alignat}
We can then either sum these in quadrature or take the naive
total envelope, obtaining
\begin{align} \label{eq:totalenv}
\text{sum of envelopes:}\qquad
\Delta_{\rm pert}
&= \sqrt{\Delta_{\rm FO}^2 + \Delta_f^2 + \Delta_{\rm resum}^2 + \Delta_{\rm match}^2}
= 2.43 \times 10^{-3}
\,,\nn \\
\text{total envelope:}\qquad
\Delta_{\rm pert} &= 1.73 \times 10^{-3}
\,.\end{align}
Regardless of the approach, we find that the perturbative uncertainty estimated
via scale variations at the currently highest known order, N$^4$LL, is large --
much larger than our desired level of precision for $\as(m_Z)$

\subsection{Theory nuisance parameters}
\label{sec:tnps_perturbative}

As explained in \sec{theory_unc}, to properly account for the point-by-point
correlations across the $q_T$ spectrum, the uncertainty must be decomposed into
well-defined components corresponding to the underlying sources of uncertainty.
The TNP approach has been developed to do just that.
In this section, we explore different methods using TNPs to estimate the perturbative
uncertainty in the extraction of $\as(m_Z)$. Our setup is as described in
\sec{asimov_setup}, with the default theory model at N$^{3+1}$LL and the Asimov
pseudodata at N$^{3+1}$LL and later also at N$^4$LL.

\subsubsection{Scanning}
\label{sec:tnps_perturbative_scanning}

\begin{figure}
\includegraphics[scale=0.4]{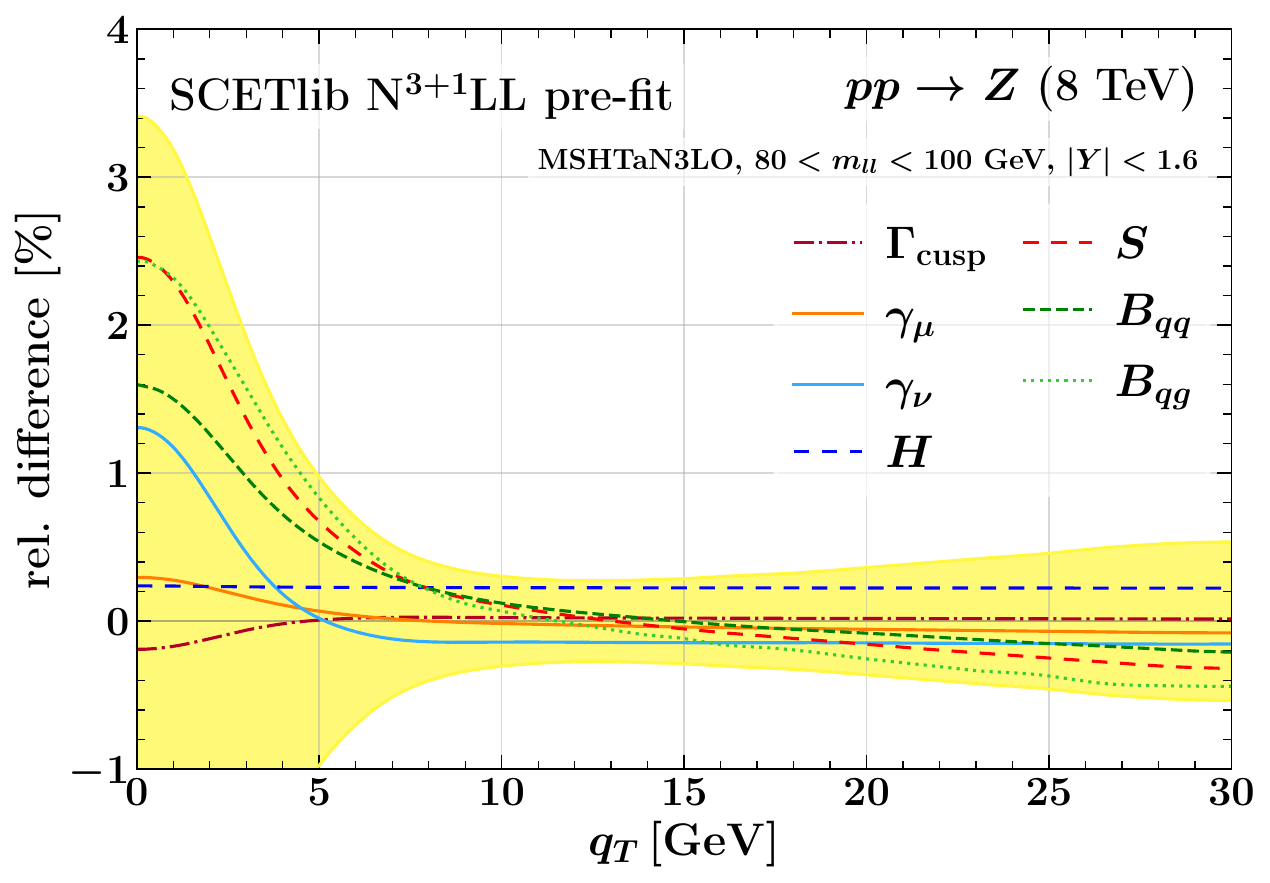}%
\hfill
\includegraphics[scale=0.4]{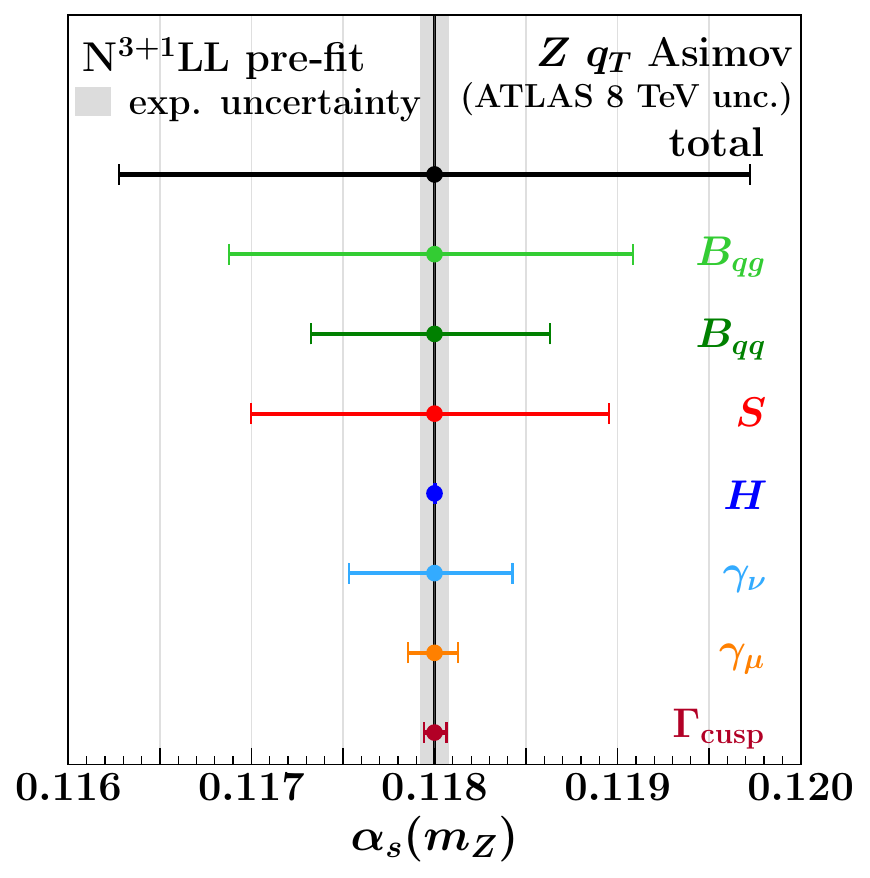}%
\caption{Left panel: Relative uncertainties in the $q_T$ spectrum with TNPs at N$^{3+1}$LL.
The different lines show the impact of varying the corresponding
TNP by $+1$ or $-1$, corresponding to 68\% theory CL.
The yellow band shows their sum in quadrature.
Right panel: Corresponding uncertainty on $\as(m_Z)$ from performing the Asimov fit
to the central N$^{3+1}$LL prediction for each TNP variation. The grey band
shows the fit uncertainty for comparison.}
\label{fig:N3p1LL_scanning}
\end{figure}

The first method we consider is scanning, which follows an analogous procedure to the one used for scale variations. However, the crucial difference is that we have now decomposed the uncertainty into its distinct sources, parameterized in terms of well-defined theory nuisance parameters, which correctly capture and propagate the bin-by-bin correlations in the $q_T$ spectrum.

The left panel of \fig{N3p1LL_scanning} presents the breakdown of the individual TNP
variations for the $q_T$ spectrum at N$^{3+1}$LL accuracy.
Each TNP is varied by $\Delta\theta_n = 1$.
Since the resulting variations in the spectrum are nearly symmetric, for clarity, we display only the
up variations for the anomalous dimensions, hard, and soft functions, and the
down variations for the beam functions.
Each TNP corresponds to an independent source of uncertainty and is thus varied independently,
capturing the correct correlations across the $q_T$ spectrum.
We can now sum in quadrature the seven independent TNP variations to obtain the
total uncertainty band, depicted in yellow.

To perform the scanning,
we use the central N$^{3+1}$LL prediction as Asimov data and
repeat the fit for N$^{3+1}$LL theory models with each TNP varied separately up
and down by $\Delta\theta_n = 1$.
The right panel of \fig{N3p1LL_scanning} shows the results, where
the error bars represent the difference between the fitted $\as(m_Z)$
and the nominal expected value of $0.118$, i.e., the
uncertainty on the fitted $\as(m_Z)$ due to a given TNP. As for the spectrum,
we can now sum the individual uncertainties in quadrature to obtain
the total uncertainty on $\as(m_Z)$, which yields
\begin{align}
\Delta_{\rm pert} = 1.75 \times 10^{-3}
\,.\end{align}
Although this estimate accounts for the correct theory correlations in the $q_T$ spectrum,
it does not reach our desired precision on $\as(m_Z)$.

The scanning approach however does not fully exploit the high precision
available in the data, which could in principle impose strict constraints on the
parameters of the theory model.
As each parameter is separately varied but otherwise held fixed in the fit,
the fit must compensate the resulting change in the theory model as best it
can solely into a change of $\as (m_Z)$. However there is no feedback whether a given variation of the
theory model is actually favored or disfavored by the data.
This is a key limitation of the scanning approach.
Furthermore, since we explore the parameter space separately in each
direction rather than all together, the global minimum
of the multi-dimensional parameter space is not explored, which means the quality
of the minimum and the precision obtained can be suboptimal.

Rather than relying on scanning -- and its inherent limitations -- we would like
to be able to fully explore the parameter space and allow the fit to select
the regions where the theory model best matches the data while
avoiding regions that are inconsistent with it. 
This challenge can be effectively addressed by profiling the TNPs in the fit
as we discuss next.

\subsubsection{Profiling}
\label{sec:profiling}

A more advanced approach to estimate systematic uncertainties is by profiling them.
In this approach, the nuisance parameters associated with the systematic
uncertainties are simultaneously fitted along with the parameters of interest,
rather than scanning them one-by-one. In this way the full
parameter space is explored, avoiding the limitations of the scanning approach and enabling
the data to constrain variations within these uncertainties to agree with the measurements.
This therefore enables a reduction in the size of the associated uncertainties for the parameters of interest.
This is a standard approach used for experimental systematic uncertainties (see e.g.~\Refscite{Cowan:2010js, Cousins:2024gkj})
as well as for PDF uncertainties \cite{Paukkunen_2014,camarda2015qcdanalysiswzboson,Willis:2018yln},
though several caveats must be considered in general. In particular, one must be careful
in the application to PDF uncertainties -- we leave this discussion to a future dedicated paper~\cite{pdf_paper}.

We now apply this approach to the nuisance parameters encoding
the perturbative uncertainties, i.e.\ the TNPs, by fitting them together
with $\as(m_Z)$. This allows the fit to find the optimal balance between
adjusting one or more TNPs, and consequently the theory model, or
adjusting $\as(m_Z)$ to find the best description of the data. We stress that
(in contrast to scale variations) fitting the TNPs is theoretically consistent
and allowed because the TNPs are proper parameters, which encode the correct
theory correlations. As a result, this method enables constraining the TNPs by
the experimental data, thereby (potentially) reducing the theory uncertainties.
The allowed range of variation for the TNPs is also constrained by the prior theory
constraint, as explained in \secs{tnps_for_qT}{asimov_setup}, for which we
use $\Delta \theta_n = 1$ by default.

\begin{figure}
\includegraphics[scale=0.4]{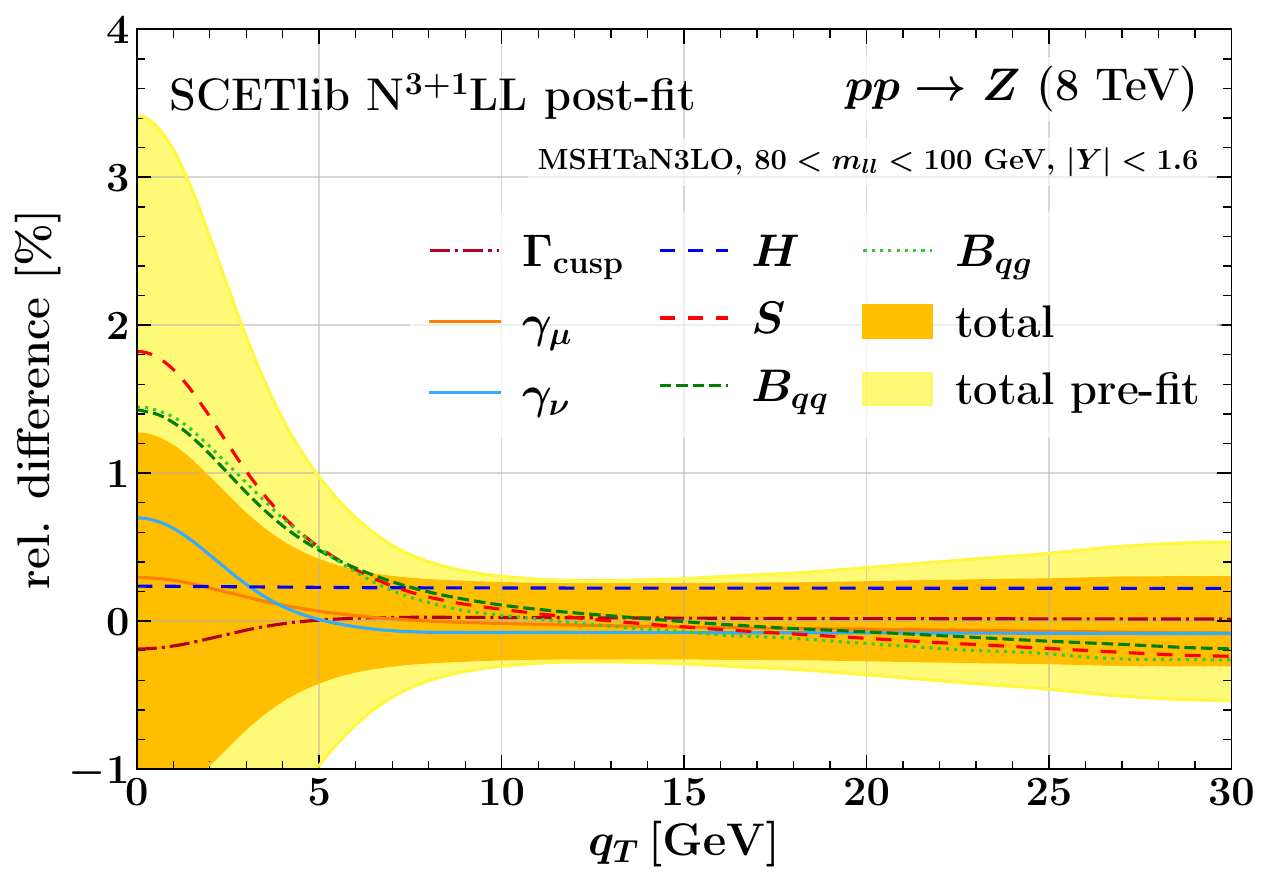}%
\hfill%
\includegraphics[scale=0.4]{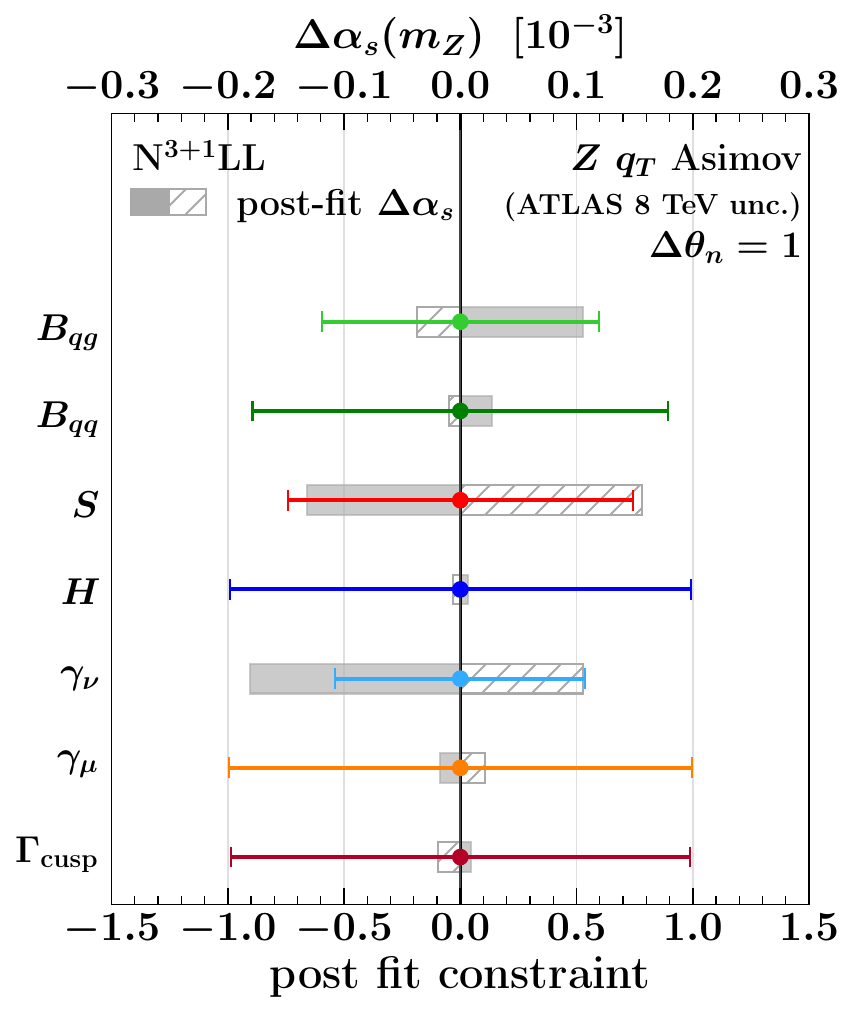}%
\caption{
Left panel: Relative uncertainties in the $q_T$ spectrum at N$^{3+1}$LL before
and after profiling the TNPs. The different lines show the post-fit relative impact of each TNP and
the orange band the total post-fit uncertainty. The yellow band shows the
pre-fit uncertainty corresponding to the yellow band in \fig{N3p1LL_scanning}.
Right panel: Post-fit constraints on the TNPs (error bars) and their impact on $\as(m_Z)$,
with the solid (dashed) grey band showing the impact of the post-fit downward (upward) TNP variations.}
\label{fig:N3p1LL_fits}
\end{figure}

To start, we perform an Asimov fit with both
pseudodata and theory model given by the N$^{3+1}$LL prediction.
The resulting perturbative uncertainty on $\as(m_Z)$ is%
\footnote{%
This value is the total fit uncertainty, which in principle includes the experimental
uncertainties as well. However, the pure fit uncertainty
when only fitting $\as(m_Z)$ is tiny, $0.06 \times 10^{-3}$, so we simply
refer to this uncertainty as the perturbative one, regardless of the use of profiling.}
\begin{align}
\Delta_{\rm pert} = 0.45 \times 10^{-3}
\,.\end{align}
The left panel of \fig{N3p1LL_fits} presents the post-fit prediction for the $q_T$
spectrum with a breakdown of the post-fit uncertainties, showing the upward
variation for each TNP, except for the
beam functions which display the downward variation.
As a general comment, which applies to this and all analogous plots in the following,
note that the individual TNP post-fit variations do not
necessarily have to lie within the total post-fit uncertainty, as there are now nontrivial correlations between the TNPs imposed by the data. In particular, if two TNPs are negatively correlated, their combined
uncertainty on the spectrum and $\as(m_Z)$ can be smaller than their individual impacts.

After profiling the TNPs, the uncertainties for both $\as(m_Z)$ and the $q_T$
spectrum are significantly smaller compared to simply
scanning the TNPs and the corresponding pre-fit uncertainties.
However, focusing solely on the final uncertainty after profiling can also be
misleading and may lead to incorrect conclusions.
An important aspect when profiling systematic uncertainties is to examine to what extent
the data actually pulls and constraints the nuisance parameters in order to ensure
that the constrained post-fit theory model is still adequate and allowed. In our case,
we have to ensure that the theory model is not overconstrained to the point that neglected
yet higher-order terms become relevant for determining the theory uncertainty~\cite{Tackmann:2024kci}.
The right panel of \fig{N3p1LL_fits} presents a pull/impact plot for all TNPs. The error
bars refer to the bottom $x$-axis and show the post-fit constraints for each TNP.
The solid (dashed) grey bars refer to the top $x$-axis and show the
corresponding impact on $\as(m_Z)$ due to each TNPs downward (upward) post-fit uncertainty.%
\footnote{%
The impacts are evaluated by repeating the fit while fixing each TNP one at a time
to its post-fit value plus/minus its post-fit uncertainty. The difference of the obtained
$\as(m_Z)$ to the result of the nominal fit gives the impact on $\as(m_Z)$.
}
The post-fit central value for all TNPs is $0$, in perfect agreement with their
values in the pseudodata, as should be the case here.
The post-fit constraints on the individual TNPs are overall not strong.
The individual impacts provide further insight into which TNPs are most important
(after profiling) for determining $\as(m_Z)$. Here, $\gamma_{\nu}$, $S$, and $B_{qg}$
have the biggest post-fit impact on $\as(m_Z)$ and they are also the ones that are
somewhat constrained.
This indicates that these parameters have the strongest correlation with $\as(m_Z)$.
Of course, to study the exact correlations between all parameters in detail, one has to
examine the post-fit covariance matrix.

\subsubsection{Profiling against \texorpdfstring{N$^4$LL}{N4LL}}
\label{sec:profiling_vs_N4LL}

\begin{figure}
\includegraphics[width=0.60\textwidth]{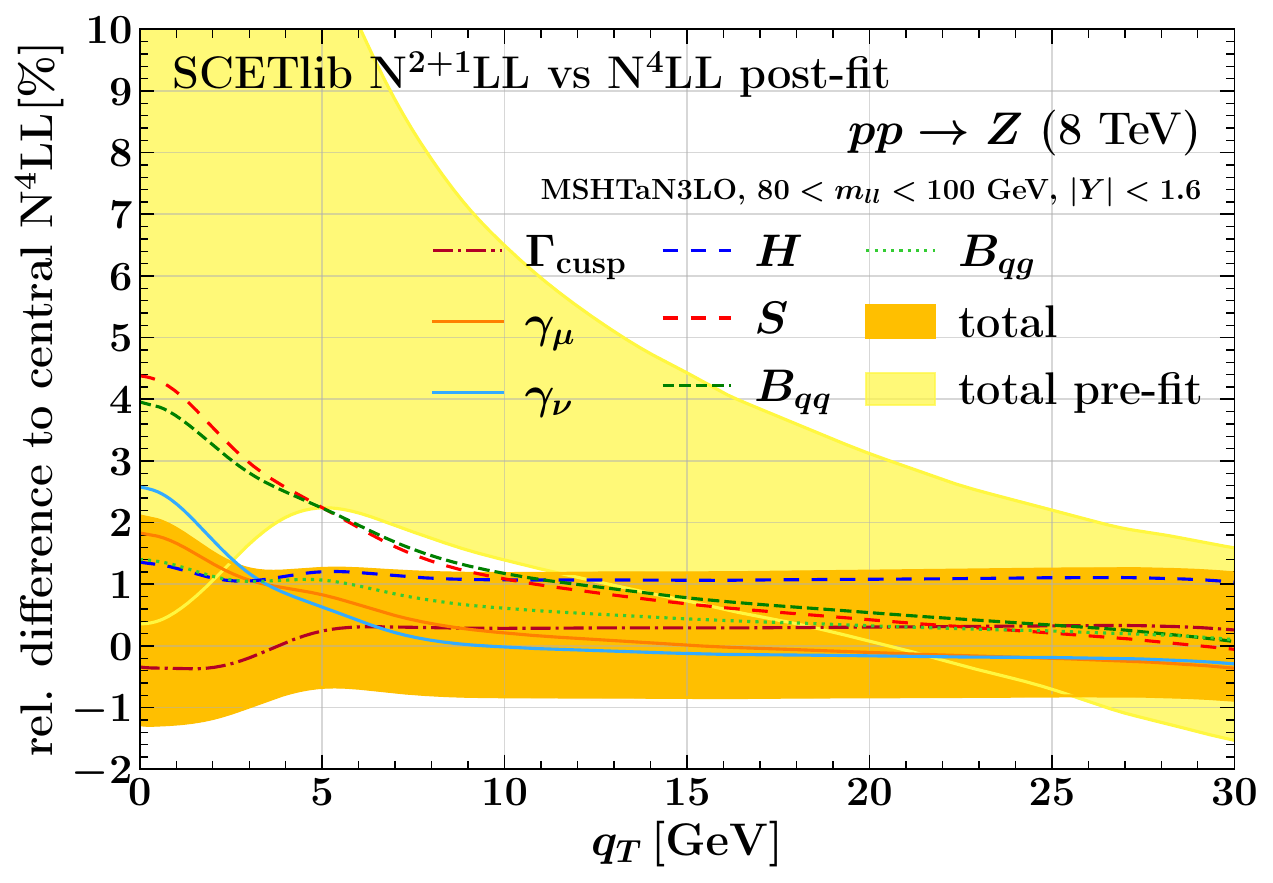}%
\hfill%
 \includegraphics[width=0.38\textwidth]{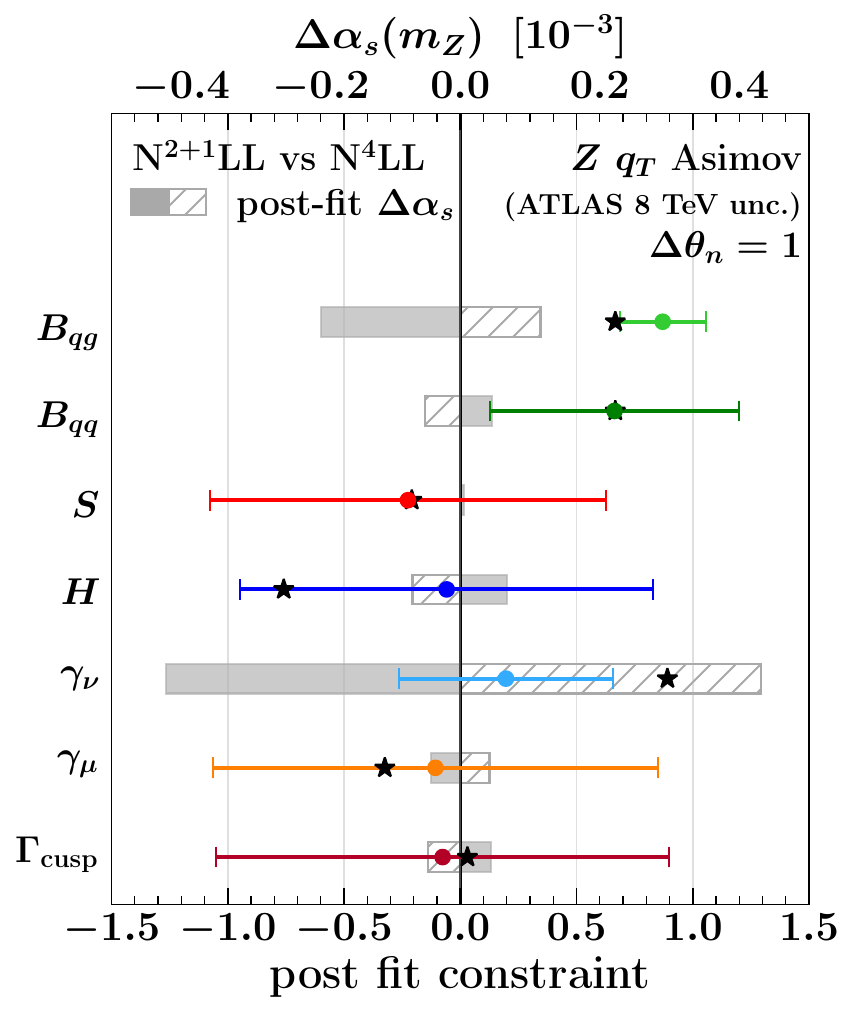}%
\caption{
Left panel: Uncertainties in the $q_T$ spectrum at N$^{2+1}$LL relative to the N$^4$LL,
before (yellow band) and after (orange band) profiling the TNPs. 
The different lines show the post-fit relative impact of each TNP.
Right panel: Post-fit constraints on the TNPs (error bars) and their impact on $\as(m_Z)$,
with the solid (dashed) grey band showing the impact of the post-fit downward (upward) TNP variations.
The stars indicate the true values of the TNPs.
}
\label{fig:N2p1LL_vs_N4LL}
\end{figure}

To further explore the profiling with TNPs, we next perform more general Asimov fits
where the pseudodata is generated from the highest N$^4$LL central
prediction, still evaluated at $\as(m_Z) = 0.118$, while using different orders for
the theory model, specifically N$^{2+1}$LL and N$^{3+1}$LL.
Since the N$^4$LL result contains the true values of the TNPs, this setup
mimics the fit to real data, which contain nature's all-order result, and provides a realistic test
of whether the TNPs are correctly constrained and whether we can recover their true
values at this order.

In \fig{N2p1LL_vs_N4LL}, we consider the case where the theory model is given by the N$^{2+1}$LL prediction.
The left panel shows the pre-fit and post-fit uncertainties for the $q_T$ spectrum
analogous to the left panel of \fig{N3p1LL_fits} but now relative to the true N$^4$LL result.
The profiling again leads to a significant reduction in the uncertainty. Furthermore,
we see that the post-fit prediction is now pulled toward the true result, with a
significant adjustment of the shape of the spectrum. Whilst the shape of the pre-fit prediction
differs from the true shape, the post-fit prediction closely reproduces it.

The right panel of \fig{N2p1LL_vs_N4LL} shows the corresponding pull plot for
the TNPs (analogous to the right panel of \fig{N3p1LL_fits}) with the true
values of the TNPs marked by stars. The fitted TNPs at N$^{2+1}$LL now exhibit
strong pulls toward their true values, showing that, as for the $q_T$ spectrum,
the fit is moving the theory model in the right direction also at the level of
individual TNPs. This is exactly what we would like to happen in this situation
where the data has a higher precision than the pre-fit theory constraint. However, some
of the TNPs, such as $\gamma_{\nu}$, $B_{qq}$, and in particular $B_{qg}$, are
quite strongly constrained now. A tight post-fit constraint implies that the
corresponding parameter at the next higher-order becomes relevant. In other
words, we are starting to ``overfit'' the N$^{2+1}$LL theory model. The
precision of the data is starting to exceed the inherent precision of the theory
model, suggesting that it is becoming insufficient and using a higher-order
theory-model might be warranted.

\begin{figure}
\includegraphics[width=0.60\textwidth]{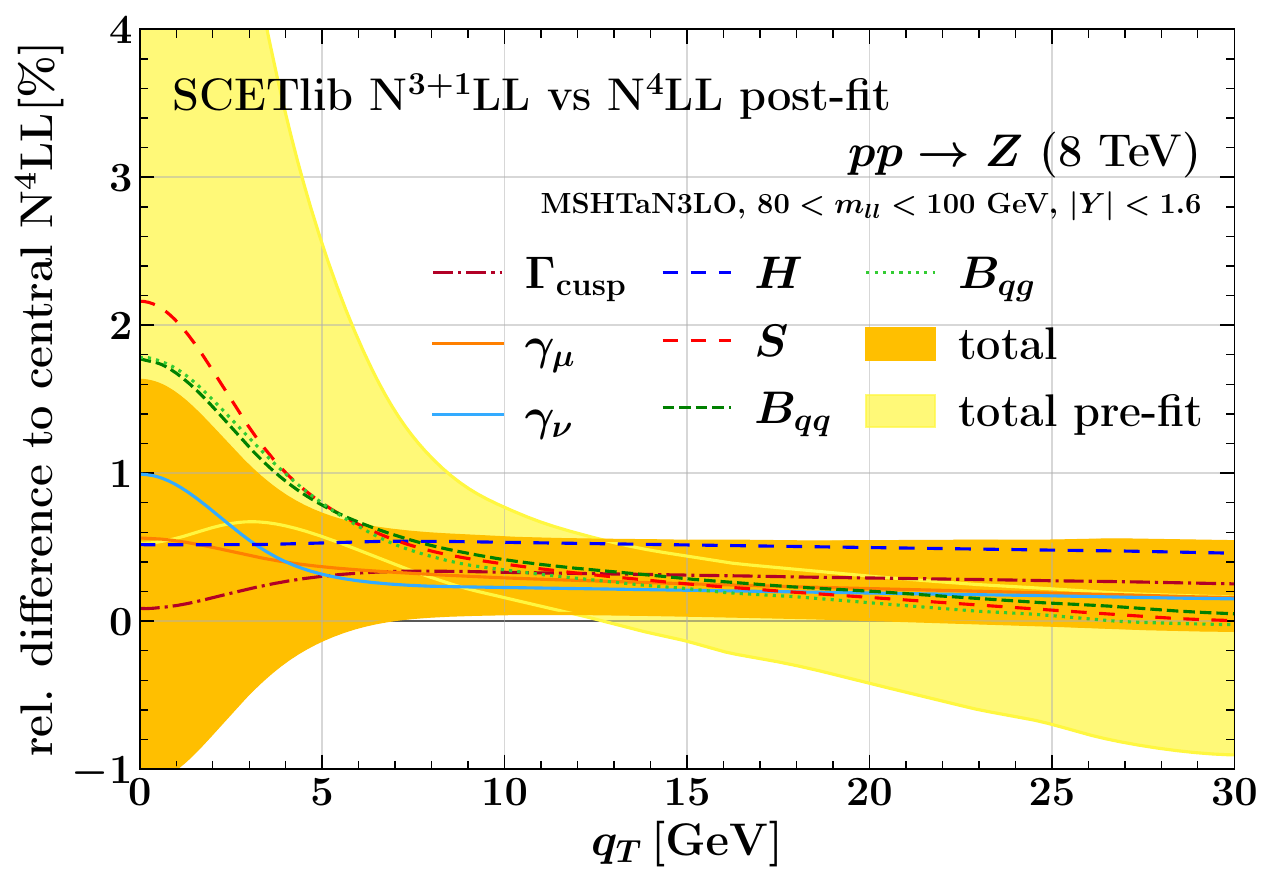}%
\hfill%
\includegraphics[width=0.38\textwidth]{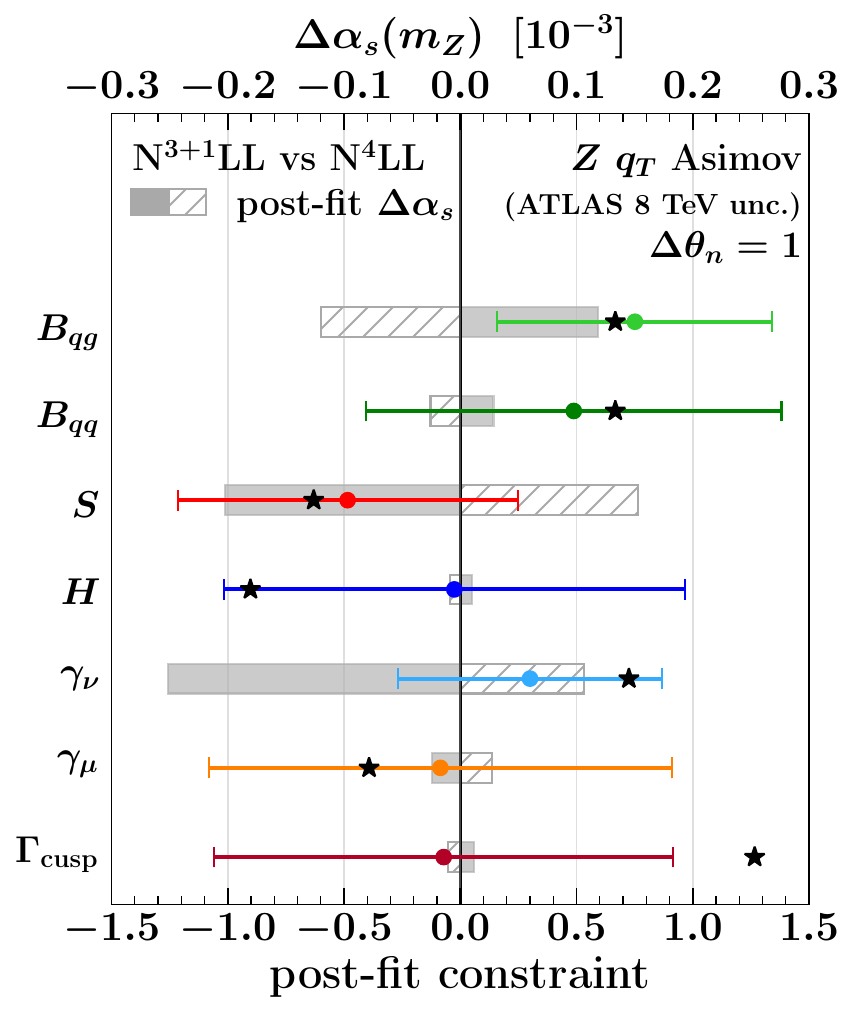}%
\caption{
Left panel: Uncertainties in the $q_T$ spectrum at N$^{3+1}$LL relative to the N$^4$LL,
before (yellow band) and after (orange band) profiling the TNPs. 
The different lines show the post-fit relative impact of each TNP.
Right panel: Post-fit constraints on the TNPs (error bars) and their impact on $\as(m_Z)$,
with the solid (dashed) grey band showing the impact of the post-fit downward (upward) TNP variations.
The stars indicate the true values of the TNPs.
}
\label{fig:N3p1LL_vs_N4LL}
\end{figure}

In \fig{N3p1LL_vs_N4LL} we repeat the same exercise with the N$^{3+1}$LL theory model.
The post-fit uncertainties for the $q_T$ spectrum are very similar to those
in \fig{N3p1LL_fits} when profiling N$^{3+1}$LL against itself, except that now
the central value is shifted toward the true result. The same holds at the level
of individual TNPs. Their post-fit uncertainties are very similar to before,
while their central values are shifted toward their true values, which is again
as we would like it to be.
Compared to the N$^{2+1}$LL case, the TNPs are now less constrained, as expected,
with no indications yet of overfitting the theory model.

\begin{figure}
\centering
\includegraphics[scale=0.4]{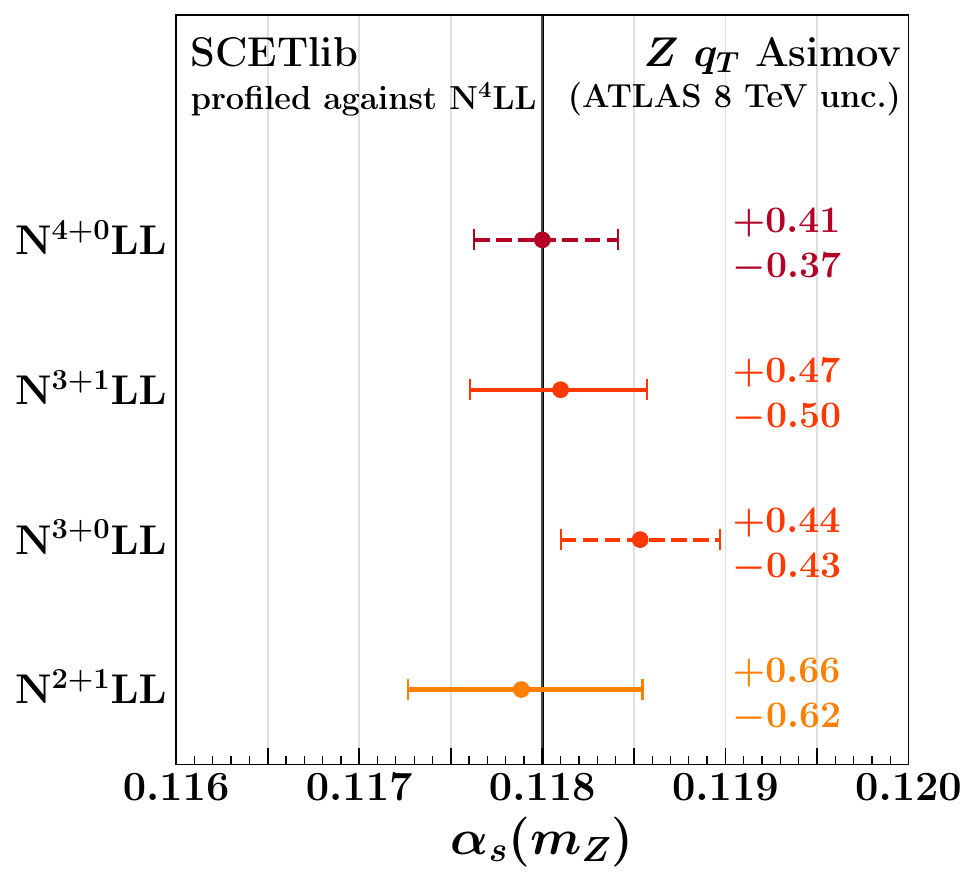}
\caption{Perturbative uncertainties on $\as(m_Z)$ when profiling the TNPs
at different orders against the central N$^4$LL prediction.
The numerically displayed uncertainties are in units of $10^{-3}$.
}
\label{fig:pullplotsasmZ_2}
\end{figure}

In \fig{pullplotsasmZ_2}, we present the results for $\as(m_Z)$
when profiling the TNPs at various orders against the N$^4$LL pseudodata.
The uncertainty on $\as(m_Z)$ is only slightly larger at N$^{2+1}$LL than at N$^{3+1}$LL,
which highlights the fact that after profiling it is largely driven by the
constraints imposed by the measurement on the theory model.
In addition, we also include results at N$^{3+0}$LL and N$^{4+0}$LL --
these orders serve as approximations of N$^{3+1}$LL
and N$^{4+1}$LL, respectively, see \app{plus0orders} for their definition and
a more detailed discussion.
While these approximations lack the correct formal structure of their corresponding higher orders,
we include them here for completeness.
Note that although N$^{4+0}$LL is technically our currently highest available order, we prefer
to use N$^{3+1}$LL as our nominal theory model, because it has the correct theory correlation
structure. Furthermore, its theory precision is sufficiently high compared to the measurement
precision and after profiling it yields uncertainties comparable to N$^{4+0}$LL.

At both N$^{2+1}$LL and N$^{3+1}$LL we observe a small bias in the
fitted central value of $\as(m_Z)$.
First note that when fitting the lower-order theory model against the higher-order
pseudodata, we intentionally introduce a mismatch. Without profiling, this naturally
leads to a potential bias in $\as(m_Z)$ as it
potentially absorbs some of the deficiencies of the theory model. This type of bias
is actually what is estimated and supposed to be covered by the scanning approach
(for either TNPs or scale variations).
Indeed, fitting the central N$^{2+1}$LL and N$^{3+1}$LL models against N$^4$LL
and evaluating the uncertainty by scanning over TNPs,
we find $\as(m_Z) = 0.1211 \pm 0.0044$ and $\as(m_Z) = 0.1195 \pm 0.0018$, respectively.
The biases compared to the correct value of $0.118$ are thus indeed covered by their
respective scanning uncertainties.

When profiling the TNPs, the fit is allowed to improve the theory model by moving
the TNPs toward their true values, which reduces this type of bias.
Nonetheless, in doing so the fit must balance the theory and data constraint.
The true values of some of the parameters, shown as stars in the pull plots,
can always happen to be of order the size of the theory constraint
away from 0. As a result, the fit picks up penalties as it tries to move the parameters
to their true values, which induces a residual bias in $\as(m_Z)$ due to
the imposed theory constraint. Importantly, we observe that
this type of residual bias is much smaller and practically negligible
compared to the output uncertainties.

To summarize, our profiling results highlight the full power of the TNP approach:
Our Asimov tests demonstrate that when profiling the TNPs, they are properly
constrained by the data and pulled toward their true values, leaving only
a negligible residual bias.
The high precision of the data thus allows us to consistently reduce the theory
uncertainties and to achieve the sensitivity required
for an extraction of $\as(m_Z)$ at the (sub)percent level.

We have to caution, however, since as discussed in \sec{tnps_for_qT}, with
the minimal set of TNPs we use here, there still remain neglected sources of
uncertainty, which could still enlarge the perturbative
uncertainty on $\as(m_Z)$. Whilst some of these can be anticipated to be
negligible, such as the singlet contributions to the hard function given the
observed minor impact of the hard function on $\as(m_Z)$, others may
yield nonnegligible additional contributions. In particular, given the notable
impact of the already included beam function uncertainties, the impact of higher
order structures in the beam functions are such an example. Another example are
the perturbative uncertainties due to PDF evolution, which are not yet accounted
for by the TNPs, but which could be of relevance (as also indicated by the
nontrivial size of $\Delta_f$ in the scale-variation approach).
The same caveats of course also apply to our earlier scanning results.

\subsubsection{Dependence on pre-fit theory constraints}
\label{sec:relaxing_constraints}

The final aspect we would like to discuss concerns the pre-fit theory constraints
imposed on the TNPs, as discussed in \secs{tnps_for_qT}{asimov_setup}.
So far, our default choice has been $\Delta\theta_n = 1$. As the adopted theory
constraint ultimately depends on some theoretical judgement, we might question whether
it is overly restrictive and to what extent the
uncertainties we find depend on it. These issues have already been discussed in
\Refcite{Tackmann:2024kci}. Here, we provide a practical application allowing us
to explicitly answer these questions.

When the theory constraints on TNPs and the precision of the data are of comparable
size -- such that neither clearly dominates -- both central value and uncertainty
of the parameters of interest will in principle depend on the theory constraint.
As we have seen, the profiling already essentially eliminates the bias
in the central value.
One key message to emphasize is that profiling also substantially reduces the
dependence of the uncertainties on the theory constraint.%
\footnote{%
This was also discussed in \Refcite{McGowan:2022nag} in the context of
the theory nuisance parameters employed in their approximate N$^3$LO PDF fit.
}
In contrast, during scanning, the uncertainty on $\as(m_Z)$ directly depends
on the chosen theory constraint.

To demonstrate this, we can relax the theory constraint by increasing $\Delta \theta_n$.
This effectively reduces the weight of the theory constraint and allows the data to play a larger
role in constraining the TNPs and thus determining the final uncertainty.
We again use the N$^{3+1}$LL prediction as theory model and profile it against
the N$^4$LL pseudodata and now vary the pre-fit constraints by factors of $2$,
setting $\Delta \theta_n = 1, 2, 4$.

\begin{figure}
\includegraphics[width=\rescalethreeplots]{figs/pullplot_tnps_N4p0LL_vs_N3p1LL_asmZimpact_PDFvariations.pdf}%
\hfill \includegraphics[width=\rescalethreeplots]{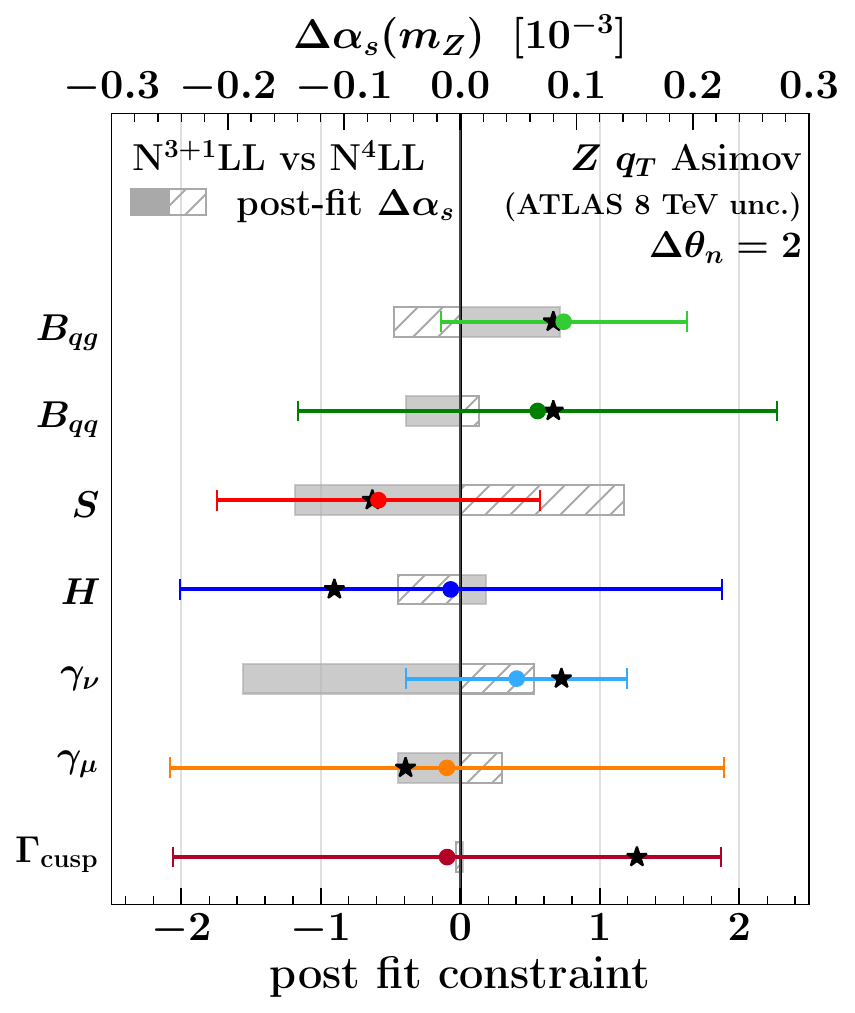}%
\hfill%
\includegraphics[width=\rescalethreeplots]{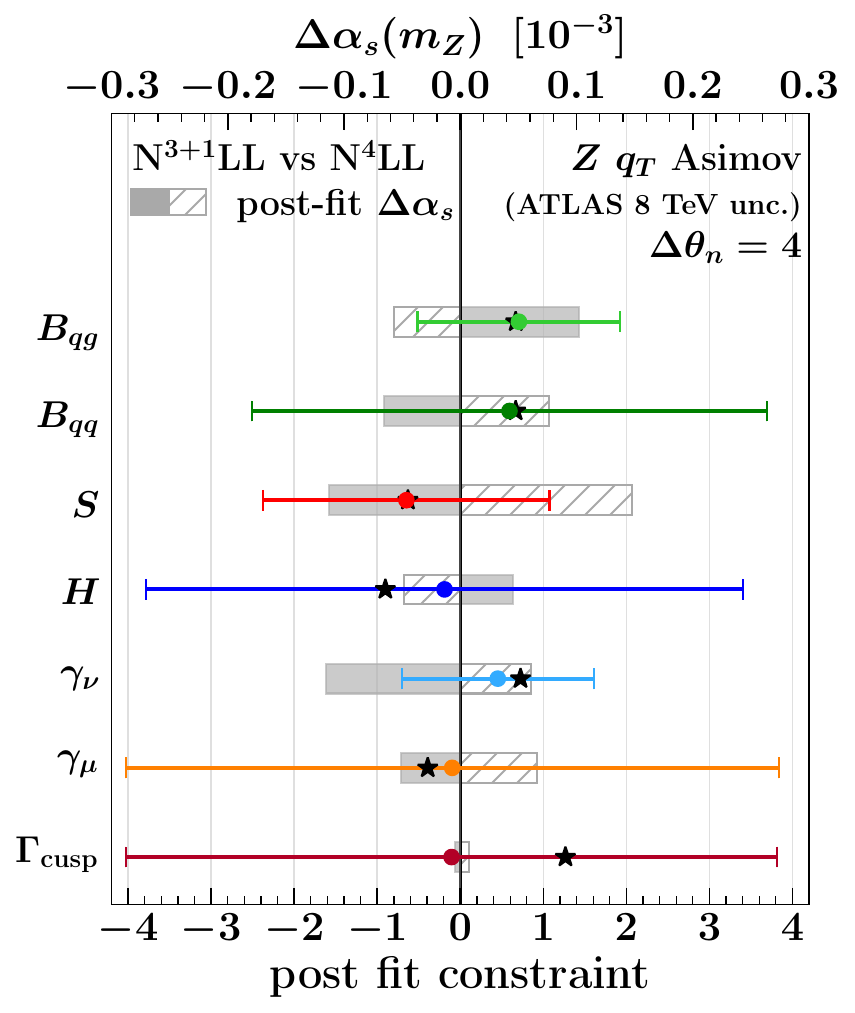}%
\caption{Post-fit constraints on TNPs at N$^{3+1}$LL profiled against the
N$^4$LL central prediction using different pre-fit theory constraints, left panel:
$\Delta \theta_n = 1$, central panel: $\Delta \theta_n = 2$ and right panel: $\Delta \theta_n = 4$.
The stars indicate the true values of the TNPs.
}
\label{fig:pullplottnps_diffTNPsconstr_N4LL}
\end{figure}

In \fig{pullplottnps_diffTNPsconstr_N4LL}, we show the resulting pull plots.
As the pre-fit uncertainty on the TNPs is increased, also the post-fit uncertainty on the TNPs increases but
much less so for some TNPs. Overall, the post-fit values become more and more compatible within
uncertainties with the TNPs' true values, because it becomes easier for the fit to move the TNPs
to their true values.
With $\Delta \theta_n = 4$, the constraints on certain TNPs become
stronger relative to the pre-fit constraint and the less constrained TNPs.
This is expected since with $\Delta \theta_n = 4$ the TNPs become essentially
unconstrained, which makes the differing sensitivities to the various TNPs more clear as the data acts to constrain the most relevant TNPs.
When relaxing the theory constraint, the TNPs having the strongest impact on $\as(m_Z)$ remain the same,
namely $\gamma_{\nu}$, $S$, and $B_{qg}$.
However, their individual impacts reduce while the impact of the other TNPs increases.

\begin{figure}
\centering
\includegraphics[scale=0.4]{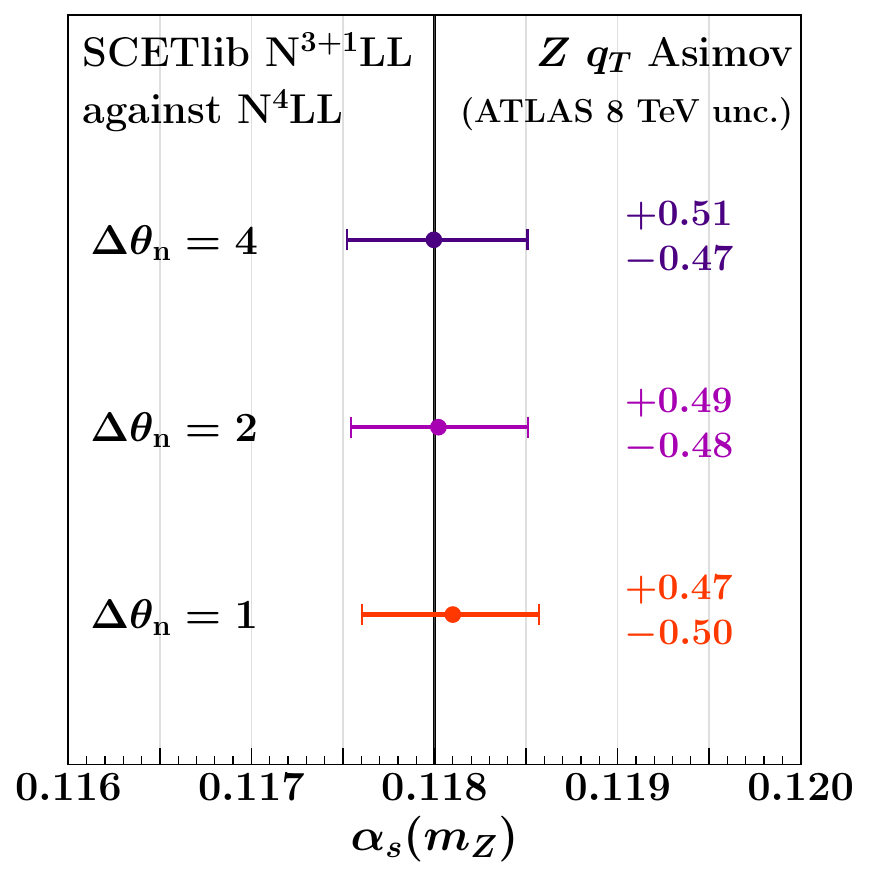}
\caption{Perturbative uncertainties on $\as(m_Z)$ when profiling TNPs
at N$^{3+1}$LL against the N$^4$LL central prediction using different pre-fit
theory constraints $\Delta \theta_n = 1, 2, 4$. The displayed uncertainties
are in units of $10^{-3}$.
}
\label{fig:pullplotsasmZ_diffTNPsconstr_N4LL}
\end{figure}

In \fig{pullplotsasmZ_diffTNPsconstr_N4LL}, we show the results for $\as(m_Z)$ for all three cases.
Interestingly, despite the shifts in the impacts of individual TNPs, the total uncertainty on
$\as(m_Z)$ remains essentially unchanged.
This shows that the total uncertainty after profiling is essentially driven by the constraints
imposed by the data. The precise theory constraint does not affect the total uncertainty
but only the relative distribution of the uncertainties among the TNPs.
This is the case in a situation like ours here where the uncertainties of the data are $\lesssim$
the pre-fit theory uncertainties, i.e., in a situation where without profiling we
would be limited by theory uncertainties.
In the opposite situation, where the data uncertainties are larger than the pre-fit theory uncertainties,
the theory uncertainty will depend on the theory constraint also after profiling simply
because the data will be too weak to constrain the theory uncertainties. However, in this
situation the theory uncertainties are subdominant to begin with, and whilst the theory uncertainty
itself will depend more on the adopted theory constraint, the total uncertainty will not.
Therefore, as discussed before, the only situation to be avoided is that the inherent
precision of the theory model, i.e., the order to which TNPs are included is too low
compared to the data precision such that it becomes overconstrained. In this case, the
theory model can be improved by including additional TNP orders.

As already discussed in the context of \fig{pullplotsasmZ_2}, there is a
minor bias in the central value of $\as(m_Z)$ due to the $\Delta\theta_n = 1$
theory constraint.
Now, as the theory constraint is weakened, the true values of the TNPs lie well within
it, so the associated penalties are reduced. As a result, the bias further reduces
for $\Delta\theta_n = 2$ and disappears completely for $\Delta\theta_n = 4$.

Based on these results, we are confident in our default choice of $\Delta \theta_n = 1$.
In case one is worried about the associated theoretical prejudice, it can always be
relaxed. Even in the extreme case of $\Delta \theta_n = 4$, which here amounts to practically
removing the theory constraint, the data are still able to sufficiently constrain the TNPs
on their own. Most importantly, the conclusions of our study do not depend on the precise
theory constraint.

\section{Nonperturbative Effects}
\label{sec:nonperturbative}

In this section we investigate another major source of uncertainty:
$q_T$-dependent nonperturbative effects as discussed in \sec{nonp_theory}. We
consider the nonperturbative contributions from both the CS kernel and the TMD
PDF boundary condition, with two parameters for each, which we find to be the
minimal necessary set.

Throughout this section, we use the
central N$^4$LL prediction at $\as(m_Z) = 0.118$ as pseudodata, while the theory model
is given by the N$^{3+1}$LL prediction.
We adopt the nonperturbative models for the CS kernel and TMD PDF discussed
in \sec{nonpert_model}. Their parameters $\lambda_{2,4}$ and $\Lambda_{2,4}$
directly correspond to the leading quadratic and quartic OPE coefficients,
which are expected to be the dominant effects.
In the pseudodata, they are fixed to their central values,
while in the theory model they are now treated as additional fit parameters.
The parameters $\Lambda_\infty$ and $\lambda_\infty$, which control the large
$b_T$ behaviour, are always  kept fixed as the fit has practically no sensitivity to them.
This confirms the expectation that the dominant nonperturbative sensitivity
indeed comes from the small-$b_T$ behaviour, where the OPE applies and
determines the $b_T$ dependence of nonperturbative contributions.

Our nominal result thus corresponds to profiling the TNPs at N$^{3+1}$LL
along with $\lambda_{2,4}$ for the CS kernel and $\Lambda_{2,4}$ for the TMD PDFs.
For $\lambda_{2,4}$ we impose the representative lattice constraints
in \Eqs{cs_constraint}{csmatrix} as prior Gaussian constraints.
As we will see, including the lattice constraints for $\lambda_{2,4}$ helps guide
the exploration of parameter space, allowing the data to
determine the TMD parameters $\Lambda_{2,4}$, which contribute to the uncertainty
but for which we otherwise lack constraints from first-principles calculations.

\begin{figure}
\begin{center}
\includegraphics[width=0.60\textwidth]{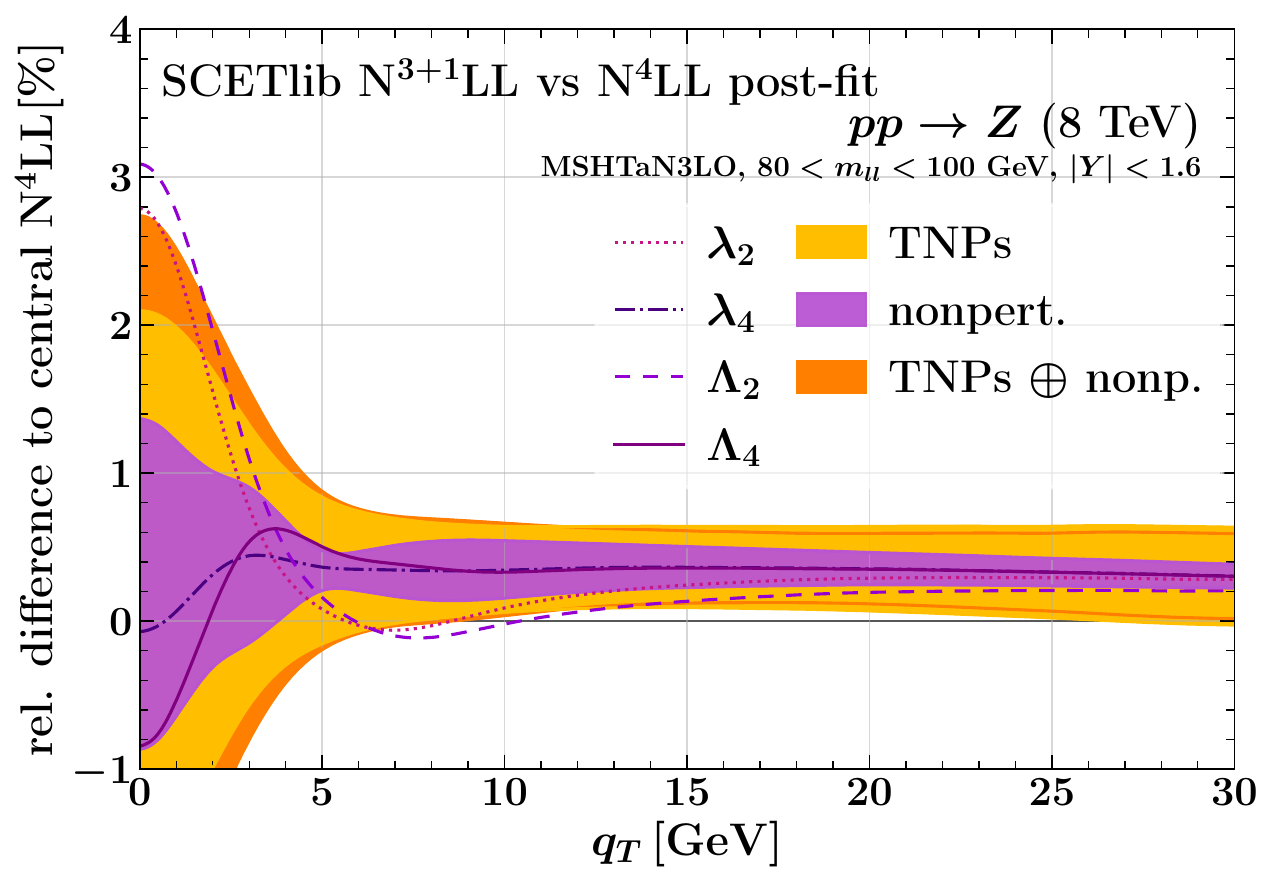}
\includegraphics[width=0.38\textwidth]{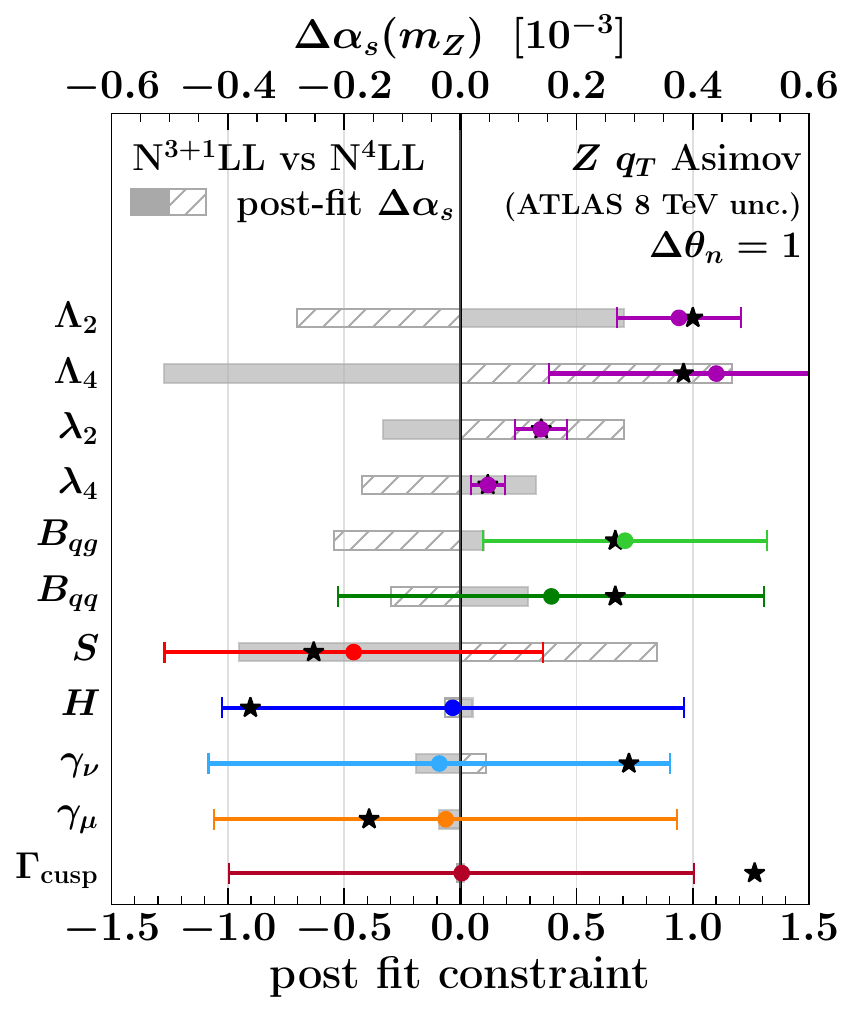}
\end{center}
\caption{\label{fig:pull_plots_nonp_pf}
Left panel: Uncertainties in the $q_T$ spectrum at N$^{3+1}$LL relative to the N$^4$LL,
due to the nonperturbative parameters (purple band), the TNPs (orange band) and their
sum in quadrature band (in dark orange).
The different lines show the post-fit relative impact of each nonperturbative parameter in the
spectrum.
Right panel: Post-fit constraints on the TNPs and nonperturbative parameters (error bars)
and their impact on $\as(m_Z)$,
with the solid (dashed) grey band showing the impact of the post-fit downward (upward)
variations of each parameter. The stars indicate the true value of the parameters.
The post-fit constraints for $\Lambda_2$ and $\lambda_2$ ($\Lambda_4$ and $\lambda_4$)
are normalized to $0.25\GeV^2$ ($0.06\GeV^4$).
}
\end{figure}

In \fig{pull_plots_nonp_pf}, we show the result of our nominal fit including nonperturbative
parameters.
The left panel displays the breakdown of the post-fit uncertainties in the $q_T$ spectrum.
The uncertainty due to the nonperturbative parameters is shown in violet,
the perturbative uncertainty due to TNPs is shown in orange, and
their combined uncertainty is represented by the dark orange band,
which corresponds to the total post-fit uncertainty for this Asimov test.%
\footnote{%
When profiling multiple sources of uncertainty, the breakdown of the total uncertainty
into individual components is somewhat ambiguous. Here, we obtain the individual bands
on the $q_T$ spectrum by considering the relevant sub-matrices of the total post-fit
covariance matrix.
One benefit of the Asimov approach however is that we can also consider different sources
in isolation, as was done for the TNP component in \sec{tnps_perturbative}.
}
We do not show a pre-fit band here, since we do not have pre-fit constraints for $\Lambda_{2,4}$.
The perturbative pre-fit uncertainty corresponds to the one shown in \fig{N3p1LL_vs_N4LL}.
Overlaid on the plot are the post-fit relative impacts of each nonperturbative parameter.
A strong negative correlation emerges between $\lambda_2$ and $\Lambda_2$,
as evident from their significant impact on the spectrum despite the relatively small size of the
overall nonperturbative uncertainty.
In the right panel of the same figure, we show the constraints on all fit parameters, along with their
individual impacts on $\as(m_Z)$.
Compared to the purely perturbative case discussed in \sec{profiling_vs_N4LL}, the
post-fit constraints on the TNPs are still consistent with their true values
(indicated by stars), but some are now not as strongly constrained by the data.
Due to the additional fit parameters introduced, the data has additional freedom now to determine which parameters are most relevant in reducing the uncertainty on the parameter of interest.
It is clear that some nonperturbative parameters, especially $\Lambda_{2,4}$,
have a large impact on $\as (m_Z)$.

\begin{figure}
\begin{center}
\includegraphics[scale=0.45]{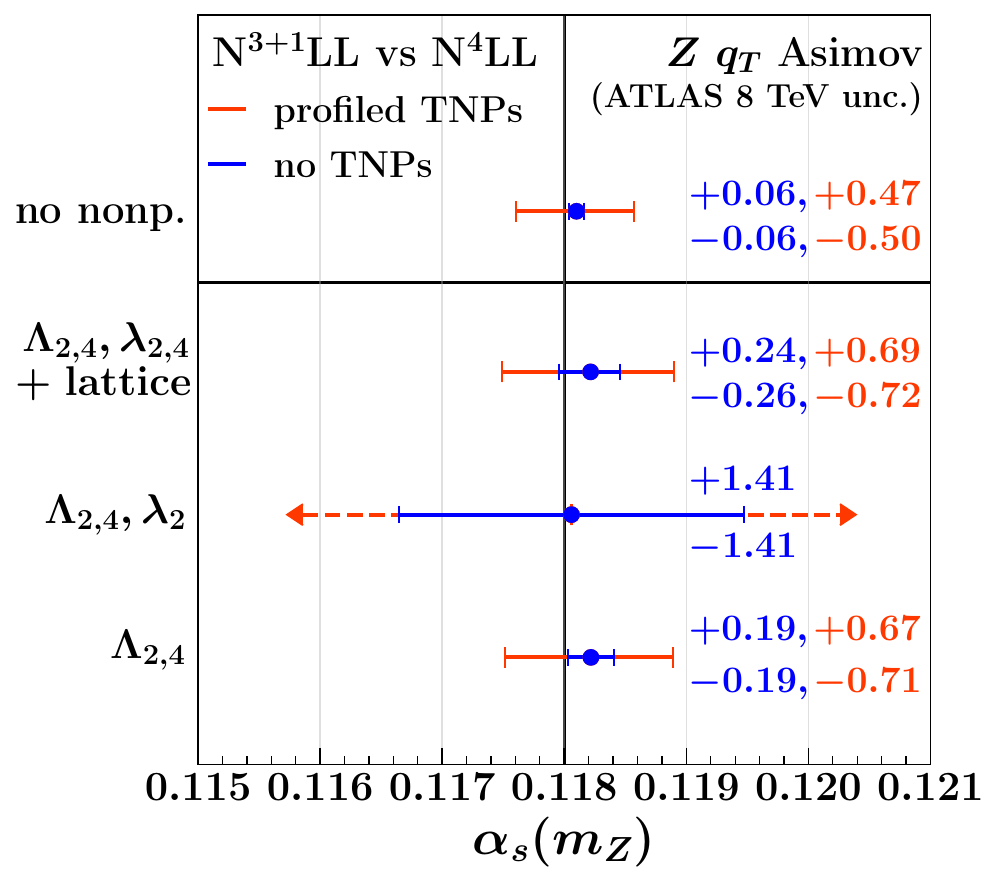}
\end{center}
\caption{%
Uncertainty on $\as(m_Z)$ from various Asimov tests.
Red bars include TNP profiling (i.e., fitting $\as(m_Z)$ together with all theory
nuisance parameters).
Below, we show cases excluding TNP profiling, in blue, with different sets
of fitted parameters.
From top to bottom: fit only $\as(m_Z)$, fit $\as(m_Z)$, $\lambda_{2,4}$,
$\Lambda_{2,4}$ including lattice QCD constraints for CS kernel parameters,
fit $\as(m_Z)$, $\lambda_2$ and $\Lambda_{2,4}$, fit $\as(m_Z)$ and $\Lambda_{2,4}$.
All the uncertainties are in units of $10^{-3}$.
}
\label{fig:pull_plots_nonp}
\end{figure}

Next, we explore the relationship between the nonperturbative effects
and the uncertainty on $\as (m_Z)$, and how the TNPs behave in this context.
\Fig{pull_plots_nonp} presents the results of several Asimov fits as discussed next.
For each case, we display two uncertainty bars. The outer red bars show the total
uncertainty of the nominal fit simultaneously profiling TNPs and nonperturbative parameters.
The inner blue bars show the uncertainty associated with fitting $\as (m_Z)$ alongside the
nonperturbative parameters only. They are obtained by repeating the fit fixing the TNPs to their
nominal post-fit central values.

At the top of the plot, we show the baseline result previously discussed in \sec{profiling_vs_N4LL},
with no nonperturbative parameter included in the fit.
In this case, the inner blue uncertainty is the uncertainty due to only fitting $\as(m_Z)$
with all the other theory parameters fixed, yielding an uncertainty of
$\Delta_{\rm fit} = 0.06 \times 10^{-3}$.
Including the TNPs in the fit increases the uncertainty to
$\Delta_{\rm pert} = ^{+0.47}_{-0.50} \times 10^{-3}$.
(As before, we ignore the very small contribution from $\Delta_{\rm fit}$ and
simply refer to $\Delta_{\rm pert}$ here and similarly for $\Delta_{\rm nonp}$ below.)

We now investigate what happens when also fitting the nonperturbative parameters.
Starting from the bottom of the plot, the first test we perform involves fitting only the
$\Lambda_{2,4}$ parameters of the TMD PDF, while keeping the parameters of the CS kernel,
$\lambda_{2,4}$, fixed at their central values. This yields a total uncertainty
of $\Delta_{\rm pert\oplus nonp} = ^{+0.67}_{-0.71} \times 10^{-3}$.

In addition to the TMD PDF, we also have to account for the nonperturbative uncertainties
due to the CS kernel. We find that including in addition $\lambda_{2,4}$ in the fit without
any further constraints on them does not yield a stable fit. Including only $\lambda_2$ the
full fit including TNPs is barely able to converge and does not yield sensible uncertainties.
We are only able to obtain a trustworthy result when including $\Lambda_{2,4}$ together
with $\lambda_2$ while fixing the TNPs. It yields a very large nonperturbative
uncertainty of $\Delta_{\rm nonp} = \pm 1.41\times 10^{-3}$.
This results suggests that without additional constraints, the Drell-Yan data may not provide
sufficient sensitivity to reliably extract $\as (m_Z)$ together with all relevant
nonperturbative parameters.

Fortunately, as discussed in \sec{nonp_theory}, the CS kernel can be constrained using
lattice QCD data. Incorporating these constraints -- via both the central values from \Eq{cs_constraint}
and the associated covariance matrix in \Eq{csmatrix} -- enables us to include both $\lambda_{2,4}$
in the fit, alongside the TMD PDF parameters $\Lambda_{2,4}$. This yields our nominal
fit already shown in \fig{pull_plots_nonp_pf}. The $\as(m_Z)$ uncertainties
resulting from this fit are
\begin{align}
\Delta_{\rm nonp} &= ^{+0.24}_{-0.26} \times 10^{-3} \, , \nonumber \\
\Delta_{\rm pert \oplus nonp} &= ^{+0.69}_{-0.72} \times 10^{-3}
\,.\end{align}

While this outcome happens to match closely the case where only the TMD parameters
were varied, this is somewhat of a coincidence as it depends on the relative size of
the constraints provided by the data and lattice results. Nevertheless, it is
very encouraging as it demonstrates that it is
possible to consistently account for all relevant nonperturbative parameters in this scenario.

Given the sensitivity of the low-$q_T$ region to nonperturbative effects,
and the resulting dependence of $\as(m_Z)$ on the nonperturbative parameters,
one could consider performing fits using only data above a certain $q_T$ threshold,
such as $q_T > 5\GeV$ or $10\GeV$.
This would reduce the dependence on nonperturbative effects,
potentially allowing for a cleaner determination of $\as(m_Z)$ at the expense of
increased fit uncertainties. A detailed investigation of this approach
is left to future work.

\section{Conclusions}
\label{sec:conclusions}

We have presented an analysis of the expected theory uncertainties in the
determination of the strong coupling constant from the $Z$-boson
transverse momentum ($q_T$) spectrum. The $Z$ $q_T$ spectrum is a benchmark
observable of the LHC precision programme and has been measured to subpercent
precision. This offers the potential for a high-precision determination of the strong
coupling constant, with a percent level change of $\as(m_Z$) inducing a similar
level change in the shape of the $q_T$ spectrum.

A competitive determination of $\as(m_Z)$ depends crucially on the
control of the theoretical uncertainties. However, to determine whether a 1\%
theory effect causes a 0.5\% or 2\% effect on $\as(m_Z)$, a reliable, precise
description of the \emph{shape} of the theory uncertainties is required. Therefore,
a robust treatment of theory correlations across the relevant region of
$q_T$ is critical.

As the dependence on $\as$ arises out of the perturbative expansion, one major
source of theory uncertainty is the perturbative uncertainty from missing
higher orders. Furthermore, with the largest $\as$ sensitivity originating from the
region around $5\GeV$, nonperturbative effects are also important. In this work
we focus largely upon the perturbative uncertainty, in addition to analysing the
nonperturbative effects. To do so, we utilise an Asimov fit setup,
taking the binning, experimental uncertainty and correlations of the
ATLAS 8~TeV inclusive $Z$ $q_T$ measurement~\cite{ATLAS:2023lsr} as an example. This
allows us to focus our efforts on the dominant expected theory uncertainties on
$\as(m_Z)$ and analyse them in a fully controlled setting without having to worry about
statistical fluctuations or possible biases due to unaccounted effects in the data.

First, considering the perturbative uncertainty, we find that the traditional
approach of varying unphysical scales is insufficient. In particular, in this
context it suffers from fundamental limitations as it offers only partial
information on the theory uncertainty itself and no insight on its correlations.
As a result, the uncertainty one might obtain on $\as(m_Z)$ from this approach
strongly depends on the specific choices and variations performed, demonstrating
that it cannot be relied upon to propagate the uncertainty. Our main
focus is therefore on exploring the new theory nuisance parameter (TNP) approach
to assess this vital source of uncertainty. A key advantage of this approach is
that the TNPs encode the higher-order structure of the theory prediction. As
such, they are true parameters of the perturbative series and their variation
correctly encapsulates the point-by-point correlations of the theory
uncertainty. This allows for the first time for a consistent profiling approach
to perturbative theory uncertainties, enabling the fit to the experimental data
to constrain the theory uncertainty. We utilise our Asimov fit setup to explore
this possibility in this context, also serving to demonstrate the benefits of
the novel TNP approach in general. We find that profiling the theory
uncertainties is vital to reach the level of precision desired on $\as(m_Z)$.

Secondly, we explore the nonperturbative uncertainty in the
extraction of $\as(m_Z)$. We utilise a nonperturbative parameterization based
on a systematic OPE expansion of nonperturbative effects
including both the nonperturbative effects of the Collins-Soper kernel and of
the TMD PDF, accounting for the quadratic and quartic OPE coefficients of each.
We find that reaching a competitive precision on $\as(m_Z)$ necessitates the
incorporation of nonperturbative constraints on the Collins-Soper kernel from
lattice QCD.%
\footnote{At least when only using the experimentally most precise Drell-Yan
data at the $Z$ pole.}

Whilst these constitute two of the largest sources of uncertainty for the
determination of $\as(m_Z)$, there are a variety of further effects relevant at
the desired level of precision. PDF uncertainties are
also large, and whilst they can in principle be treated in a straightforward fashion, there are
various subtleties which we leave to future work. The TNPs used to account for the
perturbative uncertainty are also not yet complete: the $x$ dependence of the
beam functions currently utilises the known shape, whilst uncertainties arising
from the PDF evolution within the resummation are yet to be incorporated.
At the same time, a variety of effects that could be neglected in the Asimov
test case need to be included in a fit to real data. These
include nonsingular power corrections, heavy quark mass effects, QED/EW
corrections, and the $x$ and flavor dependence of nonperturbative
TMD effects.

Overall, our analysis demonstrates the crucial role of theory uncertainties and
their correlations in the extraction of $\as(m_Z)$. Reliably propagating these
from the $q_T$ spectrum to the strong coupling is a complex task, which nonetheless
can be achieved thanks to the TNP approach. Furthermore, the TNP approach allows
for a data-driven reduction of perturbative uncertainties. Meanwhile, the
nonperturbative uncertainty can be reduced with the help of lattice QCD data.
Together, this facilitates the reduction of these key uncertainties to a level
which we anticipate will allow for an extraction of $\as(m_Z)$ from the Drell-Yan
$q_T$ spectrum that is competitive with other measurements.

\acknowledgments
We thank Peter Ploessl and Bahman Dehnadi for providing results with quark-mass
effects from \Refcite{flavorthr_cs}, and Johannes Michel and Iain Stewart for
useful discussions and comments on the manuscript.
We also thank our experimental colleagues for many helpful discussions.
Last but not least, we thank Johannes Michel, Georgios Billis, and Markus Ebert
for their many contributions to the \scetlib\ code base.
This project has received funding from the European Research Council (ERC)
under the European Union's Horizon 2020 research and innovation programme
(Grant agreement No. 101002090 COLORFREE) and from the Deutsche
Forschungsgemeinschaft (DFG, German Research Foundation) under Germany’s
Excellence Strategy -- EXC 2121 ``Quantum Universe'' -- 390833306.
TC acknowledges funding by Research Foundation-Flanders (FWO) project number: 12E1323N.

\appendix

\section{Comments on Recent \texorpdfstring{$\alpha_s$}{alphas} Extraction by ATLAS}
\label{app:atlas}

In this appendix, we comment on the treatment of theory uncertainties in
\Refcite{ATLAS:2023lhg}, which performed a fit to the Drell-Yan $q_T$ spectrum
measured in \Refcite{ATLAS:2023lsr} to extract $\as(m_Z)$.
In the next three subsections we discuss the three uncertainties of perturbative origin
that according to our estimates are dominant.%
\footnote{%
Additional perturbative uncertainties are due to QED ISR and the matching to
fixed order. They are estimated in \Refcite{ATLAS:2023lhg} to not be significant,
which we also believe to be the case, so we do not discuss them here.
}
In \app{nonpert_model}, we comment on the nonperturbative model used in
\Refcite{ATLAS:2023lhg}.

In \tab{unc_summary}, we summarize our estimates of the
expected size of perturbative uncertainties based on fits to Asimov data and
compare them to the values
quoted in \Refcite{ATLAS:2023lhg}.
For scale variations, we use the quadrature sum of envelopes here. Using instead the
naive total envelope ($1.73 \times 10^{-3}$) leads to a slightly smaller total perturbative
uncertainty of $\pm2.30 \times 10^{-3}$.
Our uncertainty estimates are based on fits to Asimov data, so while
the precise numbers will differ in a fit to the real data their typical size
will be the same. For the uncertainties due
to scale variations, we have explicitly checked in preliminary fits to real data that
the size of the obtained uncertainty is indeed very similar. We thus have
reason to believe that the total theory
uncertainty quoted in \Refcite{ATLAS:2023lhg} is underestimated by
a factor of 4 to 5. This would make it by far the dominant uncertainty and as a result
also the \emph{total} uncertainty of $\pm 0.9 \times 10^{-3}$ quoted in \Refcite{ATLAS:2023lhg}
would be underestimated by about a factor of 3.

\begin{table}[t]
\centering
\begin{tabular}{l|c|c}
\hline\hline
& \multicolumn{2}{c}{Absolute uncertainty on $\as(m_Z)$ in units of $10^{-3}$}
\\
Perturbative uncertainty &  \Refcite{ATLAS:2023lhg} & Our estimate of expected size \\
\hline
Scale variations       & $\pm 0.42$       & $\pm 2.43$   \\
N$^4$LL$^*$ approximation  & $\pm 0.04$    &  \\
N$^4$LL$'$ approximation  &  & $\pm 0.75$   \\
Flavor/quark masses   & $+0.40$ $-0.29$  & $\pm 1.32$ \\
\hline
Total                  & $+0.58$ $-0.51$  & $\pm 2.87$
\\
\hline\hline
\end{tabular}
\caption{Dominant uncertainties of perturbative origin on $\as(m_Z)$ in units of $10^{-3}$.
The table compares the values quoted in \Refcite{ATLAS:2023lhg} with our estimate
of their expected size based on fits to Asimov data. See text for more
details.
}
\label{tab:unc_summary}
\end{table}

\subsection{Perturbative uncertainties from scale variations}

The analysis in \Refcite{ATLAS:2023lhg} uses the method of enveloped scale variations
discussed in \sec{scale_variations} to estimate the perturbative
theory uncertainty on the resummed $q_T$ spectrum. This uncertainty is
propagated to the fitted $\as(m_Z)$ by scanning over individual scale variations.
This yields maximum deviations in $\as(m_Z)$ of $+0.23\times 10^{-3}$ and $-0.61 \times 10^{-3}$,
which are symmetrized to the quoted $\pm 0.42 \times 10^{-3}$ scale-variation uncertainty
on $\as(m_Z)$.

As we have argued in the main text, extracting $\as(m_Z)$ from the small-$q_T$ spectrum crucially
relies on shape effects and is therefore extremely sensitive to the theory uncertainty
on the shape of the spectrum and the corresponding bin-by-bin theory correlations.
As discussed in detail in \sec{scale_variations}, these are not correctly captured by
scale variations, which means the resulting variations in the fitted
$\as(m_Z)$ cannot be regarded as a trustworthy uncertainty estimate.
Scanning over scale variations in \sec{prof_scale_var} we find significantly
larger maximum variations in the fitted $\as(m_Z)$,
as seen in the right panel of \fig{N4LL_scanning}. The envelope over all
scale variations yields $1.73\times 10^{-3}$, while adding in quadrature the envelopes
of different classes of scale variations yields $2.43\times 10^{-3}$, see
\Eqs{subclass_env}{totalenv}.

Our scale-variation study is performed at N$^4$LL, which is the currently highest known
(essentially) complete perturbative order. It yields uncertainties of $0.5-1\%$ in the
range $5\GeV < q_T < 30\GeV$ and growing up to $3-4\%$ for $0 < q_T < 5\GeV$,
as seen in the left panel of \fig{N4LL_scanning}.
Unfortunately, from the limited information given in \Refcite{ATLAS:2023lhg} it is not
clear what the exact size of the scale-variation based uncertainties on the $q_T$ spectrum
are at their highest used order (denoted N$^4$LL$^*$ in the next subsection). From
our own runs of the public version of \dyturbo\ (v1.4.2)
we find spectrum uncertainties of $0.5-1\%$ for $q_T < 30\GeV$ and growing
to $2-3\%$ for $q_T\to 0$.%
\footnote{%
Similarly, in \Refcite{Camarda:2023dqn}, numerical results for the $q_T$ spectrum
at this N$^4$LL$^*$ order are given for $\Ecm = 13\TeV$ and including fiducial lepton cuts,
in which case the spectrum uncertainties are also close to around $1\%$.
}
Therefore, the envelope of scale variations used in \Refcite{ATLAS:2023lhg} at the level of the
$q_T$ spectrum should be of similar size to ours here. This means, for
similar-size scale variations in the $q_T$ spectrum we obtain up to five times larger
variations in $\as(m_Z)$.

One difference which could (at least partially) contribute to this large discrepancy is that the
scale-variation recipes used in the \scetlib~and \dyturbo~resummation frameworks differ.
Whilst both use the conventional factor of two for varying scales up and down
and avoid simultaneous variations that would compound to factors of four in any scale ratios,
the specific types of scale variations differ substantially and are more limited in \dyturbo.
In particular, the low
scale $\sim  q_T \sim 1/b_T$ is varied as part of the scale variation setup in \scetlib, while it is kept
fixed in \dyturbo. We stress that there are no compelling theoretical arguments
for keeping this low scale fixed when estimating uncertainties using scale variations.%
\footnote{%
This low scale is genuine to the resummation problem, as discussed in the following subsection,
and therefore must effectively appear in any resummation framework.
It is also kept variable in the original CSS
framework~\cite{Collins:1984kg} on which \dyturbo~is based.
In the context of an ongoing benchmarking study of various $q_T$ resummation
frameworks within the LHC electroweak working group, the need for varying this low scale
has been widely recognized by now and, as far as we can tell,
has been gradually adopted also by most other resummation frameworks.}

As discussed in \sec{theory_unc}, in a situation like this where theory correlations are critically
important, for the same overall uncertainty in the spectrum, the resulting uncertainty
on $\as(m_Z)$ entirely depends on the specific scale variations that
happen to be included in the scanning.
A similar breakdown into individual scale variations as in our
\fig{N4LL_scanning} of the results in \Refcite{ATLAS:2023lhg} might shed some
more light on this.
Ultimately, however, the large discrepancies we observe between different resummation frameworks using
different scale variation recipes likely just reflect the fact that
the scale-variation method cannot be relied upon for
propagating the perturbative uncertainty from the $q_T$ spectrum to $\as(m_Z)$.

Nevertheless, if scale variations are the method of choice, one must be aware of their
limitations and at minimum it must be ensured that a given scale-variation recipe
does not underestimate the resulting uncertainty on the parameter of interest. For this purpose,
we have clear evidence that the type of scale variations performed in \scetlib~yield
a more realistic uncertainty for $\as(m_Z)$. This is
because when scanning over TNPs, which do encode the correct theory correlations,
spectrum uncertainties of comparable (or even somewhat smaller) size translate into an
uncertainty on $\as(m_Z)$ of
$1.75\times 10^{-3}$ (see \sec{tnps_perturbative_scanning} and \fig{N3p1LL_scanning}),
which is similar to what we find with scale variations.

We also stress that the much reduced perturbative uncertainty on $\as(m_Z)$ that
we find after profiling the TNPs \emph{cannot} be taken to justify the much smaller
scale-variation based uncertainties quoted in \Refcite{ATLAS:2023lhg}, because the
TNP profiling allows the data to improve the overall precision of the theory description
(and for example also induces nontrivial post-fit correlations between different TNPs).
In other words, the resulting theory description obtained as an output after
profiling the TNPs has an intrinsically higher precision than what
can be ascribed to a traditional input theory prediction based on scale variations.

We finally note that another possible difference is that we are
fitting only $\as(m_Z)$ when scanning over scale variations in \sec{prof_scale_var},
whereas in \Refcite{ATLAS:2023lhg} also the PDFs and some nonperturbative
parameters are included in the fit. In principle, it could happen that other fit
parameters absorb some of the scale variations such that their impact on $\as(m_Z)$ is reduced.
For this to be the case, the effect of a particular
scale variation would have to match the shape change induced by adjusting
another fit parameter better than that induced by adjusting $\as(m_Z)$, which would
make the sensitivity to the
precise shape of the theory uncertainty even more relevant.
By including nonperturbative parameters or PDFs in our fit
while scanning over TNPs (as they provide the correct theory uncertainty shapes),
we find that this is not the case, i.e., including these additional fit parameters does
not alter the impact of the perturbative uncertainties on $\as(m_Z)$.

\subsection{Different \texorpdfstring{N$^4$LL}{N4LL} countings and approximation uncertainties}

Different nomenclatures for resummation orders involving Sudakov double-logarithms
appear in the literature. A common nomenclature, which we also use here, is
such that at N$^n$LL one includes the $n$-loop
evolution (i.e.\ $n$-loop noncusp and $(n+1)$-loop cusp anomalous dimensions)
together with $(n-1)$-loop fixed-order boundary conditions of the hard, beam, and
soft functions in our notation. At N$^n$LL$'$ order the boundary
conditions are included at one order higher, i.e., at $n$-loop order and as they
appear at N$^{n+1}$LL. As a result, the perturbative precision at N$^n$LL$'$ is often
close to N$^{n+1}$LL and thus much higher than N$^n$LL. This is summarized in
\tab{order_counting}.

The N$^n$LL nomenclature used in \Refscite{ATLAS:2023lhg, Camarda:2023dqn} also
includes $n$-loop boundary terms, and for clarity we will refer to it as
N$^n$LL$^*$ here. According to \Refcite{Camarda:2023dqn}, this N$^4$LL$^*$
supposedly provides a reliable approximation of N$^4$LL$'$, with the implied higher
perturbative precision compared to standard N$^4$LL. However, there is a crucial
difference in how the boundary terms are included at N$^n$LL$^*$, making it
rather different from N$^4$LL$'$ and with a perturbative precision more
comparable to standard N$^4$LL, as indicated in \tab{order_counting} and as we
will discuss now.

\begin{table}[t]
\centering
\begin{tabular}{c|cccc}
\hline\hline
& \multicolumn{2}{c}{boundary conditions} & \multicolumn{2}{c}{anomalous dimensions}
\\
resummation order & $h$ & $\tilde b$, $\tilde s$ & $\gamma^\mu, \gamma^\nu, P_{ij}$ & $\GammaC, \beta$
\\\hline
N$^3$LL$'$ & $\alpha_s^3(Q) $ & $\alpha_s^3(b_0/b_T)$ & $\alpha_s^3$ & $\alpha_s^4$
\\
N$^4$LL & $\alpha_s^3(Q)$ & $\alpha_s^3(b_0/b_T)$ & $\alpha_s^4$ & $\alpha_s^5$
\\
N$^4$LL$^\star$ & $\alpha_s^4(Q)$ & $\alpha_s^3(b_0/b_T)+\alpha_s^4(Q)$ & $\alpha_s^4$ & $\alpha_s^5$
\\\hline
N$^4$LL$'$ & $\alpha_s^4(Q) $ & $\alpha_s^4(b_0/b_T)$ & $\alpha_s^4$ & $\alpha_s^5$
\\
N$^5$LL & $\alpha_s^4(Q) $ & $\alpha_s^4(b_0/b_T)$ & $\alpha_s^5$ & $\alpha_s^6$
\\\hline\hline
\end{tabular}
\caption{Order counting conventions. Here, N$^4$LL$^\star$ refers to what is called
N$^4$LL in \Refscite{ATLAS:2023lhg, Camarda:2023dqn}. In practice, the first three
rows have a comparable perturbative precision, as have the last two rows. See the
text for further details.}
\label{tab:order_counting}
\end{table}

Let us first consider the soft function. As discussed in \sec{qT_resummation}, the full
soft function $\tS(b_T, \mu, \nu)$ is a perturbative series in $\as(\mu)$ and
contains Sudakov logarithms $\ln^n(\mu b_T/b_0)$ and $\ln^n(\nu b_T/b_0)$. Its resummation
proceeds by evaluating it at its canonical scales
$\mu_S^{\rm can} = \nu_S^{\rm can} = b_0/b_T$
[see \Eqs{func_depend}{canonicalscales}], which defines its boundary condition,
\begin{equation} \label{eq:soft_bc}
\tilde s[\as(b_0/b_T)] \equiv \tS(b_T, \mu = b_0/b_T, \nu = b_0/b_T)
= \sum_{n = 0} \tilde s_n\,\biggl[\frac{\as(b_0/b_T)}{4\pi}\biggr]^n
\,,\end{equation}
where the $\tilde s_n$ are the so-defined $n$-loop soft boundary coefficients.
The boundary series in \Eq{soft_bc} is by construction free of large logarithms
and can thus be evaluated at fixed order in $\as$. It is the starting point
of the evolution from the low scale $1/b_T \sim q_T$ to the high scale $Q$,
which resums the logarithms of $Q b_T \sim Q/q_T$.

The analogous beam boundary condition was already defined in \Eq{B_boundary_condition},
\begin{equation} \label{eq:beam_bc}
\tilde b_i[x, \alpha_s(b_0/b_T)] \equiv \tB_i(x, b_T, \mu = b_0/b_T, \nu/Q = 1)
= \sum_{n = 0} \tilde b_{i,n}(x)\,\biggl[\frac{\as(b_0/b_T)}{4\pi}\biggr]^n
\,,\end{equation}
where as shown in \Eq{ope_beam_soft_tmd} the beam boundary coefficients
$\tilde b_{i,n}(x)$ are computed in terms of PDFs evaluated at $\mu_f = b_0/b_T$.

Note that the beam and soft boundary conditions in \Eqs{soft_bc}{beam_bc}
\emph{genuinely} appear as a low-scale perturbative series in $\as(b_0/b_T)$.
The appearance of a perturbative series at the low scale $\mu \sim b_0/b_T \sim q_T$
is a fundamental and universal feature of $q_T$ resummation and not specific to a particular
resummation framework.
(It is also the reason why the beam and soft functions become
nonperturbative at sufficiently small $q_T \sim 1/b_T$, see \sec{nonp_theory}.)

By contrast, the canonical scale of the hard function is $\mu_H^{\rm can} = Q$,
so its boundary condition is defined as
\begin{equation} \label{eq:hard_bc}
\tilde h[\as(Q)] \equiv H(Q^2, \mu = Q) = \sum_{n = 0} \tilde h_n\,\biggl[\frac{\as(Q)}{4\pi}\biggr]^n
\,,\end{equation}
and thus it genuinely appears as a high-scale perturbative series in $\as(Q)$.

We can now state the difference between N$^n$LL and N$^n$LL$'$
more explicitly: At N$^n$LL, the hard, beam, and soft boundary series in \Eqss{soft_bc}{beam_bc}{hard_bc}
are included in \Eq{func_depend} up to fixed $\ord{\as^{n-1}}$
with their respective canonical $\as$. At N$^n$LL$'$, they are instead included
up to $\ord{\as^n}$ \emph{exactly} as they would be at N$^{n+1}$LL. Thus, the only
difference between N$^n$LL$'$ and the full N$^{n+1}$LL is that the latter further
includes the anomalous dimensions in the evolution factor in
\Eq{func_depend} to one order higher, see \tab{order_counting}.

Toward higher orders in $q_T$ resummation, the contributions from boundary terms
become the dominant effect while the anomalous dimensions become
less important. The N$^n$LL$'$ result is therefore numerically close to the full
N$^{n+1}$LL result.
For example, the \scetlib~results at N$^3$LL$'$ and N$^4$LL are almost
indistinguishable in both central value and scale variation uncertainties~\cite{Billis:2024dqq}.%
\footnote{This applies when using common PDFs. Switching from NNLO to approximate N$^3$LO
PDFs, as also formally required for the PDF evolution when switching from N$^3$LL$'$
to N$^4$LL, does make a noticeable difference.}
Furthermore, as the beam and soft boundary coefficients come with a low-scale
$\as$, they are much more important than the hard function. We have seen this
explicitly in \sec{tnps_perturbative}.

In \dyturbo, following \Refscite{Catani:2000vq, Bozzi:2005wk},
the low-scale boundary series in the CSS formalism~\cite{Collins:1984kg}, which
corresponds to our soft and beam boundary series, is rewritten to
superficially appear as high-scale perturbative
series as follows (taking the soft function as an example):
\begin{align} \label{eq:soft_rewrite}
\tilde s[\as(b_0/b_T)]
&= \tilde s[\as(Q)] \times \frac{\tilde s[\as(b_0/b_T)]}{\tilde s[\as(Q)]}
= \tilde s[\as(Q)] \times \exp\biggl\{\ln \frac{\tilde s[\as(b_0/b_T)]}{\tilde s[\as(Q)]} \biggr\}
\nn \\
&= \tilde s[\as(Q)] \times \exp\biggl\{-\int_{\as(b_0/b_T)}^{\as(Q)}\!\df \as\,
\frac{\df\ln\tilde s(\as)}{\df\as} \biggr\}
\nn \\
&= \tilde s[\as(Q)] \times \exp\biggl\{-\int_{b_0/b_T}^Q\!\df\ln\mu\,
\frac{\df\as}{\df\ln\mu}\, \frac{\df\ln\tilde s(\as)}{\df\as}
\biggr\}
\,.\end{align}
The first line is clearly trivial, and in the remaining lines the exponential
is further cast into a form that resembles the exponential evolution factor in
\Eq{func_depend}. As long as
$\tilde s(\as)$ is consistently included to the same order in $\as$ everywhere it appears,
\Eq{soft_rewrite} is a strict identity, i.e., there is no formal or numerical difference
between the left-hand side and the various versions on the right-hand side.%
\footnote{For the last line in \Eq{soft_rewrite}, this also requires the running of
$\as(\mu)$ to be treated in the same way on both sides of the equation.
Otherwise there will be formally higher-order differences which are however
irrelevant for our discussion.}
This is the case at N$^n$LL and N$^n$LL$'$.

However, in the order-counting adopted at N$^n$LL$^*$, the last line of \Eq{soft_rewrite}
is used with the factor $\tilde s[\as(Q)]$ considered as a fundamental
high-scale boundary series, and akin to the treatment of the hard boundary series
at N$^n$LL$'$, it is included to $n$-loop order (i.e.\ $\tilde s_n$).
At the same time, the exponential factor
is considered to be part of the overall evolution factor, which is included
to $n$-loop order (akin to the evolution order used at N$^n$LL$'$).
As a result, the $\tilde s(\as)$ boundary series in the exponent
is only included to $(n-1)$-loop order (i.e.\ to $\tilde s_{n-1}$).
This is because the running of $\as$
starts at one loop, so the $(n-1)$-loop boundary coefficient $\tilde s_{n-1}$ only appears
as an $n$-loop contribution as $\beta_0\, \tilde s_{n-1}$.

However, rewriting the boundary series as in \Eq{soft_rewrite} does
not change the fundamental fact that it is a low-scale
series. This fact is merely hidden on the right-hand side.
Reversing the steps in \Eq{soft_rewrite} with the N$^n$LL$^*$ counting applied to
the last line on the right-hand side, it results in effectively using
the following $n$-loop boundary series on the left-hand side,
\begin{equation}
\tilde s^{(n)}[\as(Q)] \times \frac{\tilde s^{(n-1)}[\as(b_0/b_T)]}{\tilde s^{(n-1)}[\as(Q)]}
= \sum_{k = 0}^{n-1} \tilde s_k\,\biggl[\frac{\as(b_0/b_T)}{4\pi}\biggr]^k
+ \tilde s_n \biggl[\frac{\as(Q)}{4\pi}\biggr]^n
\,,\end{equation}
where the superscripts denote the order in $\as$ to which each series is kept.
Hence, due to the mismatch in the treatment of $\tilde s(\as)$ between prefactor and
exponent, the $n$-loop coefficient $\tilde s_n$ remains multiplied by the high-scale
$\as^n(Q)$, as indicated in \tab{order_counting}, while its correction to the low
scale is postponed to N$^{n+1}$LL$^*$. By contrast, it is directly included at its
correct low scale at both N$^n$LL$'$ and full N$^{n+1}$LL.

The analogous discussion applies to the beam boundary series, which is rewritten in
the same fashion to appear as a high-scale series, including also PDFs
evaluated at the high scale, multiplied by a corresponding correction factor that
also includes the evolution of the PDFs back down to their natural low scale.
At N$^n$LL$^*$, the beam boundary terms are included at their correct low scale
only to $\tilde b_{i,n-1}$, while $\tilde b_{i,n}$ is included in terms of $\as$
and PDFs at the high scale.

As a side remark, we comment on the counting of towers of logarithms $\alpha_s^n L^m$ in the
perturbative series, with some choice of the large logarithm $L$, which traditionally
is often used to define resummation orders.
When counting logarithms in the cross section (formally counting $\alpha_s L^2 \sim 1$),
N$^n$LL$^*$ and N$^n$LL$'$ capture the same complete logarithmic towers
and one more than N$^n$LL. When counting logarithms in the exponent (formally counting
$\alpha_s L \sim 1$), all three orders capture the same complete logarithmic
towers. This does not mean that the orders are equivalent for all purposes.
The three orders differ by including different subsets of higher logarithmic towers, which are well defined and numerically relevant. This simply means that their relevant distinction is not (or not easily) visible from a traditional counting of logarithms.
Instead, here we define the resummation order (or formal accuracy) at
the level of the RGE and its ingredients, which allows us to make a formal
distinction between the orders as shown in \tab{order_counting}.

This rather peculiar treatment of the low-scale boundary conditions at N$^n$LL$^*$
has several consequences.
Since $\alpha_s(b_0/b_T)/\alpha_s(Q)$ can become $\gtrsim 2$ at small $1/b_T\sim q_T$,
and this mismatch is raised to the $n$th power, the above-mentioned dominant effect
of the $n$-loop soft and beam boundary coefficients is not yet present at N$^n$LL$^*$.
For example, the effect of $\tilde s_4$ and $\tilde b_{i,4}$ can be easily
suppressed by an order of magnitude or more at N$^4$LL$^*$, which is indeed what we
find below.
Their full impact then only appears at the next order in the disguise of a large
evolution effect.
Moreover, since they are only included as a high-scale boundary condition like
the hard function, also their effect on the shape of the spectrum,
which is particularly relevant for extracting $\as$, is not yet present.
For these reasons, the N$^4$LL$^*$ result does not
provide a tangible improvement, formally or practically, compared to N$^4$LL.
This is confirmed by the comparable scale variations we observed
at the spectrum level in the previous subsection.
In particular, and regardless of being approximate or exact, it does not provide
a reliable approximation of N$^4$LL$'$.

The analysis in \Refcite{ATLAS:2023lhg} is performed at the so-called
approximate N$^4$LL (``N$^4$LLa'') order of \Refcite{Camarda:2023dqn},
meaning at approximate N$^4$LL$^*$.
There are two types of approximations involved: First, the 4-loop evolution is
formally approximate, because the required 4-loop splitting functions
and 5-loop cusp anomalous dimensions are only known approximately. These
approximations are also present at N$^4$LL. They are expected to cause more
minor effects and are not discussed here further.%
\footnote{%
Although the available approximations for 4-loop splitting functions
are increasingly accurate for phenomenology~\cite{Cooper-Sarkar:2024crx},
their effect could in principle be of some relevance and deserves further study.}
The second type of approximation comes from the fact that the
4-loop beam and soft coefficients, $\tilde s_4$ and $\tilde b_{i,4}$,
are completely unknown.

In \Refscite{Camarda:2023dqn, ATLAS:2023lhg} the additional uncertainty due to
the 5-loop cusp anomalous dimension and the unknown 4-loop beam and soft coefficients
are estimated to be
$0.1-0.2\%$ for the $q_T$ spectrum with a resulting negligible $\pm 0.04\times 10^{-3}$
effect on $\as(m_Z)$. Since at N$^4$LL$^*$ the latter are effectively treated
like a hard-function contribution, this
is consistent with the rather negligible impact of the hard function and cusp anomalous dimension
we observe. However, as explained above, this does not provide a
realistic estimate of the actual possible impact of the missing 4-loop beam and soft
coefficients.

The dominant effect of the beam and soft functions is clearly evident in our TNP studies,
where as expected they cause by far the largest uncertainties. From our
TNP scanning at N$^{3+1}$LL, we know that the 3-loop beam and soft functions
each cause an effect of order $\sim 1\times 10^{-3}$ on $\as(m_Z)$.
We might expect the corresponding uncertainties due to the 4-loop beam and soft
functions to be between two and a few times smaller than that, but certainly not
30 to 40 times smaller.
We can obtain a more realistic estimate of the impact of the missing 4-loop
beam and soft functions from their associated TNPs at N$^{4+0}$LL
(see \app{plus0orders}).%
\footnote{Ideally, we would use N$^{4+1}$LL, which we have not yet available,
but N$^{4+0}$LL provides a sufficient approximation for our purposes here.}
From the analogous TNP scanning at this order we find a combined perturbative
uncertainty on $\as(m_Z)$ due to the missing 4-loop beam and soft functions
of $0.75\times 10^{-3}$, as given in \tab{unc_summary}, so indeed a reduction by
roughly a factor of 2 compared to N$^{3+1}$LL in line with expectation. Their
full effect would appear at N$^5$LL$^*$ and already by itself can be twice as large
as the quoted scale-variation uncertainty at N$^4$LL$^*$, providing
further evidence that the latter is strongly underestimated.

\subsection{Effects due to finite bottom and charm quark masses}
\label{app:quark_masses}

The masses of charm and bottom quarks, $m_c$ and $m_b$, are usually neglected,
which is an appropriate
description in the limit $m_{b,c} \ll q_T$. On the other hand, for
$q_T \sim m_{b,c}$, which in the case of  $m_b$ is right in the peak region of the $q_T$ spectrum,
they cause nontrivial effects at the few-percent
level. We thus expect them to be important at our level of precision.

For $q_T \sim m_b$, the mass effects genuinely change the resummation structure~\cite{Pietrulewicz:2017gxc}.
They can be classified into primary
and secondary mass effects. Primary mass effects are due to the quark masses of the incoming
quarks in the $b\bar b\to Z/\gamma^*$ and $c\bar c\to Z/\gamma^*$ hard processes.
Secondary mass effects are due to the quark masses in virtual charm and bottom quark
loops as well as final-state gluons splitting into massive quarks.

It is important to note that switching from a massless 5-flavor description
to a massless 4-flavor description at some threshold scale $\mu_b \sim m_b$ is not sufficient to
correctly account for bottom mass effects, since in the region
$q_T\sim m_b$ neither description is actually valid. A correct description of the
bottom threshold actually requires the correct treatment of the bottom
mass effects. The same holds for the charm threshold, which due to the smaller charm
mass is less relevant in practice.

\Refcite{ATLAS:2023lhg} uses a massless 5-flavor description by default.
The uncertainties due to neglecting quark-mass effects
are estimated by changing to a variable-flavor
massless description in various parts of the calculation and by varying the associated
threshold scales. From the above discussion, this can only serve as a rough indication
of the expected order of magnitude of mass effects but it may or may not reflect the actual
size of the effects. Another estimate comes from including the effect of final-state
gluon splitting into massive quark pairs. Unfortunately, \Refcite{ATLAS:2023lhg} provides
no further information than that, so it is unclear to us to what extent
this provides a complete description of secondary mass effects. The final uncertainty
on $\as(m_Z)$ due to mass effects is taken as the envelope of all observed differences
yielding ${}^{+0.40}_{-0.29}\times 10^{-3}$.

A full treatment of quark mass effects in the resummed $q_T$ spectrum at NNLL$'$
based on \Refcite{Pietrulewicz:2017gxc} has been implemented in \scetlib~and will
be presented elsewhere~\cite{flavorthr_cs}. Here, we use it to estimate the bias
induced in the extracted $\as(m_Z)$ from using a massless description.
To do so, we generate Asimov data including bottom and charm mass effects and fit it
with a 5-flavor massless theory model at the same resummation order. We find a
bias on $\as(m_Z)$ of $1.32\times 10^{-3}$, as given in \tab{unc_summary}.
The effect is entirely driven
by the bottom quark mass. This rather strong sensitivity
to mass effects in $\as(m_Z)$ is not unexpected, given that the dominant
sensitivity to $\as(m_Z)$ arises from the peak region of the spectrum $q_T\sim 5\GeV$,
which is precisely where bottom mass effects are important. And as we have seen before,
order percent shape changes in the $q_T$ spectrum can easily cause changes of order $\sim 1\times 10^{-3}$ in
$\as(m_Z)$.

\subsection{Nonperturbative model}
\label{app:nonpert_model}

In \Refcite{ATLAS:2023lhg}, nonperturbative effects are incorporated via the
following phenomenological model
\begin{align} \label{eq:atlas_nonp_model}
S_{\rm NP}(b_T) &= \exp\biggl[-g_j(b_T) - g_K(b_T) \ln \frac{Q^2}{Q_0^2} \biggr]
\,,\nn\\
g_j(b_T) &= \frac{g\, b_T^2}{\sqrt{1 + \lambda\, b_T^2}}
+ \mathrm{sign}(q) \Bigl[1 - \exp(-\abs{q} b_T^4)\Bigr]
\,,\nn\\
g_K(b_T) &= g_0 \biggl\{
1 - \exp\biggl[-\frac{1}{g_0}\,\frac{C_F \as(b_0/b_*)}{\pi}\,\frac{b_T^2}{b_{\rm lim}^2}\biggr]
\biggr\}
\,,\end{align}
which must account for the product $\tilde f_i^\nonp(x, b_T) \tilde f_j^\nonp(x, b_T)$
in our notation of \sec{nonp_theory},
and where $g_j(b_T)$ and $g_K(b_T)$ encode the nonperturbative models for the TMD PDF
and CS kernel, respectively.

This model originates from \Refcite{Collins:2014jpa}, and while it has traditionally
been used in the past, it has several limitations
that are potentially important at the high level of precision required here.
First, the model does not account for any flavor or $x$ dependence in the nonperturbative
TMD parameters, which in general cannot be neglected~\cite{Bacchetta:2018lna, Moos:2023yfa, Bacchetta:2024qre}.
As discussed in \sec{nonpert_model}, since the fit to real data
includes multiple rapidity bins, the nonperturbative TMD model should at minimum
allow for an effective rapidity dependence.

Secondly, the model does not reproduce the OPE expansion in \Eq{tmdpdf_ope}. While
it has quadratic and quartic TMD parameters, $g$ and $q$ (equivalent to
our $\Lambda_2$ and $\Lambda_4$), which are included as fit parameters,
it does not have parameters to encode the genuine nonperturbative
contributions to the CS kernel. In particular, the $\ord{b_T^2}$ piece of
$g_K(b_T)$, only cancels the
$\ord{b_T^2}$ contribution induced by the quadratic $b_*$ prescription used
in the perturbative part of the CS kernel. This leaves the genuine nonperturbative
OPE contribution due to $\lambda_2^\zeta$ unaccounted for, which as discussed
in \sec{nonp_theory} is the parametrically \emph{leading} nonperturbative effect.

Furthermore, the canonical $\ln(b_TQ)$ multiplying the CS
kernel is replaced by a fixed $\ln(Q/Q_0)$. This amounts to shifting
a corresponding $\ln(b_T Q_0)$ dependence into the
nonperturbative TMD, which however is not accounted for in $g_j(b_T)$.
This effectively neglects a nontrivial $b_T$ (and thus $q_T$) dependence, which
could easily spoil the point-by-point correlations of the nonperturbative
uncertainties.

In summary, the model does not provide a correct parameterization of
nonperturbative effects in the important region of moderately small $q_T$ or
$1/b_T$. The effect of this on the extraction of $\as(m_Z)$ is a priori unclear
and needs further study. For example, as indicated by our results in
\sec{nonperturbative}, including $\lambda_2$ as a free parameter in the fit can
significantly enlarge the fit uncertainties. Whilst the model has several
additional parameters ($\lambda$, $g_0$, $b_{\rm lim}$, $Q_0$), which are
being varied to estimate additional nonperturbative uncertainties, they do not
assess the impact of these limitations. \Refcite{ATLAS:2023lhg} also performed a
fit excluding the region of $q_T \leq 5\GeV$, where nonperturbative effects are
expected to have their dominant impact. This provides a potentially useful test,
which however is only necessary and not sufficient. It would invalidate the
model if it gave inconsistent results, but the reverse is not true -- if it
gives consistent results this does not validate the model.

\section{Additional Results}
\label{app:addresult}

\subsection{Nonsingular power corrections}
\label{app:nonsingular}

In this appendix, we briefly investigate the impact of including the nonsingular
contribution in \Eq{matching}, which has been neglected in our main analysis.
It is suppressed by $\ord{q_T^2/Q^2}$ relative to
the leading-power resummed contribution $\df\sigma^\zero$, which
captures the leading behaviour of the $q_T$ spectrum in the $q_T \to 0$ limit.
Since the nonsingular is subdominant for $q_T\ll Q$, it is typically
evaluated at fixed order in $\as$ at the hard scale $\mu\sim Q$ as the difference
\begin{equation} \label{eq:nons}
\frac{\df \sigma_{\rm nons}}{\df q_T}
= \biggl[\frac{\df \sigma_{\rm FO}}{\df q_T} - \frac{\df \sigma^\zero_{\rm FO}}{\df q_T}\biggr]_{q_T > 0}
\,,\end{equation}
where $\df\sigma_{\rm FO}$ is the full fixed-order result and $\df \sigma^\zero_{\rm FO}$
is the fixed-order expansion of the leading-power resummed contribution.

The full $\ord{\as}$ term, called LO$_1$, is implemented in \scetlib,
while the full $\ord{\as^2}$ contribution, referred to as NLO$_1$, is obtained
numerically using \dyturbo~\cite{Camarda:2019zyx}.
We are primarily interested in checking whether the nonsingular can influence
the uncertainty in $\as(m_Z)$ due to the dominant leading-power
uncertainties. We therefore add the nonsingular terms directly to the resummed contribution,
without introducing additional nuisance parameters for them.
For computational efficiency, here we include the nonsingular terms only up to
NLO$_1$.
We follow the setup described in \sec{asimov_setup}, with one minor modification:
The difference in \Eq{nons} is subject to large numerical cancellations for $q_T\to 0$.
To avoid the associated numerical instabilities we begin the $q_T$ binning at $0.5 \GeV$.

\begin{table}[t]
\centering
\begin{tabular}{c|l|l}
\hline\hline
Asimov data         & \multicolumn{1}{c|}{theory model} & \multicolumn{1}{c}{$\as(m_Z)\,[10^{-3}]$ } \\
\hline
 N$^4$LL$+$NLO$_1$  & N$^{3+1}$LL             & $116.34 \pm 0.41$ \\
 N$^4$LL$+$NLO$_1$  & N$^{3+1}$LL$+$LO$_1$    & $118.81 \pm 0.47 $ \\
 N$^4$LL$+$NLO$_1$  & N$^{3+1}$LL$+$NLO$_1$   & $118.09_{-0.49}^{+0.46}$\\
\hline\hline
\end{tabular}
\caption{Results for $\as(m_Z)$ in units of $10^{-3}$ and its perturbative
uncertainty from Asimov tests including different nonsingular contributions.}
\label{tab:nonsing_effects}
\end{table}

We perform a set of Asimov tests following the procedure in \sec{profiling_vs_N4LL}.
For simplicity, we do not fit the nonperturbative parameters here, since doing so does
not change our conclusions. The nominal perturbative uncertainty for this case obtained in
\sec{profiling_vs_N4LL} is $\Delta_{\rm pert} = _{-0.50}^{+ 0.47} \times 10^{-3}$.
The Asimov data are now defined as the
N$^4$LL$+$NLO$_1$ prediction at $\as(m_Z) = 0.118$, which we fit with three
different theory models that differ in the nonsingular terms they include.
The results are shown in \tab{nonsing_effects}.
In the case where no nonsingular term is included (top row), we find an uncertainty
$\Delta_{\rm pert} = 0.41 \times 10^{-3}$, slightly smaller than the
nominal one.
Including the LO$_1$ and NLO$_1$ terms (middle and bottom rows), the
uncertainty increases again marginally and closely reproduces the nominal one.
We conclude that under the assumptions and setup of our pseudodata analysis,
the nonsingular contributions do not affect the uncertainty of $\as(m_Z)$.
They can therefore be safely neglected for our purposes.
However, as also seen in \tab{nonsing_effects}, neglecting the nonsingular
corrections in the theory model induces a substantial bias in the central value of
$\as(m_Z)$, which will necessitate their inclusion in the fit to real data.

\subsection{Approximate TNP orders}
\label{app:plus0orders}

Here, we briefly discuss an approximate TNP implementation, following
\Refcite{Tackmann:2024kci}, that may be useful in cases where incorporating the
full next-order structure is not feasible for practical reasons.
As an approximation to the N$^{m+1}$LO implementation, 
one can consider reusing the structure of the existing N$^m$LO
prediction, absorbing the uncertainty term into the highest known order.
For example, compared to N$^{1+1}$LO and N$^{2+1}$LO in \Eq{f_pred_tnps}, we have
\begin{alignat}{9} \label{eq:approximate_tnps}
\text{N$^{1+0}$LO:}\qquad &&
f(x, \alpha, \theta_2) &= \hat f_0(x) + \bigl[\hat f_1(x) + \alpha_0 f_2(x, \theta_2) \bigr]\,\alpha
\,,\nn\\
\text{N$^{2+0}$LO:}\qquad &&
f(x, \alpha, \theta_3) &= \hat f_0(x) + \hat f_1(x)\,\alpha + \bigl[\hat f_2(x) + \alpha_0 f_3(x, \theta_3) \bigr]\, \alpha^2
\,.\end{alignat}
This approximation is expected to roughly preserve the overall size of the uncertainty
due to $\theta_2$ or $\theta_3$,
while its shape (and thus the correlation) is approximated based on the lower-order structure.

Here, $\alpha_0$ is a chosen fixed number, which does not depend on a renormalization
scale and does not participate in counting powers of $\as$. For a single perturbative
series with no further internal structure, this approximation is somewhat academic. If
$\as$ is evaluated at a fixed central scale and $\alpha_0$ is chosen to agree with that value, then
the central N$^{m+0}$LO result would reproduce the central N$^{m+1}$LO result.

However, as soon as the prediction in question is more complicated, e.g., $\as$ is evaluated
at a dynamic scale or the total prediction involves the combination of several perturbative
series expanded against each other, this approximation becomes nontrivial. This is the case
with resummation, for which the N$^{m+0}$LL result is defined by absorbing the highest
TNP-parameterized terms of all anomalous dimensions and boundary
conditions into their respective previous order terms analogous to \Eq{approximate_tnps}.
For example, compared to N$^{3+1}$LL in \Eq{F_n3p1ll}, we now have
\begin{alignat}{9}
\text{N$^{3+0}$LL:}\quad &&
F(\as^{\rm can}, L)
&= \biggl[1 + \hat F_1\, \frac{\as^{\rm can}}{4\pi}
   + \Bigl(\hat F_2 + a_0\, F_3(\theta_3^F) \Bigr) \Bigl(\frac{\as^{\rm can}}{4\pi}\Bigr)^2 \biggr]
\exp \biggl\{ \int_{0}^{L} \df L' \biggr[
\nn\\ &&& \qquad
\sum_{k = 0}^2 \hat\Gamma_k \Bigl[\frac{\as(L')}{4\pi}\Bigr]^{k+1} L'
+ \Bigl(\hat\Gamma_3 + a_0\, \Gamma_4(\theta_4^\Gamma) \Bigr) \Bigl[\frac{\as(L')}{4\pi}\Bigr]^4 L'
\nn\\ &&& \quad
+ \sum_{k=0}^1 \hat\gamma_{F,k} \Bigl[\frac{\as(L')}{4\pi}\Bigr]^{k+1}
+ \Bigl(\hat\gamma_{F,2} + a_0\,\gamma_{F,3}(\theta_3^{\gamma_F}) \Bigr) \Bigl[\frac{\as(L')}{4\pi}\Bigr]^3
\biggl]\biggr\}
\,,\end{alignat}
where $a_0 \equiv \alpha_0/(4\pi)$ and $\as^{\rm can}$ is again
evaluated at the canonical scale of $F$, which is usually a dynamic scale. At
a more generalized boundary scale, as used in the actual resummation, the full $\mu$-dependent
boundary condition as predicted by the RGE in terms of $\hat F_1$, $\hat F_2$, $\hat\Gamma_0$, $\hat\Gamma_1$, $\hat\gamma_{F,0}$, $\hat\gamma_{F,1}$ is used with $\hat F_2$ replaced by
$\hat F_2 + a_0\, F_3(\theta_3^F)$ as above. The products of boundary conditions
in the full cross section are expanded keeping terms up to $\ord{\as^2}$ (without counting
powers of $\alpha_0$). Hence, at N$^{3+0}$LL the TNP-dependent uncertainty terms associated
with the missing N$^4$LL are absorbed into the N$^3$LL structure, while at N$^{3+1}$LL they
appear in the correct N$^4$LL structure.
As emphasized before, using the full N$^{m+1}$LL result should always be preferred if possible.
However, in case it is unavailable or impractical, the N$^{m+0}$LL result can serve as a viable
approximation for it.

\begin{figure}
\includegraphics[width=0.60\textwidth]{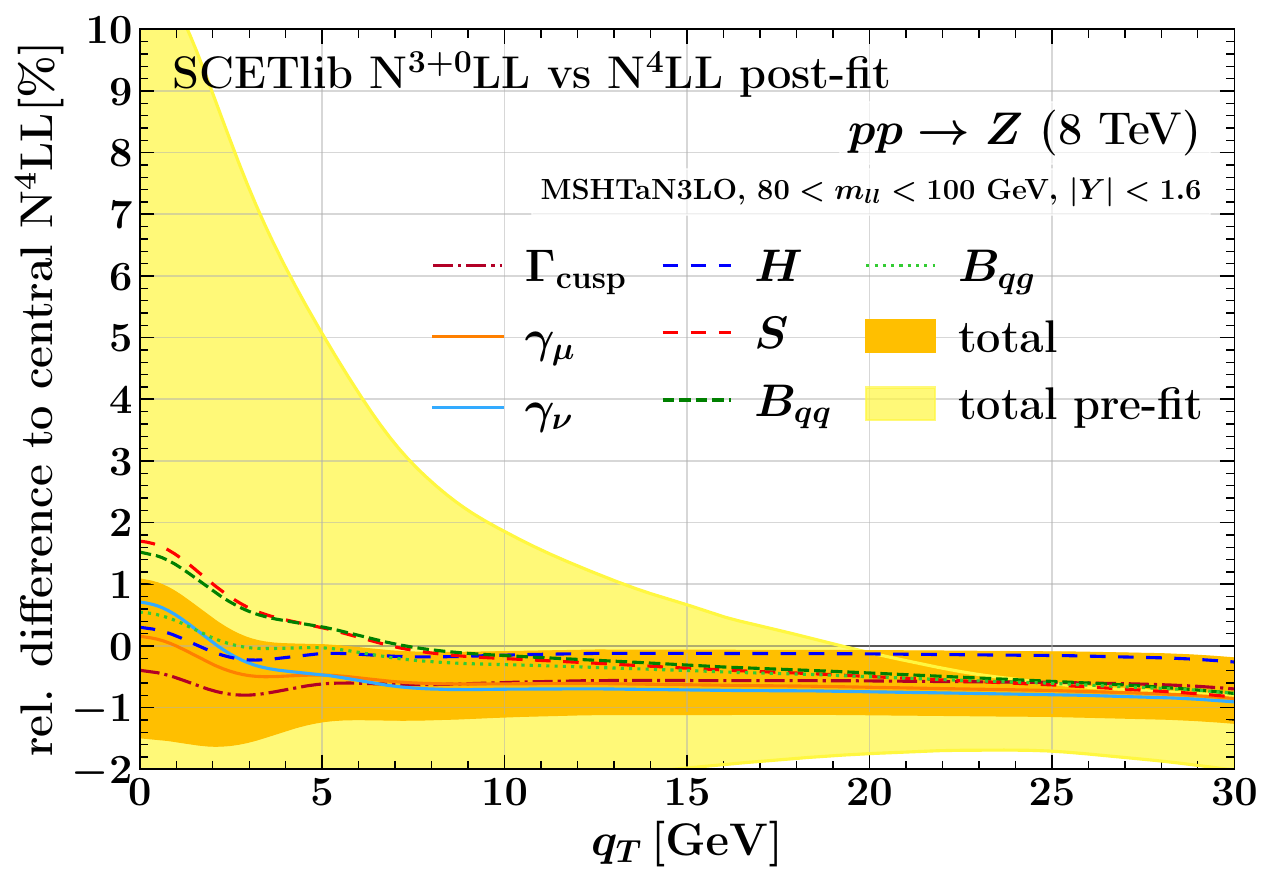}%
\hfill%
\includegraphics[width=0.38\textwidth]{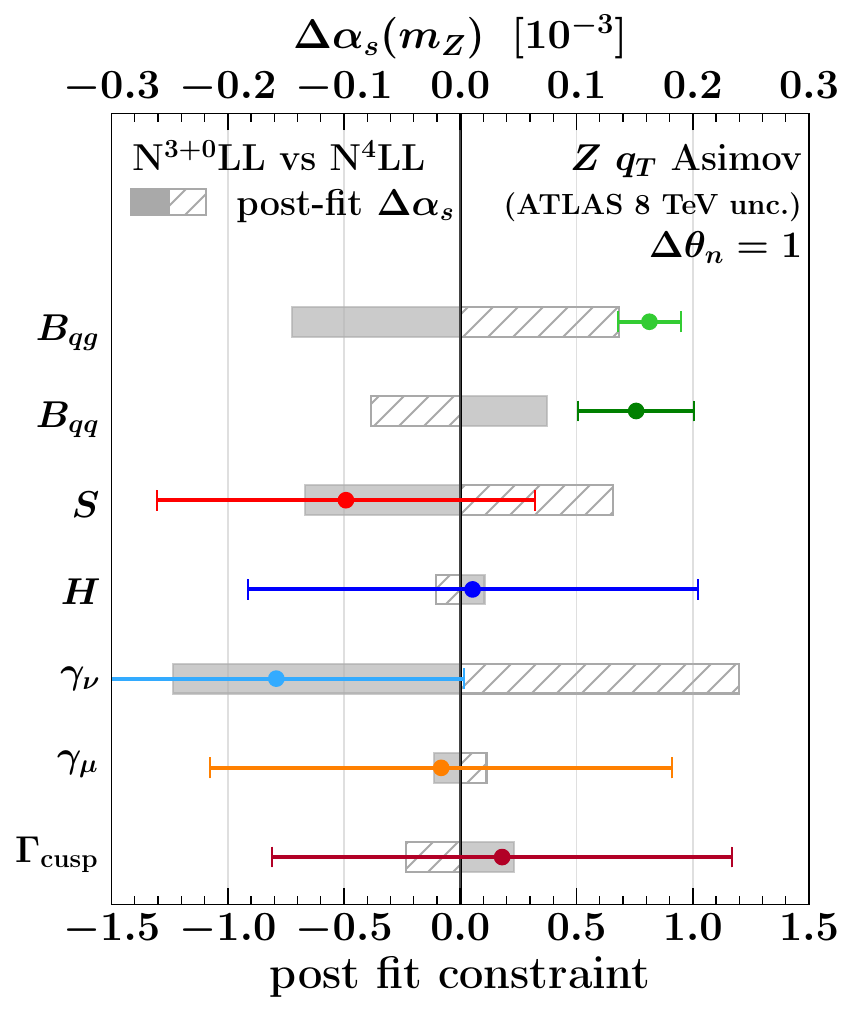}%
\caption{
Left panel: Uncertainties in the $q_T$ spectrum at N$^{3+0}$LL relative to the N$^4$LL,
before (yellow band) and after (orange band) profiling the TNPs.
The different lines show the post-fit relative impact of each TNP.
Right panel: Post-fit constraints on the TNPs (error bars) and their impact on $\as(m_Z)$,
with the solid (dashed) grey band showing the impact of the post-fit downward (upward) TNP variations.
}
\label{fig:tnps_pullplots_N3p0LL_vs_N4LL}
\end{figure}

In \fig{pullplotsasmZ_2}, we already included the results for $\as(m_Z)$ at
N$^{3+0}$LL and N$^{4+0}$LL.
Here, we provide the corresponding plots for the $q_T$ spectrum and
TNP pull plots. The Asimov fit follows the same setup as in \sec{profiling_vs_N4LL}.
The results at N$^{3+0}$LL are shown in \fig{tnps_pullplots_N3p0LL_vs_N4LL}.
In the left panel, we observe a significant reduction in uncertainty from the pre-fit to the
post-fit bands. 
This reduction is particularly strong due to the high degree of correlation in the fit.
Such a reduction may be suspicious, and indeed, the TNP pull plot in the right panel
provides insights into the reasons for this behaviour.
Some TNPs, particularly the beam functions, become strongly constrained by the data.
When the constraints on these parameters become this tight, it suggests that the
considered order may not be sufficient to correctly account for the remaining theory
uncertainties.
This is an example of overfitting the theory model, as previously discussed in \sec{profiling_vs_N4LL}.
Regarding the TNPs' impacts on $\as(m_Z)$, we find that $\gamma_{\nu}$, $S$ and $B_{qg}$ have the strongest
impact on $\as$. This is consistent with what we observed in the N$^{3+1}$LL case.
In contrast, at N$^{2+1}$LL, the pattern of TNP constraints and pulls was 
quite different.

\begin{figure}
\includegraphics[width=0.60\textwidth]{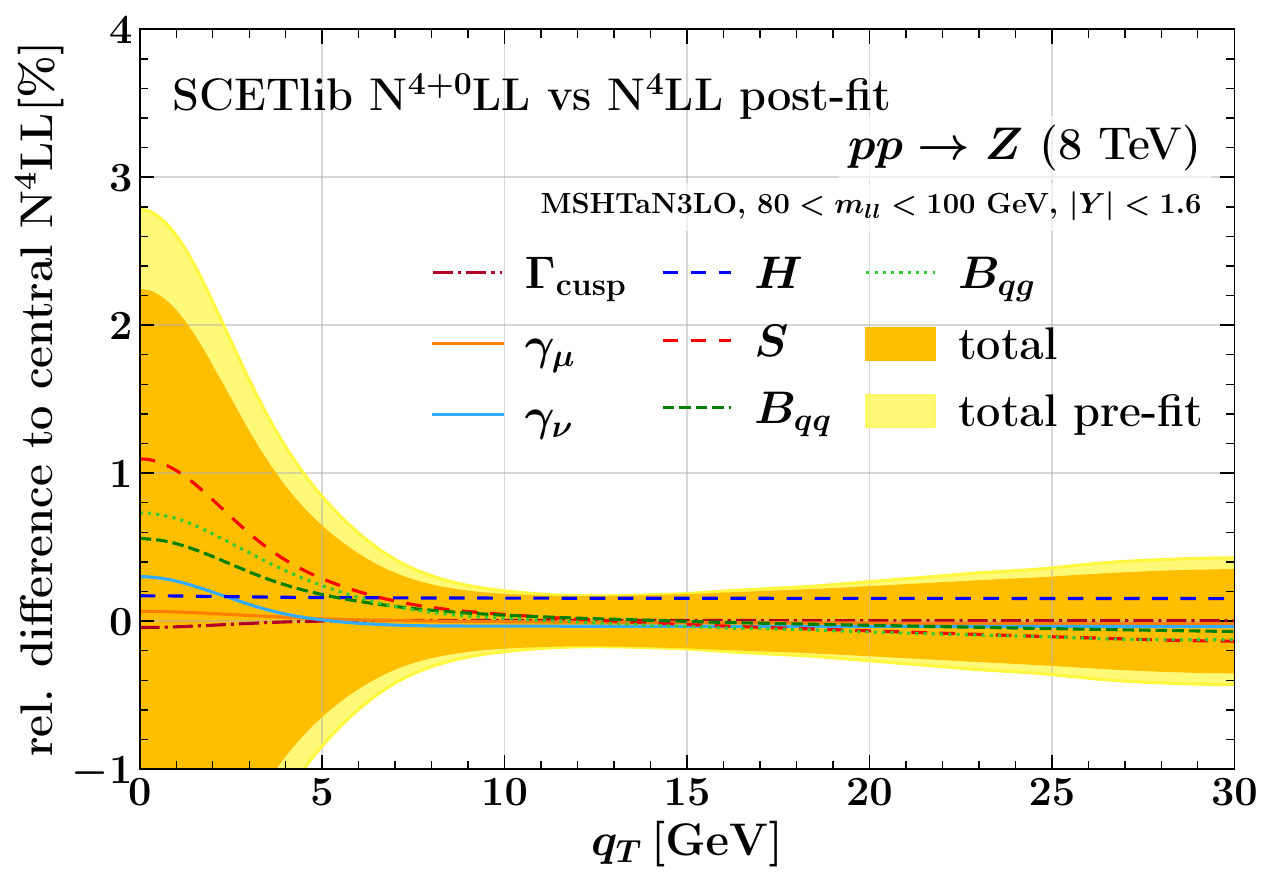}%
\hfill%
\includegraphics[width=0.38\textwidth]{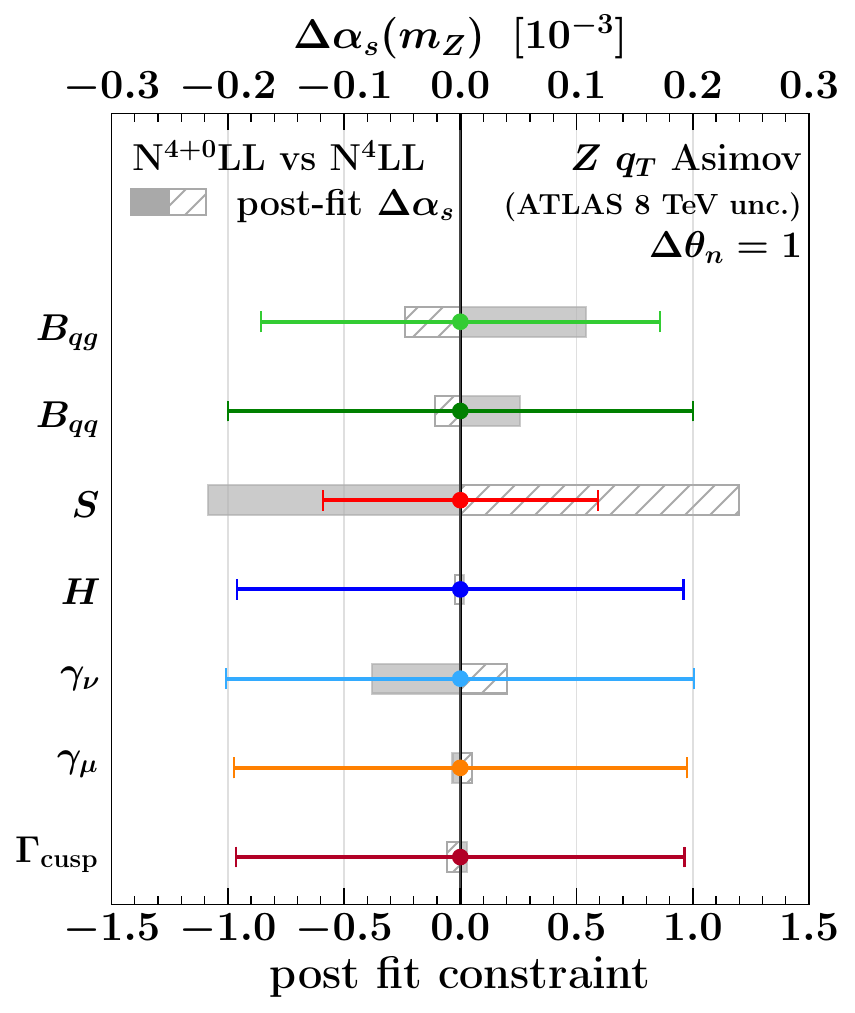}%
\caption{
Left panel: Uncertainties in the $q_T$ spectrum at N$^{4+0}$LL relative to the N$^4$LL,
before (yellow band) and after (orange band) profiling the TNPs. 
The different lines show the post-fit relative impact of each TNP.
Right panel: Post-fit constraints on the TNPs (error bars) and their impact on $\as(m_Z)$,
with the solid (dashed) grey band showing the impact of the post-fit downward (upward) TNP variations.
}
\label{fig:tnps_pullplots_N4p0LL_vs_N4LL}
\end{figure}

The results for the N$^{4+0}$LL case are shown in \fig{tnps_pullplots_N4p0LL_vs_N4LL}.
By construction, the central N$^{4+0}$LL prediction is identical to that of N$^4$LL.
The reduction in uncertainty across the $q_T$ spectrum is less pronounced compared
to the N$^{3+0}$LL case, and similarly the constraints on the TNPs are no longer
as tight as in the N$^{3+0}$LL case.

Both N$^{3+0}$LL and N$^{4+0}$LL may serve as viable alternatives to their respective N$^{m+1}$LL
counterparts.
While the difference in the resummation structure may only have a limited effect on
the overall theory uncertainty,
they can impact the theory correlations, which may not be fully captured.
For this reason, whenever possible, it is preferable to use the full N$^{m+1}$LL prescription.

\addcontentsline{toc}{section}{References}
\bibliographystyle{../jhep}
\bibliography{../scetlib,../refs,../experiments}

\end{document}